\documentclass[12pt]{report}
\usepackage[english]{babel}
\usepackage[T1]{fontenc} 
\usepackage[utf8]{inputenc} 
\usepackage{bbold}
\usepackage{amsmath} 
\usepackage{amsfonts} 
\usepackage{amssymb} 
\usepackage{booktabs} 
\usepackage{textcomp} 
\usepackage[table]{xcolor} 
\usepackage{graphicx}
\usepackage[toc,page]{appendix}
\usepackage[a4paper, margin=2.5cm]{geometry} 
\usepackage{subcaption} 
\usepackage{hyperref} 
\usepackage{dirtytalk}
\usepackage{slashed}
\usepackage{xcolor}
\usepackage{authblk}
\usepackage{mathtools}
\usepackage{float}
\usepackage{tikz, pgfplots}
\usetikzlibrary{positioning}
\usepackage[skins,theorems]{tcolorbox}

\makeatletter
\newcommand\ackname{Acknowledgements}
\if@titlepage
  \newenvironment{acknowledgements}{%
      \titlepage
      \null\vfil
      \@beginparpenalty\@lowpenalty
      \begin{center}%
        \bfseries \ackname
        \@endparpenalty\@M
      \end{center}}%
     {\par\vfil\null\endtitlepage}
\else
  \newenvironment{acknowledgements}{%
      \if@twocolumn
        \section*{\abstractname}%
      \else
        \small
        \begin{center}%
          {\bfseries \ackname\vspace{-.5em}\vspace{\z@}}%
        \end{center}%
        \quotation
      \fi}
      {\if@twocolumn\else\endquotation\fi}
\fi
\makeatother

\pgfplotsset{compat=1.18}

\begin{document}

\begin{titlepage}
	\centering
    \bfseries \LARGE{Radiative tail of solitary waves in an extended Korteweg-de Vries equation} \par
    \vspace{1cm}
    by \par
	\vspace{1.2cm}
	\large\textsc {Muneeb Mushtaq}\par
	\vspace{2cm}
	Supervised by \par
   \vspace{0.5cm}
    $\mathrm{Dr}$. \large\textsc{Gyula Fodor}\par
    \vspace{1.1cm}
    Consultant \par
   \vspace{0.5cm}
    $\mathrm{Dr}$. \large\textsc{Gyula Bene}\par
   \vspace{2.6cm}
    A thesis submitted to the institute of Physics in \\ partial fulfilment of the requirements for the \\ degree of Masters of Science Physics \par \vspace{0.3cm}
    \includegraphics[width=0.2\textwidth]{logo.pdf}\par\vspace{0.2cm}
    E\"otv\"os Lor\'and University, Budapest, Hungary
	\vfill
 May, 2023
\end{titlepage}

\begin{abstract}
    In this thesis, we solve the fifth-order Korteweg-de Vries (fKdV) equation which is a modified Korteweg-de Vries (KdV)  equation perturbed by a fifth-order derivative term multiplied by a small parameter $\epsilon^2$, with $0< \epsilon \ll 1$. Unlike the famous KdV equation, the stationary fKdV equation does not exhibit exactly localized 1-soliton solution, instead it allows a solution which has a well defined central core similar to that of the exactly localized KdV 1-soliton solution, accompanied by extremely small oscillatory standing wave tails on both sides of the core. The amplitude of the standing wave tail oscillations is $\mathcal{O}(\exp(-1/\epsilon))$, i.e. it is beyond all orders small in perturbation theory. The analytical computation of the amplitude of these transcendentally small tail oscillations has been carried out up to $\mathcal{O}(\epsilon^5)$ order corrections by using the complex method of matched asymptotics which has been developed by Segur and Kruskal in Phys. Rev. Lett. $\mathbf{58}$, 747 (1987). Also the long-standing discrepancy between the $\mathcal{O}(\epsilon^2)$ perturbative result of Grimshaw and Joshi in SIAM J. Appl. Math. $\mathbf{55}$, 124 (1995) and the numerical results of Boyd in Comp. Phys. $\mathbf{9}$, 324 (1995) has been resolved. In addition to the stationary symmetric weakly localized solitary wave-like solutions, we analyzed the stationary asymmetric solutions of the fKdV equation which decay exponentially to zero on one side of the (slightly asymmetric) core and blows up to large negative values on other side of the core. The asymmetry is quantified by computing the third derivative of the solution at the origin which also turns out to be beyond all orders small in perturbation theory, i.e. it is $\mathcal{O}(\exp(-1/\epsilon))$. The analytical computation of the third derivative of a function at the origin has also been carried out up to $\mathcal{O}(\epsilon^5)$ order corrections. We use the exponentially convergent pseudo-spectral method to solve the fKdV equation numerically. The analytical and the numerical results show remarkable agreement.
\end{abstract}

\begin{acknowledgements}
    First and foremost, all thanks to Almighty Allah for showering His innumerable blessings and mercy upon me. All glory to Him for providing assistance and support through people and places.

    Most importantly, the contents herein would never have been commenced if it were not for the supervision of Dr. Gyula Fodor who provided a great quantity of immeasurable support, guidance, and motivation. His (seemingly unlimited) patience towards my mistakes is and was well-appreciated. As my teacher and mentor, he has taught me more than I could ever give him credit for here. I would also like to express my sincere regards to Dr. Peter Forgacs for the valuable discussions, his constant suggestions and insights helped to enrich my research work. Their expertise, encouragement, and unwavering commitment to my academic and professional development have been instrumental in shaping this research. Their insightful feedback, constructive criticism, and continuous support have pushed me to strive for excellence. I would also like to express my sincere gratitude to my university consultant Dr. Gyula Bene for his administrative support throughout this project.

    To the Tempus Public Foundation for awarding me the scholarship that have made my masters studies possible in Hungary.

    Nobody has been more important to me in the pursuit of this project than the members of my family. I would like to express my sincere appreciation to my whole family especially parents, whose love and guidance are with me in whatever I pursue. Their emotional support and patience during the challenging moments of this journey have been invaluable. Most importantly, I would like to express my heartfelt gratitude and give special thanks to my beloved grandmother for her unwavering love, support, and countless prayers throughout my journey. Her presence in my life has been a source of strength and inspiration, guiding me through both the joys and challenges I have encountered.

    To Isma Irshad, thank you for being my constant cheerleader throughout this process. Your belief in my abilities, your words of motivation, and your willingness to lend an ear during the challenging moments have meant the world to me.

    To Umair Ashraf and Faizan Hilal Lone, your friendship has been a source of inspiration. Your intellectual curiosity and passion for learning have motivated me to push my boundaries and explore new perspectives. Thank you for engaging in countless discussions and offering valuable insights that have shaped the ideas presented in this thesis.

    Finally, I express my sincere gratitude to the researchers and scholars whose works have laid the foundation for this thesis. Their groundbreaking contributions, pioneering research, and insightful publications have shaped the landscape of this field and have been fundamental in shaping my own understanding and ideas.
\end{acknowledgements}

\tableofcontents

\chapter{Introduction}\label{chapter:introduction}

Scientifically, the initial observation of a solitary wave was made by Scottish engineer and naval architect, John Scott Russell in the year 1834. He observed a type of water wave which did not disperse. He also  noticed that depending on its amplitude, an initial arbitrary waveform set in motion in the channel evolves into two or more waves that move at different velocities and progressively move apart until they form individual solitary waves. Russell called this wave as \say{great wave of translation}. At that time, Russell's observations puzzled the scientists as these could not be explained by the linear wave theories of which the main feature is dispersion, i.e. inside the disturbance, different Fourier modes travel at different speeds. Although only a year later, his observations were severely criticised by the two well known scientists of that time namely Sir G. B. Airy whose main argument was that the formula derived by the J. Scott Russell did not agree with the shallow water waves (also called long waves whose wavelength is much greater than the water depth) and G. G Stokes whose main argument was that the solitary waves could not exist in a non viscous fluid. However in 1870's two other scientists Joseph Valentine de Boussinesq and Lord Rayleigh confirmed that the Russell's observations are correct. The former scientist gave the mathematical explanation while the latter gave the physical reason for Russell's observation. In 1895, Korteweg and de Vries found the basic governing non-linear equation of shallow water gravity waves and this equation got their name as Korteweg-de Vries (KdV for short) equation which describes the properties of the Russell's solitary waves. 

A solitary wave is a wave which propagates without any temporal evolution in shape or size when viewed in the reference frame moving with the wave. The core of these waves is spatially localized which means that it decays exponentially fast far from the maximum of the structure. The solitons, which are non-linear solitary waves are the solutions of the \say{integrable} non-linear partial differential equations with the additional property that the wave retains its permanent structure, even after interacting with another solitons. These soliton solutions form a special class of solutions of model equations, including the Sine-Gordon, the Nonlinear Schr{\"o}dinger and the KdV  equations.

After derivation of the KdV equation by D. Korteweg and G. de Vries in 1895, it was not until 1960's that the development in the nonlinear waves emerged when N.J. Zabusky and M.D. Kruskal \cite{zabusky} initiated an explosion of research in this field. While doing research on the propagation of nonlinear waves in plasma, they noticed that the impulses (solitary waves) of different amplitudes (and hence of different velocities) does not change their properties after collisions with each other. They introduced the term \emph{soliton} for such kind of waves as their behaviour resembles with that of the particles. Three years later, Zabusky \cite {zabusky1} in the numerical simulations observed a train of solitons of decreasing amplitudes from an initial cosine wave, evolving according to KdV equation which gave rise to an intensive research in which multi-soliton solutions for several kinds of nonlinear wave equation were discovered. The general method for construction of such multi-soliton solutions is called the \emph{Inverse Scattering Transform} method \cite{gard, lax, miura}.

The celebrated KdV equation which is only valid for weak dispersion and weakly nonlinear waves, for example the surface gravity waves, admits stationary travelling wave solutions which decays exponentially fast far from the maximum. However due to some practical concerns, for instance, to develop a model which takes into account the higher orders of dispersion and/or nonlinear effects, it is necessary to modify the KdV equation. It is known from the literature, e.g. see \cite{hs}, that the KdV equation does not exhibit the traveling solitary wave solutions when modified by higher order dispersion effects, for instance when we add a fifth order derivative term multiplied by a small parameter $\epsilon^2$ to it. This extended KdV equation is called ``fifth-order KdV'' (fKdV for short) equation which models the long wavelength waves such as capillary-gravity waves where in addition to gravity, surface tension can also be no longer neglected. Such KdV-type waves are not strictly localized, but are accompanied by co-propagating small oscillations which spreads out to infinity without decay. These KdV-type waves may occur for water waves with surface tension when the Bond number is approximately but slightly less than $1/3$. The Bond number is a dimensionless number which measures the relative importance of the two forces, gravity and surface tension, given as $B=T/(\rho g h^2)$, where $T$ is the surface tension, $\rho$ is the density of the liquid, $g$ is gravity constant and $h$ is the depth of the liquid from the undisturbed surface. The amplitude of the far field oscillating tails may be exponentially (or algebraically) small in a small parameter $\epsilon$. In the literature, see e.g. \cite{boyd1, boyd2}, these coherent structures are referred to as \emph{nanopterons} (or \emph{micropterons}).

\section{The KdV equation: A brief introduction}\label{section:KdV}
In 1895, Diederik Korteweg and Gustav de Vries formulated a mathematical model to explain Russel's observation. They derived a non-linear equation that arises as an approximate equation that models the theory of long water waves propagating in a shallow channel, such as a canal. The original form of the famous KdV equation in terms of the physical parameters is given as \cite{jager, miura1},
\begin{equation}
    \frac{\partial \eta}{\partial \tau}=\frac{3}{2}\sqrt{\frac{g}{h}}\frac{\partial}{\partial \xi}\left(\frac{1}{2}\eta^2+\frac{2}{3}\alpha \eta+\frac{1}{3}\sigma\frac{\partial^2\eta}{\partial\xi^2}\right) \ ,
\label{eqn:KdV1}
\end{equation}
where $\sigma=h^3/3-Th/(\rho g)$ with surface tension $T$ and the density of the liquid $\rho$, $\eta$ is the surface elevation of the wave above the undisturbed level $h$, $\alpha$ is the small arbitrary (dimensionless) constant which is related to the uniform motion of liquid, see \cite{jager}, $g$ is the gravitational acceleration (about $9.81 \ m/s^2$ at sea level), and $\tau$ and $\xi$ are the temporal and spatial coordinates respectively.

\begin{figure}[ht!]
	\centering
        \includegraphics[width=0.75\linewidth]{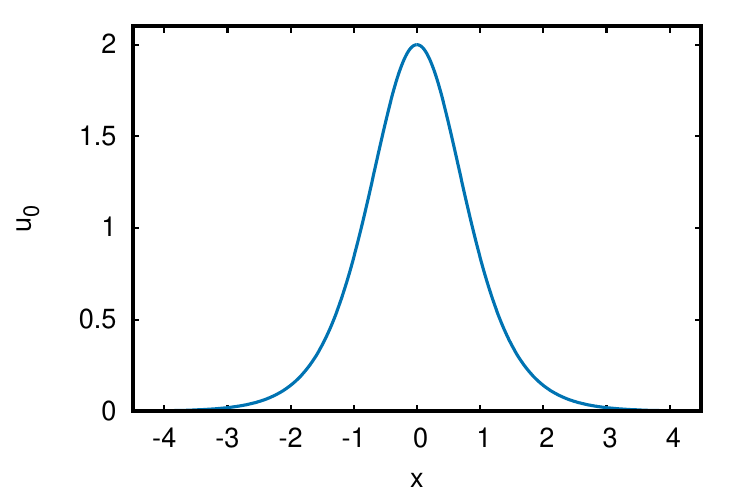}
        \caption{
            The graph of an exact 1-soliton solution $u_{0}$ of the KdV equation as a function of $x$ for $\gamma = 1$.
    \label{fig:u0fig}}
\end{figure}

Equation (\ref{eqn:KdV1}) is an approximation that holds under a set of restrictive set of conditions. The main assumptions made by Korteweg and de-Vries for the derivation of waves in a liquid having wavelength $\lambda$ are:
\begin{enumerate}
    \item Viscous effects are neglected, 
    \item bottom of the fluid is even (flat), i.e. the height of the fluid from the bottom surface is constant,
    \item weak dispersion is assumed, i.e. as compared to the undisturbed depth, the wave are long waves (also called shallow waves), $\displaystyle \frac{h}{\lambda}\ll 1$,
    \item weak non-linearity is assumed, i.e. the amplitude of the waves is small, $\displaystyle \delta = \frac{\eta}{\lambda}\ll 1$,
    \item weak dispersion and weak non-linear effects approximately balance, i.e. $\displaystyle \frac{h}{\lambda} = \mathcal{O}(\delta)$
\end{enumerate}
If we make the transformations 
\begin{equation}
    t = \frac{1}{2}\sqrt{\frac{g}{h \sigma}} \ \tau \, , \quad  y=-\frac{\xi}{\sqrt{\sigma}} \, , \quad \text{and} \quad u = \frac{1}{2}\eta+\frac{1}{3}\alpha \ ,
\end{equation}
then the KdV equation, equation (\ref{eqn:KdV1}) can be recast in dimensionless variables as 
\begin{equation}
    u_{yyy}+6uu_y+u_t=0 \ ,
\label{eqn:KdV2}
\end{equation}
 where $u$ is a dependent variable and the independent variables $y$ and $t$ respectively measures distance and time respectively. The subscripts denotes the partial derivative of the function $u$ with respective coordinates. Equation (\ref{eqn:KdV2}) is an integrable equation and its stationary solution in a frame moving with speed $c_0$ can be found analytically. Introducing the co-moving coordinate, $x=y-c_0t$, which is used as a spatial coordinate, we obtain the reduced KdV equation
\begin{equation}
   u_{xxx}+(6u-c_0)u_x=0 \ , 
\label{eqn:reducedKdV}
\end{equation}
which is a third-order nonlinear ordinary differential equation. Equation (\ref{eqn:reducedKdV}) shows that if $u$ is the solution moving with velocity $c_0$, then $\bar u = u + \mathcal{C}$, where $\mathcal{C}$ is a constant, is also a solution which moves with different velocity $\bar c_0 = c_0 + 6\mathcal{C}$. Note, however that this transformation changes the asymptotic value of the function $u$. Integrating equation (\ref{eqn:reducedKdV}) once with respect to $x$, we get
\begin{equation}
   u_{xx}+3u^2-c_0u=\zeta \ ,
\label{eqn:KdV3}
\end{equation}
where $\zeta$ is a constant which can be taken as zero if $u$ is a localized solution which tends to zero at large distances. This can also be done by assuming that the function $u$ and its second derivative vanishes at $|x|\to \infty$. Then equation (\ref{eqn:KdV3}) can be written as
\begin{equation}
    u_{xx}+3u^2-c_0u=0 \ .
\label{eqn:kdvc0}
\end{equation}
Equation (\ref{eqn:kdvc0}) admits a single soliton solution $u_0$ centered at $x=0$ which decays exponentially to zero at large $|x|$, see Figure (\ref{fig:u0fig}) and is given as
\begin{equation}
    u_0=\frac{c_0}{2} \ \mathrm{sech}^2\left(\sqrt{\frac{c_0}{4}} \ x\right) \ ,
\label{eqn:solitaryKDV}
\end{equation}
where $c_0 > 0$ is the solitary-wave speed. The constant $c_0$ cannot be negative so there are no dip-like soliton solutions of the KdV equation. If we define the positive constant $\gamma$ as 
\begin{equation}
    c_0=4\gamma^2 \ ,
\label{eqn:c0gam}
\end{equation}
then $u_0$ from equation (\ref{eqn:solitaryKDV}) can be written as,
\begin{equation}
   u_{0}=2\gamma^2\mathrm{sech}^2(\gamma x) \ ,
\label{eqn:u0}
\end{equation}
with the hyperbolic secant defined as $\mathrm{sech}(x)=1/\mathrm{cosh}(x)$ and $2\gamma^2$, the amplitude of the KdV soliton. In a lab frame, $u_0$ can be written as, $u_0(y,t)=2\gamma^2\mathrm{sech}^2(\gamma y-4\gamma^3 t)$. So, $u_0(y,t)$ has a positive propagation speed $4\gamma^2$ which is proportional to the amplitude of the wave.

The 1-soliton solutions $u_0$ given in equation (\ref{eqn:u0}) are the one-parameter family solutions characterized by the parameter $\gamma$, which controls the amplitude, the width of the soliton core and the phase speed $c_0$ of the solitary wave.

We can understand what are the requirements for the solitons to exist if we understand the physical role of different terms in equation (\ref{eqn:KdV2}). The first and second term in this equation are the dispersion and the non-linear terms respectively. What happens if we remove one of these terms from equation (\ref{eqn:reducedKdV}). If we eliminate the dispersion term, then we get $6uu_y+u_t=0$. If we compare it with the linear equation $c u_y+u_t=0$, one can consider the coefficient of $u_y$ to determine the propagation speed of the wave. As a result the larger the amplitude of the pulse the greater the speed. Hence this case favours the formation of shock waves (waves which exhibit discontinuities in a finite time) and the result is breaking. If we eliminate the non-linear term, e.g. in the far field region, we get $u_{yyy}+u_t=0$. The result in such a case is dispersion.  The existence of the soliton solutions results from the  stable equilibrium between the non-linearity, which tends to localise the wave and the dispersion, which spreads out the wave. 

For the linearized stationary KdV equation, $u_{xx}-c_0u = 0$, the two solutions (modes) are $\exp{(\pm 2\gamma x)}$ which is the exponentially growing mode which corresponds to the growing behaviour at large $x$ and the other one is the exponentially decaying mode which corresponds to the decaying behaviour at large $x$. The solution in the far field is a linear combination of these two modes. However if we assume symmetry at the origin we then actually have only one free parameter which is $u(x=0)$ as $u_x(x=0)=0$. While changing the value of this parameter at the origin there is a numerical plausibility to suppress the exponentially growing mode at large $|x|$. This suggests that there is a numerical possibility of the symmetric exponentially decaying modes for the linearized KdV equation.

Just to mention, there are many other solutions of equation (\ref{eqn:KdV2}) such as multisoliton solutions. The interactions of the solitons in such solutions is quite interesting as the equation is non-linear and hence the superposition principle is not applicable. The properties like, shape and velocity of solitons having different amplitudes in the multisoliton solutions before and after interactions are same except there is a phase change during this interaction and the joint amplitude of the interacting solitons decreases which is obviously in contradiction to what would happen if two waves overlapped linearly \cite{zabusky}. We are not going to discuss these solutions in detail here as these these type of solutions are not so relevant to the topic of this thesis. For detailed discussion of such solutions, an interested reader may refer to e.g. \cite{solitons}.

\section{The fifth-order KdV equation}\label{section:FKdV}

In order to describe the higher nonlinear and/or higher order dispersion models, we need to modify the KdV equation, equation (\ref{eqn:KdV2}) which is valid only for weakly nonlinear and weak dispersion models. When higher order effects are taken into account we get an extended KdV equation \cite{yang}
\begin{equation}
   \epsilon^2(u_{yyyyy}+\nu_1uu_{yyy}+\nu_2u_{y}u_{yy}+\nu_3u^2u_y)+u_{yyy}+6uu_y+u_t = 0 \ ,
\label{eqn:extendedKdV}
\end{equation}
where $\epsilon > 0$ is a small parameter and $\nu_n$, $n=1,2,3$ are constant coefficients whose value depends on the physical situation. The values of these coefficients decides the integrability of equation (\ref{eqn:extendedKdV}). For instance if $\nu_1 = 15$, $\nu_2 = 15$, and $\nu_3 = 45$ then equation (\ref{eqn:extendedKdV}) with these coefficients is called the Coudrey-Dodd-Gibbon (also called Sawada-Kotera) equation which is an integrable (exactly solvable) equation \cite{yang}. For small amplitude capillary-gravity waves on shallow water with a Bond number approximately equal to but slightly less than $1/3$, Hunter and Scheurle \cite{hs} have shown that all the three constant coefficients are equal to zero. In this case the resulting (non-integrable) equation is called fifth-order KdV (fKdV for short) equation \cite{boyd1, boyd2, boyd, Muneeb, gaj} which is given as
\begin{equation}
   \epsilon^2 u_{yyyyy}+u_{yyy}+6uu_{y}+u_{t}=0 \ , 
\label{eqn:tFKDV}
\end{equation} 
where $u$ is a function of spatial coordinate $y$ and time coordinate $t$ and the indices denote the partial derivatives with respective coordinates. This equation is a non-integrable fifth-order nonlinear Partial Differential Equation (PDE) and therefore one cannot find its exact solutions analytically. If higher order nonlinear terms have also been included in addition to the higher dispersion term, we may call these model equations as fKdV-like equations. The study of such model equations is beyond the scope of the present  work.

\begin{figure}[ht!]
	\centering
    \includegraphics[width=0.75\linewidth]{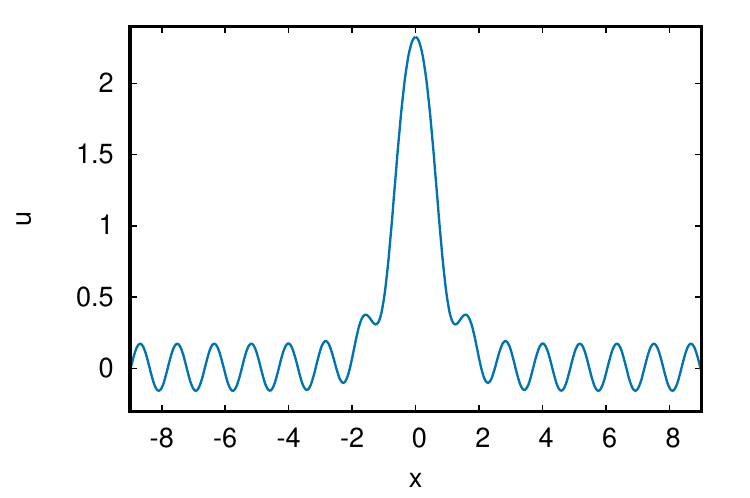}
    \caption{
        The graph of the symmetric weakly localized solution $u$ as a function of $x$. This is a true numerical solution of the fKdV equation (\ref{eqn:xFKDV}) for $\epsilon=0.2$.
    \label{fig:unum}}
\end{figure}

Equation (\ref{eqn:tFKDV}) is extensively studied in several papers including \cite{asym, boyd1, boyd2, boyd, Muneeb, gaj, prg, sun} and these works show that unlike the KdV equation, equation (\ref{eqn:tFKDV}) does not exhibit exactly localized solitary wave solutions for positive velocity, instead it allows only weakly-localized solitary wave-like solutions which has a central core almost same as the KdV 1-soliton solution, accompanied by extremely small oscillatory tail, see Figure (\ref{fig:unum}). Benilov, Grimshaw, and Kuznetsova \cite{asym}, studied the non-stationary solutions of equation (\ref{eqn:tFKDV}) analytically as well as numerically in which they demonstrated that these non-stationary solutions are not exactly localized and they also calculated the exponentially small energy loss rate of amplitude of the central core in the limit as the parameter $\epsilon\to 0$. However most of the attention has been focused on the stationary solutions of equation (\ref{eqn:tFKDV}), see the analytical studies by Pomeau, Ramani, and Grammaticos \cite{prg}, Grimshaw and Joshi \cite{gaj}, and Sun \cite{sun}, and the numerical studies by Boyd \cite{boyd1, boyd2, boyd}. If we introduce the co-moving coordinate, $x=y-ct$, then from equation (\ref{eqn:tFKDV}), we get the stationary fKdV equation which is a fifth-order Ordinary Differential Equation (ODE) given as
\begin{equation}
   \epsilon^2 u_{xxxxx}+u_{xxx}+(6u-c)u_x=0 \ ,
\label{eqn:xFKDV}
\end{equation} 
where the wave speed $c>0$. If $c<0$ then there are only oscillatory solutions for above equation for large $x$. In this case the solution having a well defined core and small oscillatory tail is not likely to exist. Equation (\ref{eqn:xFKDV}) shows that if $u$ is a solution, then $\bar u = u + \mathcal{C}$, where $\mathcal{C}$ is a constant is also a solution moving with different velocity $\bar c = c + 6\mathcal{C}$. Note, however that this transformations changes the asymptotic value of the function $u$.

Two types of stationary weakly localized solitary wave solutions have been discussed in the literature. The symmetric weakly localized solitary wave solutions has a central core, accompanied by standing wave tail of small amplitude on both sides of the core, and the asymmetric weakly localized solitary wave solutions which decays to zero on one side of the core and may possibly have a tail in the other direction. Pomeau et al. \cite{prg} examined both types and they found the leading order analytic expression for the tail amplitude asymptotically in the limit $\epsilon\to 0$. Boyd \cite{boyd1} attempted to calculate numerically  the stationary asymmetric solutions, but he was unsuccessful which led him to conclude that equation (\ref{eqn:xFKDV}) allows only symmetric stationary weakly localized solitary wave solutions. Later Benilov et al. \cite{asym} showed that the stationary asymmetric solutions with small but different amplitude tails on the two sides contradicts the conservation of energy flux, which indicates the non-existence of these type of solutions.

In this thesis, our main interest is focused on the stationary solutions of equation (\ref{eqn:tFKDV}) traveling with speed $c$ to the right. The main reason to focus on the stationary fKdV equation is that it is much simpler to solve than the non-stationary fKdV equation. However, due to identical oscillatory tails on both sides of the core, there will be in general non-zero energy flux. Such symmetric solitary wave-like solutions can not be physically realized since they require an energy source on one end of the wave and the energy sink on the other side. In practice these waves appear as non-stationary asymmetric weakly localized solitary wave solutions which has a central core, accompanied by small oscillatory travelling wave tails only on one side either behind or in front of the core. These asymmetric solutions are not steady as they are radiating by extremely small tails in the direction of propagation and hence the amplitude of the core decreases \cite{asym}. However we can calculate the amplitude of the radiation tail of these non-stationary asymmetric solutions of the more difficult time dependent fKdV equation (\ref{eqn:tFKDV}) by solving the comparatively easier time independent fKdV equation (\ref{eqn:xFKDV}). We expect that the oscillatory tail amplitude of the stationary symmetric solutions approximates the radiation tail amplitude of the non-stationary asymmetric solutions.

\begin{figure}[ht!]
	\centering
    \includegraphics[width=0.7\linewidth]{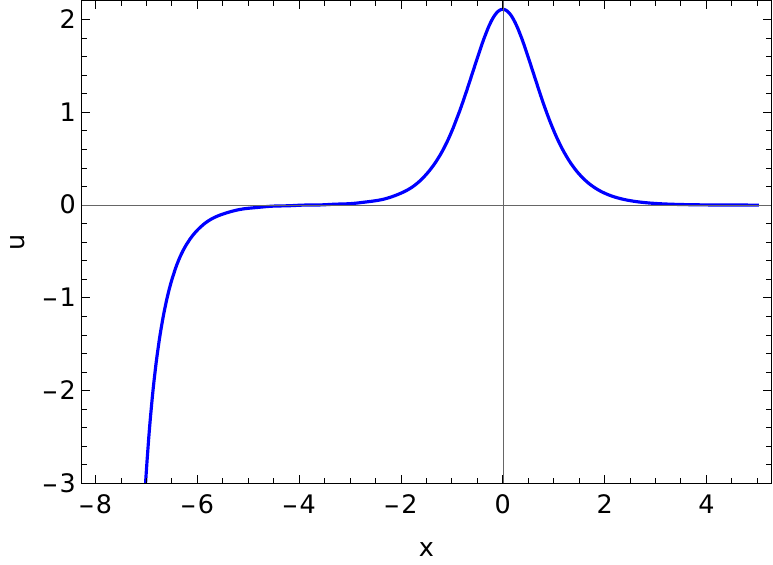}
    \caption{
        The graph of the asymmetric solution $u$ as a function of $x$. This is a true numerical steady asymmetric solution of the fKdV equation (\ref{eqn:xFKDV}) for $\epsilon=0.1$.
    \label{fig:uasym}}
\end{figure}

The stationary fKdV equation (\ref{eqn:xFKDV}) also allows asymmetric solutions in addition of the symmetric weakly localized solitary wave solutions which we discussed already, see Figure (\ref{fig:unum}). The asymmetric solutions do not have any oscillatory tail either side of the asymmetric core, instead they decay exponentially to zero on one side of the core, and  blow up at finite $x$ value on the other side, see Figure (\ref{fig:uasym}). According to our knowledge explicit numerical results for these (blow up) asymmetric solutions have not been presented. Originally, the existence of asymmetric solutions with an oscillatory tail on one side were expected.

Integrating equation (\ref{eqn:xFKDV}) once with respect to $x$, we obtain
\begin{equation}
    \epsilon^2 u_{xxxx}+u_{xx}+3u^2-cu=M \ ,
\label{eqn:1intFKDV}
\end{equation}
where $M$ is an integration constant which can be taken as zero without any loss of generality. This can be achieved either by imposing the solitary wave boundary conditions ($u\to 0$ as $|x|\to\infty$) or by translating function $u$ by some constant $\eta$. If we shift function $u$ as $\bar{u}=u+\eta$, then the new constant $\bar{M}=M+c\eta-3\eta^2$ and hence $M$ can be transformed to zero if $c^2 +12M >0$. If this condition is not satisfied then no solutions of equation (\ref{eqn:1intFKDV}) exists which goes asymptotically to some constant. From now on, we will concentrate on equation (\ref{eqn:1intFKDV}) for $M=0$
\begin{equation}
    \epsilon^2 u_{xxxx}+u_{xx}+3u^2-cu=0 \ ,
\label{eqn:1aintFKDV}
\end{equation}
where in general, the parameters $c$ and $\epsilon$ are independent of each other. 

Equation (\ref{eqn:1aintFKDV}) is invariant under the rescalings
\begin{equation}
    u=\frac{\tilde u}{\chi^2} \ , \quad x=\chi \tilde x \ , \quad c=\frac{1}{\chi^2}\tilde c \ , \quad \epsilon=\chi \tilde \epsilon \ ,
\label{eqn:rescalings}
\end{equation}
for any constant $\chi>0$. Using the above rescaling, $\epsilon^2 c$ remains invariant and this gives us freedom to scale either of the constants $\epsilon$ or $c$ to some given value, and consider equation (\ref{eqn:1aintFKDV}) as an equation containing only one free parameter. We can choose a parameter $c$ such that it is $\epsilon$ dependent as shown by R. Gimshaw and N. Joshi \cite{gaj}. We denote the parameter $c(\epsilon)$ by $c_0$ in the limit $\epsilon\to 0$. For $\epsilon=0$, equation (\ref{eqn:1aintFKDV}) becomes the KdV equation and has the well-known solitary wave solution $u_0$ given in equation (\ref{eqn:u0}), which approximates the core region of fKdV equation. This is important because in all the numerical  calculations, we have chosen the value of $c_0=4$ (and consequently $\gamma=1$). By using the rescaling, equation (\ref{eqn:rescalings}), we can set the value of $c$ (and consequently value of $\gamma$) to any other possible value.

\subsection{Linear asymptotic analysis} \label{subsection:linearasymp}

 The far field region is defined to be that region where the core of solitary wave has decayed to such small amplitude that the dynamics in this region is accurately described by linear wave equation. In this region, we can neglect the $\mathcal{O}(u^2)$ terms. For $c>0$, the linearized version of equation (\ref{eqn:1aintFKDV}), $\epsilon^2 u_{xxxx}+u_{xx}-cu=0$ has four linearly independent solutions of the form $u\sim e^{i\tilde{k} x}$, which gives the dispersion relation as
\begin{equation}
    c=-\tilde{k}^2+\epsilon^2 \tilde{k}^4 \ ,
\label{eqn:disp}
\end{equation}
which gives 
\begin{equation}
    \tilde{k}^2=\frac{1\pm \sqrt{1+4 \epsilon^2 c}}{2\epsilon^2} \ .
\label{eqn:lambda}
\end{equation}
The positive sign in equation (\ref{eqn:lambda}) gives the two real roots, while the negative sign gives the two imaginary roots. The two imaginary solutions of equation (\ref{eqn:disp}) are 
\begin{equation}
    \tilde{k}^2_{\mathrm{Img}}=\frac{1 - \sqrt{1+4 \epsilon^2 c}}{2\epsilon^2} \ ,
\label{eqn:imgktilde}
\end{equation}
and the two real solutions of equation (\ref{eqn:disp}) are
\begin{equation}
    \tilde{k}^2_{\mathrm{real}}=\frac{1 + \sqrt{1+4 \epsilon^2 c}}{2\epsilon^2} \ .
\label{eqn:realktilde}
\end{equation}
So for the linearized fKdV equation, we have four modes which are, the exponentially decaying mode, the exponentially growing mode and the two linearly independent oscillatory modes. The general solution to the linearized problem is the linear combination of these four modes. However for the symmetric solutions, we have only two free parameters at the origin which are $u(x=0)$ and $u_{xx}(x=0)$ as the first and third derivative of the function $u(x)$ at the origin are zero due to the symmetry. To get the exponentially decaying mode, we need at least three free parameters at the origin to suppress the other three modes at large $|x|$. Hence the exponentially decaying soliton like mode is not likely for the fKdV equation (\ref{eqn:1aintFKDV}). However to get the oscillatory solution at large $|x|$, we just need just one free parameter to suppress the exponentially growing mode at large $|x|$. So the oscillatory mode is the plausible solution for the linearized fKdV equation. This oscillatory solution near the far field is the linear combination of two linearly independent oscillatory modes.

Now what about if the fifth-derivative term in fKdV equation is negative, then the dispersion relation for the linearized version of the same equation of which the solution can also be written in the form $u\sim e^{i\lambda x}$ would be
\begin{equation}
    \epsilon^2 \lambda^4 + \lambda^2 + c = 0
\label{eqn:disersonnegative5}
\end{equation}
which has four solutions for $\lambda$ which are given as
\begin{equation}
    \lambda = \pm\frac{\sqrt{-1\pm p}}{\sqrt{2}\epsilon} \, , \quad \ p = \sqrt{1-4\epsilon^2 c} \, .
\label{eqn:tildeknegative5}
\end{equation}
As the parameter $\epsilon$ is very small, which implies $p < 1$ and hence all the four solutions are purely imaginary which allows the two exponentially decreasing solutions of fKdV equation, and no exponentially growing solution for large $|x|$. Hence the exactly localized solitary wave (KdV 1-soliton) solutions for the negative fifth-order derivative fKdV equation are likely to exist.

Let us go back to equation (\ref{eqn:1aintFKDV}). For the  exponentially growing and decaying solutions, $\tilde{k}$ must be imaginary. Let $\tilde{k}_{\mathrm{Img}}=\xi i$, where, $\xi \in \mathbb{R}$. We are looking for solutions $u$ which asymptotically become small. Using equation (\ref{eqn:disp}), we obtain $\xi|_{\epsilon=0}=\pm 2\gamma$, where $\gamma$ is a constant defined in equation (\ref{eqn:c0gam}). The two imaginary solutions are
\begin{equation}
    \tilde{k}_{\mathrm{Img}}|_{\epsilon=0}=\pm 2i\gamma \, .
\label{eqn:limag}
\end{equation}
The $\epsilon$ independence of $\tilde k_{\mathrm{Img}}$ is just an assumption that will determine the $\epsilon$ dependence of phase speed $c$. The imaginary solution $\tilde{k}_{\mathrm{Img}} = -2i\gamma$ give us the exponentially growing behaviour of the solution and for decaying behaviour of the solitary wave, we have $\tilde{k}_{\mathrm{Img}}=2 i\gamma$. Substituting $\tilde k_{\mathrm{Img}}$ in equation (\ref{eqn:disp}), we get an exact $\epsilon$ dependence of the phase speed $c$ which is true for both the growing and decaying solution as
\begin{equation}
    c(\epsilon)=4\gamma^2+16\gamma^4\epsilon^2 \, .
\label{eqn:exactgamc}
\end{equation}
Now, for oscillatory behaviour $\tilde{k}$ is real and we have two such solutions which can be obtained by substituting $\tilde{k}_{\text{Img}}=2 i\gamma$ into equation (\ref{eqn:imgktilde}). In this way we get $\sqrt{1+4\epsilon^2 c}=1+8\gamma^2\epsilon^2$. Substituting this into equation (\ref{eqn:realktilde}), we obtain
\begin{equation}
    \tilde{k}_{\text{real}}=\pm\frac{k}{\epsilon} \, ,
\label{eqn:lreal}
\end{equation}
where
\begin{equation}
    k=\sqrt{1+4\epsilon^2 \gamma^2} \, .
\label{eqn:kappa}
\end{equation}
We expect the frequency of the far field oscillations to be very high for $\epsilon\ll 1$, which is also clear from the equation (\ref{eqn:lreal}). This equation tells us that the wave number (spatial frequency) of tail of the nanopteron tends to infinity when $\epsilon$ tends to zero.

Using equation (\ref{eqn:lreal}), the asymptotic form of the solution of the fKdV equation, equation (\ref{eqn:1aintFKDV}) far from the core region can be written as
\begin{equation}
    u\approx \alpha_{\pm}\sin\left(\frac{k}{\epsilon}|x|-\delta_{\pm}\right)\qquad\text{(Expected tail of solution)} \ ,
\label{eqn:uwithamp}
\end{equation}
provided $0 < \epsilon \ll 1$, where $\alpha_{\pm}$ and $\delta_{\pm}$ are the oscillatory tail amplitude and far field phase factor respectively in the far region of positive and negative direction. The important point to note here is that, in general the values of $\alpha$ and $\delta$ can be different in the left and right asymptotic regions. However for symmetric solutions, $\alpha_+=\alpha_-$ and also we can make $\delta_+=\delta_-$ by shifting the variable $x$. Hence for symmetric solutions, the four unknown constants, $\alpha_\pm$ and $\delta_\pm$ in equation (\ref{eqn:uwithamp}) decreases to just two, i.e. $\alpha$ and $\delta$. The far field oscillations or the tail region of the solution $u$ in the positive and negative directions can be written as 
\begin{equation}
    u \approx \alpha\sin\left(\frac{k}{\epsilon}|x|-\delta\right)\qquad\text{(Expected tail of symmetric solution)} \ ,
\label{eqn:ualpha}
\end{equation}
where, $k/\epsilon$ is the ``far field wave number'' and $k$ is defined in equation (\ref{eqn:kappa}). The parameter $\delta$ is the ``far field phase shift''. The parameters, $\epsilon$ and $\delta$ control the shape of the solutions of fKdV equation. The amplitude of the tail $\alpha$, in general is a function of $\epsilon$ and $\delta$ and for a given $\epsilon$, $\alpha$ is a function of only $\delta$.

We will see later that for the minimum far field phase shift $\delta_m$, we get the corresponding minimal tail amplitude $\alpha_m$ of the far field oscillations. Since the minimal tail amplitude of the symmetric solutions of fKdV equation is exponentially small in terms of small parameter $\epsilon$, to the leading order it can be written as
\begin{equation}
    \alpha_m = \frac{\Lambda}{\epsilon^\mu}\exp{\left(-\frac{\rho}{\epsilon}\right)}(1+\mathcal{O}(\epsilon)) \ ,
\label{eqn:leadingorder}
\end{equation}
where $\Lambda$, $\mu$ and $\rho$ are constants. The tail amplitude $\alpha_m$ of the steady symmetric solutions of fKdV equation was first calculated analytically by Pomeau et al. \cite{prg} only up to the leading order in $\epsilon$ by using the method of matched asymptotics in the complex plane in the limit $\epsilon\to 0$ which has been developed by Segur and Kruskal in \cite{kruskal1987}. They showed that $\mu = 2$, $\rho = \pi/2$, and the proportionality constant $\Lambda = -\pi K$, where $K\approx -19.969$. So, according to \cite{prg} the leading order result looks like
\begin{equation}
    \alpha^{(PRG)}_m = -\frac{\pi K}{\epsilon^2}\exp{\left(-\frac{\pi}{2\gamma\epsilon}\right)}(1+\mathcal{O}(\epsilon)) \ .
\label{eqn:alphamPRG}
\end{equation}
Please note that there is an additional factor of $2$ and a missing division sign in \cite{prg}. This leading order result is confirmed numerically by Boyd \cite{boyd1, boyd2}. Equation (\ref{eqn:alphamPRG}) is supposed to dominate for small $\epsilon$ values for which the tail amplitude is so small it is demanding and difficult to calculate it by numerical methods.

It is desirable to compute the higher order corrections to the leading order result, equation (\ref{eqn:leadingorder}) both from a theoretical and a practical point of view. It is also necessary to compute higher order corrections in order to compare the results to the numerical results for small (and finite) values of $\epsilon$. For fKdV solitary wave-like solutions, the tail amplitude with higher order corrections can be written as
\begin{equation}
    \alpha_m = \frac{\Lambda}{\epsilon^\mu}\exp{\left(-\frac{\rho}{\epsilon}\right)}\left(1+\sum_{j=1}^N \zeta_j\epsilon^j+\mathcal{O}\left(\epsilon^{N+1}\right)\right) \ ,
\label{eqn:corrections}
\end{equation}
where $\zeta_j$ are yet some unknown constants. The analytical study of Pomeau et al. \cite{prg} was further extended by Grimshaw and Joshi \cite{gaj} up to $\epsilon^2$ order (i.e. they have computed $\zeta_1$ and $\zeta_2$) by using the same method as in \cite{prg}. According to \cite{gaj}, the tail amplitude $\alpha^{(GJ)}_m$ of the weakly localized solitary wave like solutions is given as
\begin{equation}
    \alpha^{(GJ)}_m = -\frac{\pi K}{\epsilon^2}\exp{\left(-\frac{\pi}{2\gamma\epsilon}\right)}\left(1-\pi\gamma\epsilon+\frac{1}{2}\pi^2\gamma^2\epsilon^2+\mathcal{O}(\epsilon^3)\right) \ .
\label{eqn:alphamGJ}
\end{equation}
Hence the value of first two corrections in equation (\ref{eqn:corrections}) are given as
\begin{align}
    \zeta^{(GJ)}_1 &= -\pi\gamma \ , \label{eqn:corr1} \\
    \zeta^{(GJ)}_2 &= \frac{1}{2}\pi^2\gamma^2 \approx 4.94\gamma^2 \ .
    \label{eqn:corr2}
\end{align}
These symmetric weakly localized solutions of the fKdV equation has been extensively studied by Boyd in \cite{boyd1, boyd2, boyd} by using multiple precision pseudo-spectral numerical methods where the leading order result and the value of first order perturbation coefficient $\zeta^{(GJ)}_1$ of \cite{gaj} has been also confirmed (Boyd used $c_0 = 4$ which corresponds to $\gamma = 1$). However a disagreement between the second order perturbation coefficient $\zeta^{(GJ)}_2$ in \cite{gaj} and in depth very precise numerical simulations of Boyd \cite{boyd2} has been found. According to the numerical simulations by Boyd, $-0.1 < \zeta^{(B)}_2 < 0$ which is clearly in disagreement with the result in \cite{gaj}, equation (\ref{eqn:corr2}). Boyd called this discrepancy a mathematical mystery, \emph{\say{The multiple precision calculations have left us with an open mathematical mystery: What is the true value of $\mathcal{O}(\epsilon^2)$ term in the perturbation theory? There is also an open numerical question: Given that multiple precision pseudo-spectral calculations have been pushed to the practical limit (at least with current hardware), can radically different algorithms somehow do better?}} (Boyd, 1995). One of the prime focus of the present work is to answer these open questions.

 Since there is a difference in the variables and equations used in this report and by Boyd in \cite{boyd1, boyd2, boyd}, it is necessary to discuss the connection with the variables and equations used by Boyd. This connection is important as we are going to compare our numerical results with that of Boyd. Apply the re-scaling, equation (\ref{eqn:rescalings}) with $\chi=\epsilon$ and introduce a further re-scaled function $\tilde u = \tilde v/6$. In this way the $\epsilon^2$ factor disappears in front of the fourth derivative term and the coefficient of quadratic terms also changes. In this way we get the fKdV equation used by Boyd which can be written as
\begin{equation}
    \tilde v_{\tilde x \tilde x \tilde x \tilde x}+\tilde v_{\tilde x \tilde x}+\frac{1}{2}(\tilde v -\tilde c) \tilde v = 0 \ .
\label{eqn:fkdvboyd}
\end{equation}
The only parameter in equation (\ref{eqn:fkdvboyd}) is $\tilde c = \epsilon^2 c = 4\gamma^2\epsilon^2+16\gamma^4\epsilon^4$. By using the re-scaled function, the relation between the tail amplitude is given as
\begin{equation}
    \alpha^{(B)}_m = 6\epsilon^2 \alpha_m \ .
\label{eqn:alphamgjb}
\end{equation}

The focus of the present work is the analytical and the numerical study of stationary (symmetric as well as asymmetric) solutions of the fKdV equation. The symmetric solutions has a central core associated with the exponentially small amplitude standing wave tail on both sides. The core can be described by the asymptotic series expansion in $\epsilon$, but the standing wave tail cannot due to its exponential dependence on $1/\epsilon$. In order to calculate analytically the exponentially small amplitude tail, we have applied the method of matched asymptotics in the complex plane in the limit $\epsilon\to 0$ which has been pioneered by Segur and Kruskal in \cite{kruskal1987}. In order to verify the analytical results, we have developed a pseudo-spectral code to solve fKdV equation numerically. There are also stationary asymmetric solutions which do not have oscillatory tail either side of the (asymmetric) core, but they decay exponentially to zero on one side of the core and blow up (to the negative values) on the other side. Quantitatively, this asymmetry can be described by calculating the third derivative of the asymmetric function at the origin (as the first derivative of function at the origin is zero to make the solution unique) both numerically and analytically.

We finish the introduction with an outline of the thesis: In Chapter (\ref{chapter:numerical_methods}), the numerical methods and their implementation both in the finite and the semi-infinite cases are discussed in detail. Next in Chapter (\ref{chapter:numerical_results}), some numerical results of symmetric and asymmetric solutions are presented. For instance for the weakly-localized symmetric solutions, the dependence of tail amplitude $\alpha$ on the small parameter $\epsilon$ and on the phase shift $\delta$ is discussed, and for asymmetric blow up solutions, the $\epsilon$ dependence of the third derivative of the function at the origin is given. In the same chapter the difference of the symmetric and the asymmetric solutions is also discussed. In Chapter (\ref{chapter:analytical_results}), the exponential asymptotics is discussed in which the core region of the symmetric and the asymmetric solutions is described by asymptotic series expansion in $\epsilon$. In this chapter, the method of matched asymptotics for the tail region is also discussed and the leading order result for the minimal tail amplitude $\alpha_m$ is presented. In Chapter (\ref{chapter:wkbsolution}) Wentzel-Kramers-Brillouin (WKB for short) method has been used for the solution around the core region to arbitrary order in $\epsilon$. Chapter (\ref{chapter:higher_orders}) contains the higher order generalizations of the leading order result of the minimal tail amplitude $\alpha_m$, namely the third and fifth order corrections are discussed. The comparison with the numerical simulations is also given in this chapter. In Chapter (\ref{chapter:thder}), the WKB solution is used for the calculation of the third derivative of the function at the origin. The (analytical) result is given up to $\mathcal{O}(\epsilon^5)$ order and the comparison with the  numerical simulations is also discussed. The last Chapter (\ref{chapter:conclusions}) contains the conclusions.

\chapter{Numerical methods}\label{chapter:numerical_methods}
Our main purpose is to solve numerically the once integrated stationary fKdV equation, equation (\ref{eqn:1aintFKDV}), which is a fourth-order ordinary differential equation. We are solving this equation for the symmetric solutions which have a well defined central core and an associated exponentially small amplitude standing wave tail to infinity, and the asymmetric solutions which decay exponentially to zero on one side of an asymmetric core, but blow up on the other side. Since fKdV equation is a non-integrable model and therefore numerical integration algorithms have to be used in order to find solutions. Several numerical methods have been developed for solving nonlinear differential equations such as, finite difference method, finite element method, Galerkin method, Fourier expansion method, spectral method, pseudo-spectral method and so on. Because the tail amplitude of the far field oscillations of the symmetric solutions and third derivative of the asymmetric solutions are exponentially small in $\epsilon$, we have used the pseudo-spectral method to solve equation (\ref{eqn:1aintFKDV}) for these solutions because of its exponential accuracy.

The basic idea is to assume that the unknown function $u(x)$ can be approximated as the sum of $N$ basis functions $\phi_n(x)$ as
\begin{equation}
    u(x)\approx u_N(x)=\sum_{n=0}^{N-1} l_n\phi_n(x) \ ,
\label{eqn:ux}
\end{equation}
where $l_n$ are coefficients of the basis functions $\phi_n(x)$. If we substitute this into the differential equation $\mathcal{D}u(x)=f(x)$, we get the residual function as
\begin{equation}
    R(x,l_0,\ldots, l_{N-1})=\mathcal{D}u_N-f \ ,
\end{equation}
where $\mathcal{D}$ is the operator of the differential equation. Since the residual function vanishes for an exact solution, our aim is to choose the series coefficients $l_n$ in such a way that minimizes the residual function as much as possible, so that equation (\ref{eqn:ux}) would be the best possible approximate solution of the differential equation $\mathcal{D}u(x)=f(x)$.

The pseudo-spectral method is based on evaluating the residual function only at the selected points, say at $\{x_i\}$. These selected points are called the collocation points and hence the another name for pseudo-spectral method is the collocation method. The reason why we choose the pseudo-spectral method but not the finite difference or finite element method is that the pseudo-spectral method is exponentially convergent. The pseudo-spectral error is given as \cite{boydcheb}
\begin{equation}
 \text{Pseudo-spectral error}\approx \mathcal{O}(h^N) \ ,
\label{eqn:pseudoerror}
\end{equation}
where $h=1/N$ is a space between the adjacent grid points. The pseudo-spectral error, see equation (\ref{eqn:pseudoerror}) is decreasing faster than any finite power of $N$. On the other hand, the finite element method is less accurate. For instance, for $N=10$ to have an accuracy equal to the pseudo-spectral accuracy, one would need $10$-th order finite difference method to have an error $\mathcal{O}(h^{10})$.

We have used \verb!C! and \verb!C++! codes for our numerical purposes. Due to the extremely smallness of the oscillatory tails, the usual $16$ or $19$ digits arithmetic is not enough for this calculation. In order to get the higher precision, we have used the additional external \verb!C++! library called the Class Library for Numbers (CLN for short) \cite{CLN}. This numerical package enables us to calculate the exponentially small tail amplitudes to arbitrary many digits precision. However for smaller $\epsilon$ values, we need higher resolution $N$, the number of grid points. For instance for $\epsilon=0.02$, we need $N=3000$ collocation points and around $100$ digits arithmetic in the whole numerical calculation. In such cases the calculations may take half a day on a desktop computer. In order to get rid of this difficulty, we have used an additional \verb!C! library called the arbitrary-precision ball arithmetic (ARB for short) \cite{arb} which uses advanced matrix multiplication methods and it turns out that this library is about twenty times faster than the CLN. For instance, for the previous example, $\epsilon=0.02$, the calculations using ARB library only takes an hour and a half on a desktop computer.

\section{Newton-Kantorovich method}

Applying the pseudo-spectral method to the non-linear differential equation results a non-linear system of algebraic equations equal to the number of coefficients $l_n$ of the basis functions $\phi_n$. We then need a method such as Newton's method (also called Newton-Raphson method)  which iterates from a first guess until the solution is sufficiently accurate. However this method needs the Jacobian matrix to calculate at each and every iteration which is usually time consuming. Another costly step in Newton's method is to calculate the inverse of the Jacobian matrix (which is usually a full matrix) at each iteration. However we can reduce this cost by reversing the application of the two methods, first apply the Newton's method directly to the original differential equation and then apply the pseudo-spectral method to convert the resulting iterative sequence to linear differential equations into the corresponding linear matrix problem. This method then is called Newton-Kantorovich method. For more details, see for instance Appendix C. of \cite{boydcheb}.

Because the (once integrated) fKdV equation is non-linear differential equation, see equation (\ref{eqn:1aintFKDV}), we use the Newton-Kantorovich method which can be applied directly to a nonlinear partial differential equation to reduce it to a sequence of linear differential equations. The basic idea of this method is that if at the $j$-th iteration, the approximate solution is $u^{(j)}$, then the next approximation can be written as
\begin{equation}
    u^{(j+1)}=u^{(j)}+\Delta \ .
\label{eqn:ujplus1}
\end{equation}
Substituting equation (\ref{eqn:ujplus1}) into equation (\ref{eqn:1aintFKDV}), we get
\begin{equation}
\begin{aligned}
    \epsilon^2\Delta_{xxxx}+\Delta_{xx}+3\Delta^2 +6u^{(j)}\Delta-c\Delta  =-\{\epsilon^2u^{(j)}_{xxxx}+u^{(j)}_{xx}+3(u^{(j)})^2-cu^{(j)}\} \ .
\label{eqn:newton1}
\end{aligned}
\end{equation}
If the function $u^{(j)}$ is a known approximate solution of equation (\ref{eqn:1aintFKDV}) so that $\Delta\ll 1$, we can neglect the $\mathcal{O}(\Delta^2)$ terms in equation (\ref{eqn:newton1}) and hence it becomes a linear equation which is then used to compute the correction $\Delta$,
\begin{equation}
\begin{aligned}
    \epsilon^2\Delta_{xxxx}+\Delta_{xx}+6u^{(j)}\Delta-c\Delta = R \ ,
\label{eqn:newton}
\end{aligned}
\end{equation}
where $R$ is the residual which is given as
\begin{equation}
    R =-\{\epsilon^2u^{(j)}_{xxxx}+u^{(j)}_{xx}+3(u^{(j)})^2-cu^{(j)}\} \ .
\label{eqn:Res}
\end{equation}
We repeat the iteration procedure equation (\ref{eqn:ujplus1}) and equation (\ref{eqn:newton}) until the residual $R$ is negligibly small.

There is still a remaining difficulty that the iteration requires a good first guess for the function $u(x)$ so that the iteration converges. The idea is to begin with a known solution. In the limit of a very small tail-amplitude, that is when $\epsilon \ll 1$, the KdV 1-soliton solution $u_0(x) = 2\gamma^2\mathrm{sech}^2(\gamma x)$ given in equation (\ref{eqn:u0}) is a good first guess. This solution for a given smaller $\epsilon$ is used as the first guess for slightly larger $\epsilon$. Hence through this bootstrapping procedure, we can calculate the solutions $u(x)$ of fKdV equation having small oscillatory tail amplitude to solutions having comparatively larger oscillatory tail amplitude. 

The left hand side of equation (\ref{eqn:newton}) can be considered as a linear matrix operator $L_{jk}$ multiplying the collocation values $\Delta_k$ of the unknown function (correction) $\Delta$. The collocation values $R_j$ can be calculated at each step from the previous approximation. We replace (any) two lines of linear matrix operator $L_{jk}$ by values enforcing the boundary conditions while replacing the corresponding elements in $R_j$ by the previous error in the boundary conditions. After this, the matrix equation can be solved for $\Delta_j$ and updating $u^{(j)}$ by adding the calculated $\Delta$ gives the next approximation $u^{(j+1)}$.

The once integrated fKdV equation given in equation (\ref{eqn:1aintFKDV}) is translationally invariant which means that if $u(x)$ is a solution to this equation, then $v(x)=u(x+\xi)$ is also a solution, where $\xi$ is a constant. However, we can remove this translational invariance simply by restricting the function $u(x)$ to be symmetric around $x=0$. In this way the arbitrary shift in $x$ is no longer possible as it compels the function $u(x)$ to have a local maxima (or local minima, for instance if there are more than one core) at the origin.

\section{Pseudo-spectral method}

We use the pseudo-spectral (also called collocation) method to solve the fKdV equation, equation (\ref{eqn:1aintFKDV}) numerically. We solve this problem in two domains, a finite interval $x\in [0, \, L]$ and a semi-infinite interval $x\in [0, \ \infty)$. In both these domains, the alternatives are to use the Chebyshev polynomials of different mappings. In addition to these alternatives, one must always choose some parameter $L$. The physical meaning of this parameter depends on the domain and problem we are solving. This may be the size of a truncated computational domain, or it can be just a mapping parameter for the coordinate transformation from finite domain to (semi) infinite domain and vice-versa.

\subsection{Finite case: Symmetric solution with tail} 

In this subsection, we give an algorithm and its implementation to find the solutions $u$ which are symmetric about  the centre $x = 0$. These solutions has a central core similar to KdV 1-soliton solution, accompanied by extremely small oscillatory tail to infinity on both sides. We match the numerical solution to the solution of the linearized problem, $u = \alpha \ \mathrm{sin}(k x/\epsilon-\delta)$, see equation (\ref{eqn:ualpha}), at the outer numerical boundary $x=L$, where $\alpha$ is the oscillatory tail amplitude and $\delta$ is the far field phase shift. We use property that the difference from the linearized (tail) solution decreases exponentially fast as we go away from the center. Due to the symmetry, we solve the problem only for $x > 0$. The steady symmetric solutions are two parameter ($\epsilon$ and $\delta$) families of solutions. For any chosen value of these parameters, we can compute the numerical value of oscillatory tail amplitude $\alpha$ by solving fKdV equation. At the outer boundary $x = L$, we use two boundary conditions
\begin{align}
 \frac{u_{xx}}{u} &= \frac{u_{\alpha,xx}}{u_\alpha} = -\frac{k^2}{\epsilon^2} \ ,\\
 \frac{u_{x}}{u} &= \frac{u_{\alpha,x}}{u_\alpha} = \frac{k}{\epsilon}\frac{\mathrm{cos}(k x/\epsilon -\delta)}{\mathrm{sin}(k x/\epsilon-\delta)} \ .
\end{align}
The above two equations can also be written as 
\begin{align}
 \epsilon^2u_{xx}+k^2 u = 0  \label{eqn:bc1} \ ,\\
 \epsilon \ \mathrm{sin}\left(\frac{kx}{\epsilon}-\delta\right)u_x-k \ \mathrm{cos}\left(\frac{kx}{\epsilon}-\delta\right)u = 0 \ .
\label{eqn:bc2}
\end{align}
As we are solving our numerical problem in a finite interval $x\in [0, L]$, we make a transformation in the independent variable $\vartheta$ by 
\begin{equation}
 x=L\ \mathrm{cos} \, \vartheta \ ,
\end{equation}
where $\vartheta\in[0,\pi/2]$. In this way the centre $x=0$ corresponds to $\vartheta=\pi/2$, and the outer boundary corresponds to $\vartheta=0$. We expand the function $u$ by $N$ Fourier components $U_n$,
\begin{equation}
    u=\sum_{n=0}^{N-1}\frac{1}{c_n}U_n\mathrm{cos}(2n\vartheta) \, ,
\label{eqn:spectral}
\end{equation}
where
\begin{equation}
c_n =
        \begin{dcases}
            1 & \quad \text{if}\quad 1 \leq n \leq N-2 \, , \\
            2 & \quad \text{if}\quad n=0\quad\text{or}\quad n=N-1 \, ,
        \end{dcases}
    \label{eqn:zetan}
\end{equation}
where $N$ is the number of Fourier components used in the current approximation. Because of the symmetry, we use only even indexed Fourier components. The Fourier expansion of the function $\tilde{u}(x)=\tilde{u}(L\ \mathrm{cos} \, \vartheta)=u(\vartheta)$ in the variable $\vartheta$ corresponds to the expansion in Chebyshev polynomials in $x/L$, $T_{2n}(x/L)$ as
\begin{equation}
    T_n(\mathrm{cos} \, \vartheta)=\mathrm{cos}(n\vartheta) \, .
\label{eqn:cheb}
\end{equation}
The choice of the basis functions depends on the types of solution we are interested in and domain in which one is interested in solving the problem. Since we are interested in the stationary symmetric solutions (nanopterons) of fKdV equation in a finite domain which are not periodic, the best choice is to use the Chebyshev polynomials of even degree in constructing the basis functions \cite{boydcheb}. This choice of basis functions also suppresses the translational degree of freedom in the Newton-Kantorovich equation, equation (\ref{eqn:newton}). We can write the expansion of the function $u$, equation (\ref{eqn:spectral}) in terms of Chebyshev polynomials, equation (\ref{eqn:cheb}) as
\begin{equation}
     u=\sum_{n=0}^{N-1}\frac{1}{c_n}U_n T_{2n}\left(\frac{x}{L}\right) \, ,
\label{eqn:spectralcheb}
\end{equation}
where $c_n$ is given in equation(\ref{eqn:zetan}). By using the pseudo-spectral method, the Newton-Kantorovich equation is converted into a matrix equation by substituting the expansion of the function $u$, equation (\ref{eqn:spectral}) into differential equation of the correction $\Delta$, equation (\ref{eqn:newton}). In this way the solution can be alternatively represented by its value at $N$ collocation (or interpolation) points $x_n=L \ \mathrm{cos}\vartheta_n$, where
\begin{equation}
    \vartheta_n=\frac{\pi}{2(N-1)}n\qquad\text{for}\quad n=0,1,\ldots, N-1 \, .
\label{eqn:thetan}
\end{equation}
 Hence the collocation points $x_n$ can be written as
 \begin{equation}
    x_n=L \ \mathrm{cos}\left(\frac{\pi}{2(N-1)}n\right)\qquad\text{for}\quad n=0,1,\ldots, N-1 \, .
\label{eqn:xn}
\end{equation}
For the function values at collocation points, we introduce the notation $\tilde{u}_n=u(x_n)$. By using equation (\ref{eqn:xn}), the value of function $u$ at the centre is $\tilde{u}_{N-1}$ and the value of function $u$ at the outer boundary is $\tilde{u}_0$.  In this way, equation (\ref{eqn:spectral}) can be interpreted as a matrix multiplication
\begin{equation}
    \tilde{u}_k=\sum_{j=0}^{N-1}T_{kj}U_j\qquad\text{where, }\quad T_{kj}=\frac{1}{c_j}\mathrm{cos}\left(\frac{kj\pi}{N-1}\right) \, .
\label{eqn:ucoll}
\end{equation}
The Fourier coefficients $U_j$ can be calculated by the inverse transformation which is given as
\begin{equation}
    U_k=\sum_{j=0}^{N-1}G_{kj}\tilde{u}_j\qquad\text{where, }\quad G_{kj}=\frac{2}{c_j(N-1)}\mathrm{cos}\left(\frac{kj\pi}{N-1}\right) \, .
\label{eqn:ufour}
\end{equation}
Apart from the factor $2/(N-1)$, the transformation in equation (\ref{eqn:ufour}) is same as the one in equation (\ref{eqn:ucoll}). In both the directions (collocation to Fourier and Fourier to collocation) corresponds to type I discrete cosine transform DCT-I. The coefficients $U_j$ can be considered Chebyshev coefficients in terms of $x$, or Fourier coefficients in terms of $\vartheta$.

Multiplication of functions can be easily calculated using the collocation values $\tilde{u}_j$, while derivatives can be naturally obtained using Fourier coefficients $U_j$. The Fourier coefficients $(U_{xx})_j$ of the second derivative of function $u$ represented by $U_j$ can be calculated as
\begin{equation}
    (U_{xx})_j=\sum_{k=0}^{N-1}D^{(2)}_{jk}U_k \, ,
\end{equation}
where $D^{(2)}_{jk}$ is a second derivative matrix in Fourier picture and is given as
\begin{equation}
D^{(2)}_{jk} =
        \begin{dcases}
            \frac{8}{c_k L^2} \ k(k^2-j^2) & \quad \text{if}\quad k\geq j+1 \, ,\\
            0 & \quad \text{Otherwise} \, .
        \end{dcases}
\label{eqn:derf}
\end{equation}
The second derivative matrix in collocation picture can be easily obtained from equation (\ref{eqn:derf}) as
\begin{equation}
    \tilde{D}^{(2)}_{jk} =T_{jl}D^{(2)}_{lp}G_{pk} \, ,
\end{equation}
where Einstein summation convention is used here. The even higher derivative matrices can be easily obtained by multiplying second derivative matrix with itself. For instance, the fourth derivative matrix can be obtained by multiplying the second derivative matrix with itself, $D^{(4)}_{jk} =D^{(2)}_{jl}D^{(2)}_{lk}$. We also need first derivative matrix for boundary conditions, see equation (\ref{eqn:bc2}). Due to symmetry about $x=0$, we need first derivative only at the outer boundary, $x=L$ which can be calculated by the inner product $V_j\tilde{u}_j$, where
\begin{equation}
    V_j=\sum_{k=0}^{N-1}\frac{4k^2}{L}G_{kj} \, .
\end{equation}

\subsection{Semi-infinite case: Blow-up asymmetric solution}

In this subsection, we give an algorithm and its implementation to find the solutions $u$ of the stationary fKdV equation, equation (\ref{eqn:1aintFKDV}) which are asymmetric about the origin and has an extremum at the origin $x = 0$. These solutions decay exponentially to zero on one side of an asymmetric core and blow up to negative values at finite $x$ on the other side. To solve equation (\ref{eqn:1aintFKDV}) in a finite interval, we need at the center the value of a function $u$, its first, second, and third derivative. The first derivative at the center is zero, since we impose the extremum condition at $x = 0$ in order to make the solution unique. The other three boundary conditions are that the two oscillating and one exponential blow-up mode is suppressed at infinity.  

In order to transform (compactify) semi-infinite interval $[0, \, \infty)$ into a finite interval $y\in [-1, 1]$, we introduce new independent variable $y$,
\begin{equation}
    y = \frac{x-L}{x+L} \, ,
\label{eqn:yvar}
\end{equation}
where $L$ is a constant representing the length scale. The inverse relation of equation (\ref{eqn:yvar}) is given as
\begin{equation}
    x = L\frac{1+y}{1-y} \, .
\label{eqn:xvar}
\end{equation}
We intend to expand the function $\tilde u(y)=u(x(y))$ in terms of Chebyshev polynomials which can be done most naturally by introducing a new independent variable $\theta\in [0, \pi]$ by
\begin{equation}
    y = \mathrm{cos} \, \theta \, .
\label{eqn:ycost}
\end{equation}
The Fourier expansion of function $\hat u(\theta) = \tilde u(y(\theta)) = \tilde u(\mathrm{cos} \, \theta)$ in variable $\theta$ correspond to expansion in Chebyshev polynomials in $y$, since $\mathrm{cos}(n\theta) = T_n(y)$. The boundary $x\to\infty$ corresponds to $y = 1$ and $\theta = 0$ and the center $x = 0$ corresponds to $y = -1$ and $\theta = \pi$. From equation (\ref{eqn:xvar})
\begin{equation}
    x = L\frac{1+\mathrm{cos} \, \theta}{1-\mathrm{cos} \, \theta} = L \ \mathrm{cot}^2\frac{\theta}{2} \, ,
\label{eqn:xlcot}
\end{equation}
where we have used $2 \, \mathrm{cos}^2\theta = 1+\mathrm{cos}(2\theta)$ and $2 \ \mathrm{sin}^2\theta = 1-\mathrm{cos}(2\theta)$. The inverse relation from equation (\ref{eqn:xlcot}) is given as
\begin{equation}
    \theta = 2 \ \mathrm{arccot}\sqrt{\frac{x}{L}} = \pi - 2 \ \mathrm{arctan}\sqrt{\frac{x}{L}} \ ,
\label{eqn:thetacottan}
\end{equation}
where we have used a trigonometric identity $\mathrm{arctan} \, x + \mathrm{arccot} \, x = \pi/2$. This expansion corresponds to expansion in rational Chebyshev polynomials (also called Chebyshev rational functions) $R_n(y)$ of degree $n$ defined as
\begin{equation}
    R_n(x) = L_n(y) = L_n\left(\frac{x-L}{x+L}\right) = L_n(\mathrm{cos} \, \theta) \, ,
\label{eqn:rational}
\end{equation}
where the argument in rational Chebyshev polynomial, $x$, is related to the argument of Chebyshev polynomial, $y$, which in turn is related to the argument of the trigonometric function, $\theta$, through the relations which are given above.

We choose $N$ collocation points placed uniformly in $\theta \in [0, \pi]$ as
\begin{equation}
    \theta_k = \frac{\pi}{N-1}k \, , \qquad \mathrm{where} \quad  k = 0, 1, 2, \ldots, N-1 \, .
\label{eqn:thetak}
\end{equation}
The values of function $u$ at these collocation points can be written as
\begin{equation}
    u_k = \hat u(\theta_k) = u\left(L \, \mathrm{cot}^2\frac{\theta_k}{2}\right) \, .
\label{eqn:ucollocation}
\end{equation} 
We expand the function $u$ in Fourier components $U_n$ which holds for arbitrary $\theta$ and hence can be used to calculate the function $u$ at arbitrary $x$,
\begin{equation}
    u=\sum_{n=0}^{N-1}\frac{1}{c_n}U_n\mathrm{cos}(n\theta) \, ,
\label{eqn:uFourier}
\end{equation}
where $c_n$ is given in equation (\ref{eqn:zetan}). At the collocation points, equation (\ref{eqn:uFourier}) can be written as
\begin{equation}
    u_k = \sum_{n=0}^{N-1}T_{kn}U_n \, , \qquad \mathrm{where} \quad T_{kn} = \frac{1}{c_n}\mathrm{cos}\left(\frac{k n \pi}{N-1}\right) \, .
\label{eqn:ukTnk}
\end{equation}
The coefficients $U_k$ which can be considered as Fourier coefficients with respect to variable $\theta$, Chebyshev coefficients in terms of $y$, or rational Chebyshev coefficients using $x$. These coefficients can be calculated as
\begin{equation}
    U_k = \sum_{n=0}^{N-1}G_{kn}u_n \ , \quad \mathrm{where} \quad G_{kn} = \frac{2}{c_n(N-1)}\mathrm{cos}\left(\frac{k n \pi}{N-1}\right) \, .
\label{eqn:UkGnk}
\end{equation}
Now, the first derivative of the function $u$ with respect to $x$ can be written as
\begin{equation}
    \frac{\mathrm{d}u}{\mathrm{d}x} = \sum_{k=0}^{N-1}\frac{1}{c_k}U^{(1)}_k\mathrm{cos}(k\theta) \, ,
\label{eqn:uder1}
\end{equation}
where $U^{(1)}_k$ denotes the Fourier coefficients of the first derivative which can be calculated in terms of $U_k$ by taking the first derivative of equation (\ref{eqn:uFourier}) with respect to $x$. The derivative of $\mathrm{cos}(n\theta)$ with respect to $x$ can be written as 
\begin{equation}
    \frac{\mathrm{d}}{\mathrm{d}x}\mathrm{cos}(n\theta) = \sum^{\infty}_{k=0}\frac{1}{\bar c_k}\bar D_{nk}\mathrm{cos}(k\theta) \, ,
\label{eqn:dcosnt}
\end{equation}
where $\bar c_k$ is a constant and $\bar D_{nk}$ is a first derivative matrix which are given as
\begin{equation}
 \bar c_k =
        \begin{dcases}
            1 & \quad \text{if}\quad k \geq 1 \, ,\\
            2 & \quad \text{if}\quad k = 0 \, ,
        \end{dcases}
    \label{eqn:barck}
\end{equation}
and 
\begin{equation}
 \bar D_{nk} =
        \begin{dcases}
            0 & \quad \text{if}\quad k \geq n+2 \, ,\\
            \frac{n}{4} & \quad \text{if}\quad k = n+1 \ , \\
            -n & \quad \text{if}\quad k = n \, , \\
            \frac{7}{4}n & \quad \text{if}\quad k = n-1 \ \text{and} \, n \geq 2 \ ,\\
            \frac{3}{2} & \quad \text{if}\quad k = 0 \ \text{and} \, n = 1 \, , \\
            2n(-1)^{n+k+1} & \quad \text{if}\quad k \leq n-2 \, .
        \end{dcases}
    \label{eqn:barDnk}
\end{equation}
The higher order derivatives of $\mathrm{cos}(n\theta)$ can be easily calculated by taking the products of $\bar D_{nk}$ with itself. Using equation (\ref{eqn:dcosnt}), truncate $\mathrm{cos}(n\theta)$ terms for $n\geq N$, and change the order of summation, we obtain the Fourier coefficients of the derivative as
\begin{equation}
    U^{(1)}_k = \sum_{k=0}^{N-1}\frac{c_k}{\bar c_k c_n}\bar D_{kn}^{\mathrm{T}}U_n = \sum_{k=0}^{N-1} D_{kn}U_n \, ,
\label{eqnFder}
\end{equation}
where $\bar D_{kn}^{\mathrm{T}}$ denotes the transpose of derivative matrix operator $\bar D_{kn}$ given in equation (\ref{eqn:barDnk}), and $\bar c_k$ and $c_k$ are given in equations (\ref{eqn:barck}) and (\ref{eqn:zetan}) respectively.

\begin{figure}[ht!]
\centering
    \includegraphics[width=0.58\linewidth]{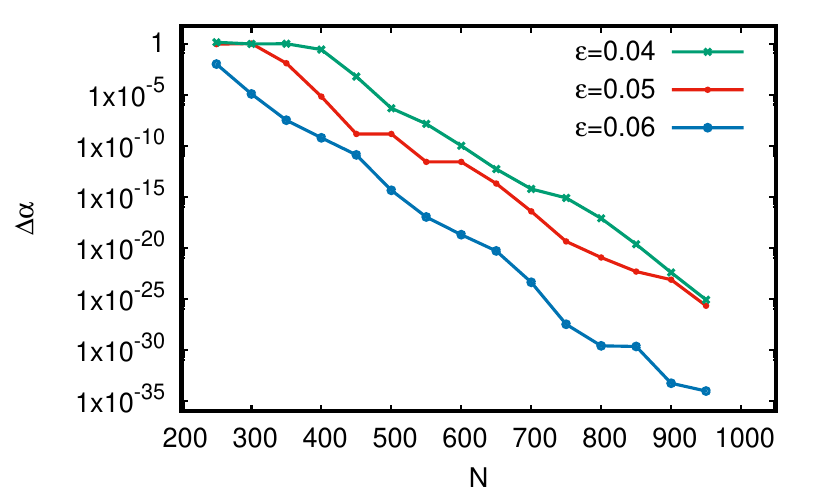}
    \caption{
        A logarithmic plot of the absolute value of the relative error $\Delta\alpha=|(\alpha-\alpha_N)/\alpha|$ of the oscillatory tail amplitude $\alpha$ of the symmetric solution as a function of the resolution $N$. We used $\delta = 6\epsilon\gamma$ for the phase shift which is a leading order approximation for the minimum phase shift $\delta_m$, for details see Chapter (\ref{chapter:wkbsolution}). This log-linear plot shows the exponential accuracy of pseudo-spectral method.
    \label{fig:acc}}
\end{figure}

The crucial point about the pseudo-spectral method is that it is exponentially accurate which means the error decreases exponentially fast as the resolution $N$, the number of grid points increases, see Figure (\ref{fig:acc}). The accuracy is really important for our numerical calculations as the oscillatory tail amplitude $\alpha$ is exponentially small in $\epsilon$ and hence many decimal places of precision is needed. The relative error of the tail amplitude $\alpha$ can be written as
\begin{equation}
    \Delta\alpha=\left|\frac{\alpha-\alpha_N}{\alpha}\right| \, ,
\label{eqn:accuracy}
\end{equation}
where $\alpha_{N}$ represents the value of tail amplitude at resolution $N$, and $N=250,\ldots,950$. We take the value of $\alpha$ at $N=1000$ as the best possible value of $\alpha$ for given $\epsilon$ values and calculated the relative error by using equation (\ref{eqn:accuracy}). 

\chapter{Numerical results}\label{chapter:numerical_results}

This chapter concerns with the numerical simulations of the solutions of the stationary fKdV equation (\ref{eqn:1aintFKDV}). We discuss two types of solutions in this chapter. The \emph{nanopeteron} -- symmetric weakly localized solitary wave-like solution which has a long standing wave tail associated with a well defined central core and the asymmetric solutions which do not have any oscillatory tail either side of the (asymmetric) core. These asymmetric solutions decay exponentially to zero on one side of the core and blow up to the negative values on the other side of the core.

We use $c_0=4$ (and consequently $\gamma=1$) in all our numerical calculations and for the phase $\delta$, we use the minimum phase $\delta_m$ every time unless stated otherwise. The minimum phase $\delta=\delta_m$ is computed numerically by using the Brent's minimization methods \cite{numeric} and this in turn gives the corresponding minimum oscillatory tail amplitude $\alpha_m$.

\begin{figure}[ht!]
\centering
    \includegraphics[width=0.75\linewidth]{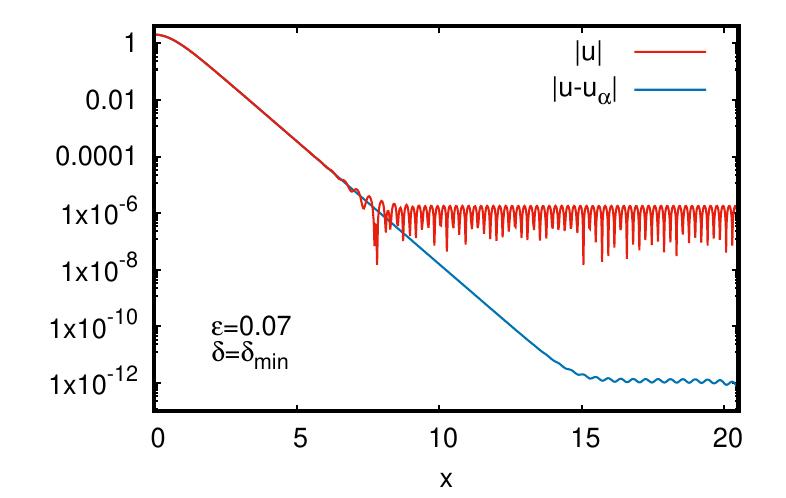}
    \caption{
        A plot of the numerical symmetric solution $u$ as a function of $x$ (red curve) and the difference $u-u_\alpha$, where $u_\alpha$ is given in equation (\ref{eqn:ualpha}), $u_\alpha = \alpha_m \ \mathrm{sin}(k x/\epsilon -\delta)$ which corresponds to the matched tail (blue curve).
    \label{fig:u0.07}}
\end{figure}

The structure of the obtained numerical (symmetric) result for $\epsilon=0.07$ can be seen in Figure (\ref{fig:u0.07}), which shows the $x$ dependence of the numerical solution $u$ logarithmically. The phase in this case is chosen to be the minimum phase of which the numerical value is $\delta_m=0.4207362$. To include a large portion of the tail, we set the outer boundary at $L=20$. The downward spikes (red curve) correspond to the zero crossings in the tail.

The numerical value of the minimum oscillatory tail amplitude corresponding to the obtained value of minimum phase $\delta_m$ is $\alpha_m = 1.802403\times 10^{-6}$. The striking agreement of the function $u$ and the tail for $x>14$ indicates that the obtained $\alpha_m$ value is precise to six digits. To obtain this precision, we need at least 300 collocation points and 19 digits of precision during the whole numerical calculation. This whole calculation was completed in about a minute using CLN code, and it took only less than quarter a minute for the ARB code on a desktop computer. We represent the matched tail with the linearized solution given in equation (\ref{eqn:ualpha}), $u =\alpha \ \mathrm{sin}(k x/\epsilon -\delta)$ and hence make an error of the order $\mathcal{O}(\alpha^2)$. So even if we increase the resolution to higher values, the difference from the matched tail will not go below $10^{-12}$. However for smaller $\epsilon$ values (the plots for smaller $\epsilon$ values are not shown here), the obtained values of minimal tail amplitude is precise to more than ten digits and for that, we need much larger number of collocation points (resolution) as well as more digits of precision.

\section{The nanopteron: symmetric weakly localized solitary wave}

\begin{figure}[ht!]
\centering
    \includegraphics[width=0.75\linewidth]{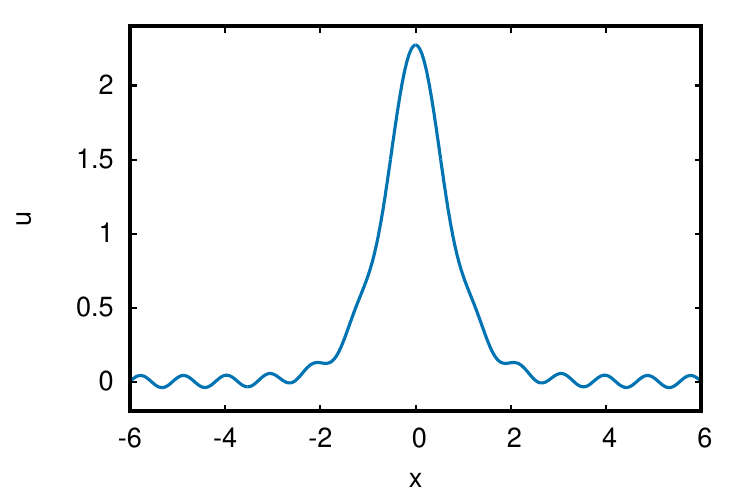}
    \caption{
        A plot of the symmetric weakly localized solitary wave solution of fKdV equation (\ref{eqn:1aintFKDV}) for $\epsilon=0.15$ and $\delta=\delta_m=0.957$.
    \label{fig:u0.15}}
\end{figure}

The numerically computed weakly localized solitary wave-like solution of fKdV equation for $\epsilon=0.15$ can be seen in Figure (\ref{fig:u0.15}). For this larger value of $\epsilon$, the oscillatory tail is easily visible and the tail amplitude is small in comparison to the single tall peak centered at $x=0$. We have used the minimum far field phase $\delta=\delta_m$ and the corresponding numerical value of the minimal-amplitude tail is $\alpha_m=0.041$. 

\begin{figure}[ht!]
\centering
    \includegraphics[width=0.75\linewidth]{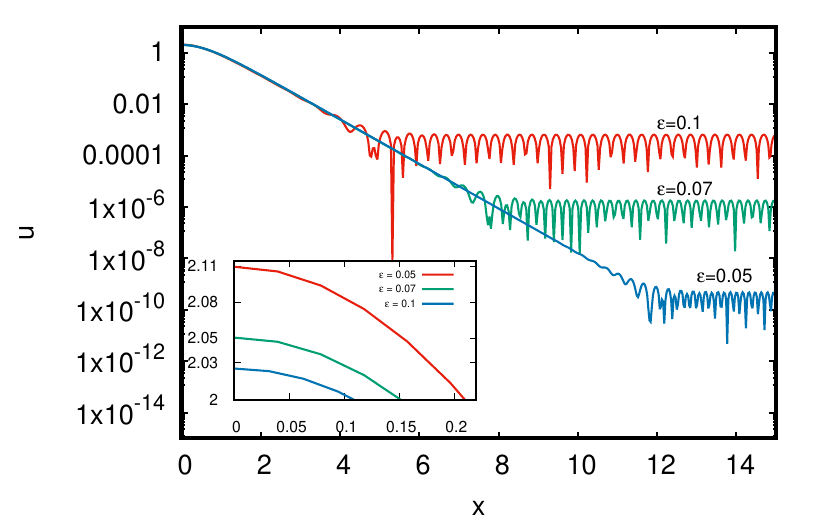}
    \caption{
        Main: A logarithmic plot of the symmetric weakly localized solitary wave solution of fKdV equation (\ref{eqn:1aintFKDV}) as a function of $x$ for $\epsilon=0.1$, $\epsilon = 0.07$ and $\epsilon=0.05$. Inset: It shows how close the amplitude of the solution of fKdV equation is to that of KdV 1-soliton solution at the origin which is $2$, for given $\epsilon$ values. The $\gamma$ is set to $1$ in the numerical simulations.
    \label{fig:uatx}}
\end{figure}

The core amplitude at $x = 0$ is always (for all $\epsilon$ values) close to the amplitude of KdV 1-soliton solution $u_0 = 2\gamma^2\mathrm{sech}^2(\gamma x)$. However, the size of the core region is different for different $\epsilon$ values, smaller the $\epsilon$ value is, the larger is the core region. What it means is that for the smaller value of $\epsilon$, the tail would appear at larger $x$ value as compared to if the $\epsilon$ value is comparatively larger, see Figure (\ref{fig:uatx})

\begin{figure}[ht!]
\centering
    \includegraphics[width=0.75\linewidth]{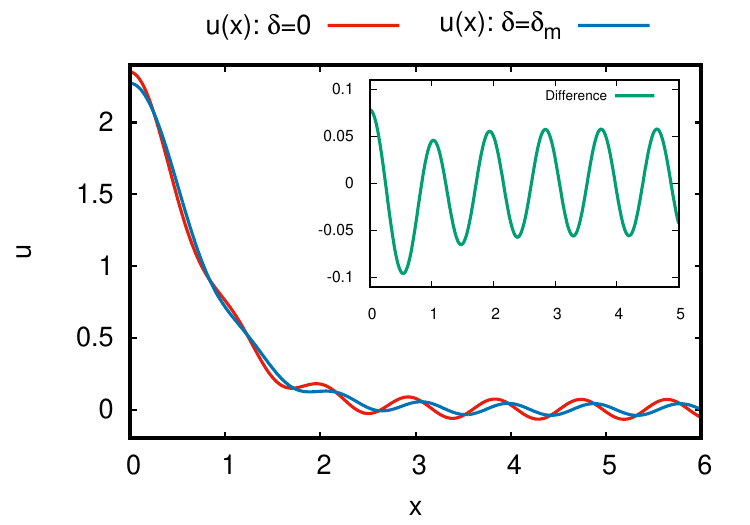}
    \caption{
        Main: A plot of the symmetric solution when $\epsilon=0.15$ for two different values of $\delta$, $\delta=0$ and $\delta=\delta_m=0.957$. Inset: The difference of the two symmetric solutions shown in the main plot. The parameter $\gamma$ is set to $1$ in the numerical simulations.
    \label{fig:u015}}
\end{figure}

For a given value of $\epsilon$, the nanopterons are a one parameter family characterized by the phase parameter $\delta$. We can calculate a particular solution $u$ by specifying the value of the far field phase $\delta$ for a given $\epsilon$. Please note that the amplitude of the oscillatory tail changes with $\delta$ in order to match the shifted oscillatory tails  with the core. Figure (\ref{fig:u015}) shows the plot of two different members of the family when $\epsilon=0.15$ for $\delta=\delta_m=0.957$ and $\delta=0$. The numerical value of oscillatory tail amplitude for $\delta=0$ is $\alpha=0.069$. 

\begin{figure}[ht!]
\centering
    \includegraphics[width=0.75\linewidth]{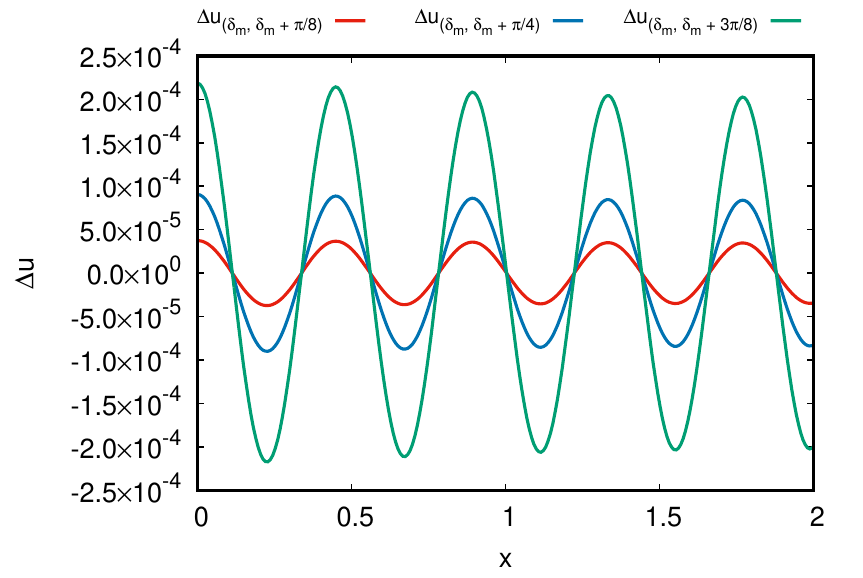}
    \caption{
        The difference of two symmetric solutions for $\epsilon = 0.07$, but for different pairs of phase shift ($\delta$) values, where we denote $\Delta u_{(\delta_1,\delta_2)}=u(x,\delta_1)-u(x,\delta_2)$.
    \label{fig:diffdelta}}
\end{figure}

The difference of the two symmetric solutions for the same $\epsilon$, but different phase shifts $\delta$ is also a symmetric function and its amplitude depends on the chosen $\delta$ values. The larger the absolute value of difference of the two $\delta$ values, the larger the amplitude of the difference and vice-versa. The inset in Figure (\ref{fig:u015}) shows the difference of the two symmetric solutions for same $\epsilon = 0.15$ and for two $\delta$ values, $\delta = 0$ and $\delta = \delta_m = 0.957$. Similar plots can be made for other values of $\epsilon$. For instance, if we take $\epsilon = 0.07$ and different $\delta$ values as $\delta_m$, $\delta_m + \pi/8$, $\delta_m + \pi/4$, and $\delta_m + 3\pi/8$, where $\delta_m = 0.4207362$, then the difference between the symmetric solutions for different pairs of $\delta$ values is shown in Figure (\ref{fig:diffdelta}).

\subsection{Oscillatory tail amplitude}

\begin{figure}[ht!]
\centering
    \includegraphics[width=0.75\linewidth]{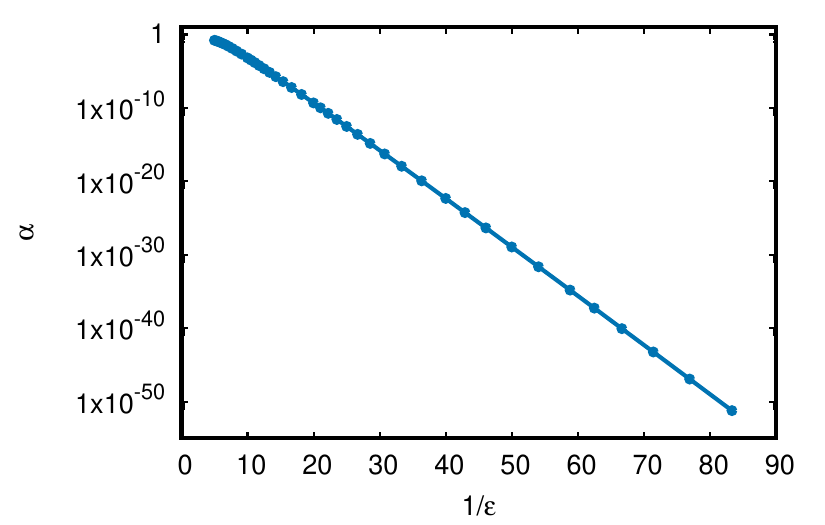}
    \caption{
       A plot of of the oscillatory tail amplitude $\alpha$ with respect to inverse of a small parameter $\epsilon$. }
    \label{fig:invepsal}
\end{figure}

The plot of the oscillatory tail amplitude $\alpha$ of the symmetric weakly localized solution of the fKdV equation (\ref{eqn:1aintFKDV}) as a function of $1/\epsilon$ can be seen in Figure (\ref{fig:invepsal}). This plot shows that the tail amplitude $\alpha$ decreases exponentially with $1/\epsilon$. The phase $\delta$ in all the calculations (for all $\epsilon$ values considered) is taken to be the minimum phase $\delta_m$.

The amplitude $\alpha$ of the oscillatory tail is very sensitive to small changes in $\epsilon$. As $\alpha$ is exponentially small in $\epsilon$, a small change in $\epsilon$ can change $\alpha$ by orders of magnitude. As $\epsilon$ decreases, $\alpha$ decreases exponentially fast and the function $u$ resembles more and more to the KdV 1-soliton solution for all $x$, see equation (\ref{eqn:u0}) and Figure (\ref{fig:u0fig}). The values of the minimum phase $\delta_m$ and the corresponding minimal tail amplitude $\alpha_m$ for several $\epsilon$ values are listed in Table (\ref{table:alpdel})
\begin{table}[ht!]
\centering
\begin{tabular}{c c c|c c c}
\hline
\hline
$\epsilon$   &  $\alpha_m$             &   $\delta_m$    &    $\epsilon$   &  $\alpha_m$ &   $\delta_m$  \\
\hline
$0.15$       &   $4.1\times 10^{-2} $               &   $0.958$   &  $0.05$       &   $4.811363375\times 10^{-10}$       &   $0.3001268310$                   \\
$0.12$       &   $5.72\times 10^{-3} $               &   $0.7383$    &   $0.04$       &   $3.024796077\times 10^{-13}$        &   $0.2400405089$                 \\
$0.1$        &   $6.572\times 10^{-4}$              &   $0.06552$   &   $0.035$      &   $1.472008979\times 10^{-15}$       &   $0.2100205651$                 \\
$0.085$       &   $6.04724\times 10^{-5}$           &   $0.512126$    &   $0.025$      &    $4.771438977\times 10^{-23}$     &       $0.1500037632$                \\ 
$0.07$       &   $1.802403\times 10^{-6}$           &   $0.4207362$   &   $0.02$      &   $1.142588001\times 10^{-29}$      &   $0.1200012259$              \\
$0.065$       &   $3.791592\times 10^{-7}$           &   $0.39049678$   &  $0.017$       &   $1.527829748\times 10^{-35}$      &   $0.1020005424$               \\ 
$0.06$       &   $6.0519239\times 10^{-8}$           &   $0.36032631$   &  $0.013$       &   $1.189881243\times 10^{-47}$      &   $0.0780001414$                \\ 
$0.55$       &   $6.7903467\times 10^{-9}$               &   $0.330207479$ &   $0.012$       &   $5.935328843\times 10^{-52}$           &   $0.0720000947$                   \\
\hline
\hline
\end{tabular}
\caption{The numerical values of minimum phase $\delta_m$ and the corresponding minimum tail amplitude $\alpha_m$ for various $\epsilon$ values.}
\label{table:alpdel}
\end{table}

\begin{figure}[ht!]
\centering
    \includegraphics[width=0.75\linewidth]{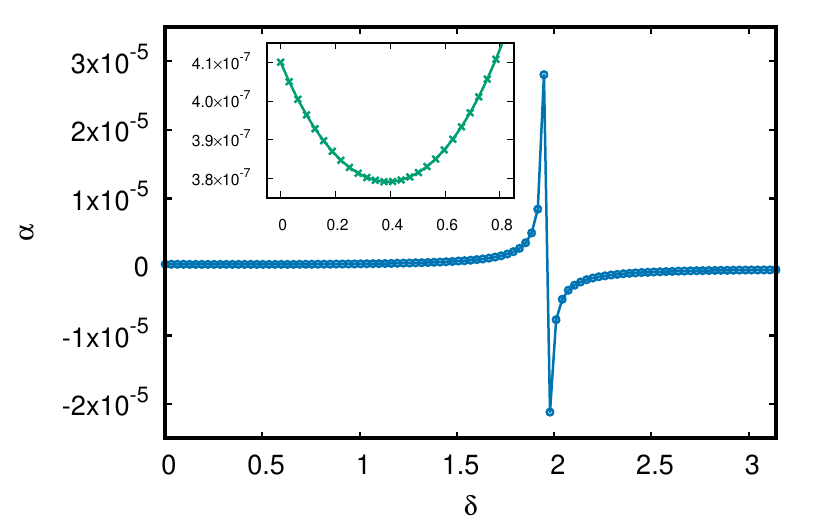}
    \caption{
       Main: A plot of the dependence of the oscillatory tail amplitude $\alpha$ on the phase parameter $\delta$ in the interval $[0,\pi]$ for $\epsilon = 0.065$.
       Inset: A plot of the oscillatory tail amplitude $\alpha$ with respect to phase $\delta$ in a smaller interval $[0,\pi/4]$.}
    \label{fig:alphadelta}
\end{figure}

Figure (\ref{fig:alphadelta}) illustrates how the oscillatory tail amplitude $\alpha$ varies with the phase parameter $\delta$. In this numerical simulation, the parameter $\epsilon$ is fixed and has value $0.065$. The main part in Figure (\ref{fig:alphadelta}) shows the dependence of the tail amplitude $\alpha$ on the phase parameter $\delta$ in the interval $[0,\pi]$. As can be seen from inset in Figure (\ref{fig:alphadelta}), there is a huge peak at some particular $\delta$ value. The reason for this is that for a given $\epsilon$, the tail amplitude $\alpha$ of any symmetric solution with phase $\delta$ is related to the minimal tail amplitude $\alpha_m$ by
\begin{equation}
    \alpha=\frac{\alpha_m}{\mathrm{cos}(\delta-\delta_m)} \ , \nonumber
\label{eqn:alpham}
\end{equation}
which will be explained in detail in Chapter (\ref{chapter:wkbsolution}). The peak (correspond to $\alpha\to\pm\infty$) occurs when the denominator of above equation becomes zero, i.e. when $\delta=\pm(2n+1)\pi/2+\delta_m$, for any positive integer $n$. For a current chosen $\epsilon$, the huge peak occurs at about $\delta\approx 1.98$. An inset in Figure (\ref{fig:alphadelta}) shows the dependence of the tail amplitude $\alpha$ on the phase $\delta$ in the interval $[0,\pi/4]$. We have chosen a shorter interval so that we could see the position of the minimum phase $\delta_m$, which otherwise is not so clear from the main part of Figure (\ref{fig:alphadelta}). For $\epsilon = 0.065$, the numerical value of the minimum phase is $\delta_m=0.3904968$ and the corresponding value of the minimum tail amplitude is $\alpha_m =3.791592\times 10^{-7}$.

\section{Asymmetric solutions}

\begin{figure}[ht!]
	\centering
        \includegraphics[width=0.7\linewidth]{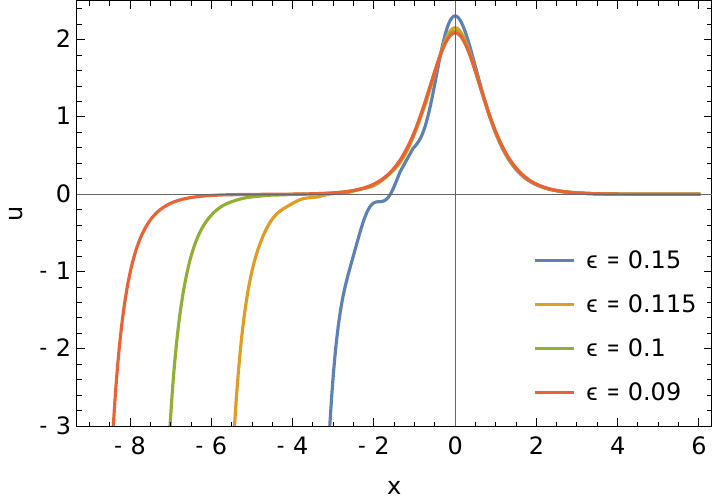}
        \caption{
            The plots of the asymmetric solutions of stationary fKdV equation (\ref{eqn:1aintFKDV}) for several $\epsilon$ values. It shows that the smaller the $\epsilon$ value is, the larger the value of spatial coordinate $x$ at which the solution blows up.
    \label{fig:asymplots}}
\end{figure}

In this section, we discuss the asymmetric solutions of the stationary fKdV equation (\ref{eqn:1aintFKDV}) which decay exponentially to zero on one side of the (slightly) asymmetric core and blow up to the negative values on the other side of the core. The stationary asymmetric solutions of the fKdV equation for several $\epsilon$ values can be seen in Figure (\ref{fig:asymplots}). This figure also shows that the smaller the value of $\epsilon$ is, the larger the value of $x$ at which the solution blows up.

\subsection{Third derivative at the origin}

The asymmetry of a solution can be quantified by computing the third derivative of the asymmetric solution $u_-$ at the origin. The numerical values of the third derivative of a function $u$ which decays exponentially to zero on one side of the core and blow up to the negative values on the other side, at the center $x = 0$ for several $\epsilon$ values can be seen in Table (\ref{table:thirdder}).
\begin{table}[ht!]
\centering
\begin{tabular}{c c|c c c}
\hline
\hline
$\epsilon$ &       $\partial^3_x u_-(0)$                  &     $\epsilon$   &   $\partial^3_x u_-(0)$                     \\
\hline
$0.1$     &       $0.592$           &     $0.01$       &   $3.66952012851177\times 10^{-57}$      \\
$0.09$    &       $0.1882$           &     $0.009$      &      $1.64049569151658\times 10^{-64}$  \\
$0.08$    &       $0.04077$           &     $0.008$      &      $9.93894226062684\times 10^{-74}$   \\
$0.07$    &       $50802\times 10^{-3}$           &     $0.007$      &      $1.27878547964615\times 10^{-85}$   \\
$0.06$    &       $2.739205\times 10^{-4}$           &     $0.006$      &      $1.58626892919443\times 10^{-101}$   \\
$0.05$    &   $3.793352409836\times 10^{-6}$    &     $0.005$      &      $7.21255715711993\times 10^{-124}$   \\
$0.04$    &   $4.684749893695\times 10^{-9}$    &     $0.004$      &      $1.71646090989488\times 10^{-157}$   \\
$0.03$    &   $4.240077936084\times 10^{-14}$   &     $0.003$      &      $1.02732441063580\times 10^{-213}$   \\
$0.02$    &   $1.425312199619\times 10^{-24}$   &     $0.002$      &      $1.56862990474708\times 10^{-326}$  \\
\hline
\hline
\end{tabular}
\caption{The numerical values of the third derivative of an asymmetric function $u_-$ which decays exponentially to zero on one side of the core, at the origin $x = 0$.}
\label{table:thirdder}
\end{table}

\begin{figure}[ht!]
	\centering
        \includegraphics[width=0.75\linewidth]{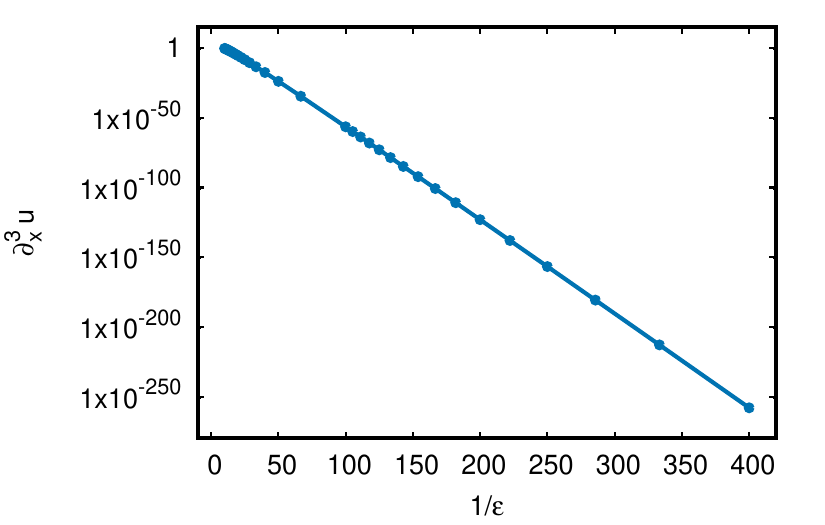}
        \caption{
            The plot of the third derivative as a function of the inverse of the small parameter $\epsilon$.
    \label{fig:inv3der}}
\end{figure}

The plot of the third derivative of the asymmetric function $u_-$ at the origin as a function of $1/\epsilon$ is shown in Figure (\ref{fig:inv3der}). The plot shows that similar to the oscillatory tail amplitude $\alpha_m$, the third derivative of an asymmetric function at the origin is also  an exponential function of $1/\epsilon$.

\section{Difference between the symmetric and the asymmetric solutions}
\begin{figure}[ht!]
	\centering
        \includegraphics[width=0.6\linewidth]{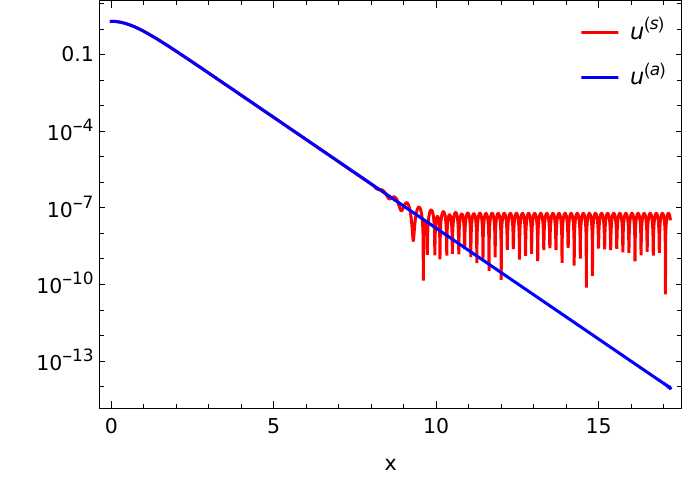}
        \caption{
            A logarithmic plot of the symmetric and the asymmetric solutions as a function of $x$ for $\epsilon = 0.06$. The red curve shows the tail of the symmetric solution and the blue curve shows the exponential decaying behaviour of the asymmetric solution without any oscillatory tail at large $x$. 
    \label{fig:udifflog}}
\end{figure}

The symmetric solution (lets us call it here as $u^{(s)}$) has small standing wave tail oscillations on both sides of the core, see Figure (\ref{fig:u0.15}). The asymmetric solution (lets call it here as $u^{(a)}$) decays exponentially to zero on one side of the core and blow up to the negative values on the other side of the core, see Figure (\ref{fig:asymplots}). The plot of the symmetric and the asymmetric solutions logarithmically as a function of $x$ can be seen in Figure (\ref{fig:udifflog}).

\begin{figure}[ht!]
	\centering
        \includegraphics[width=0.6512\linewidth]{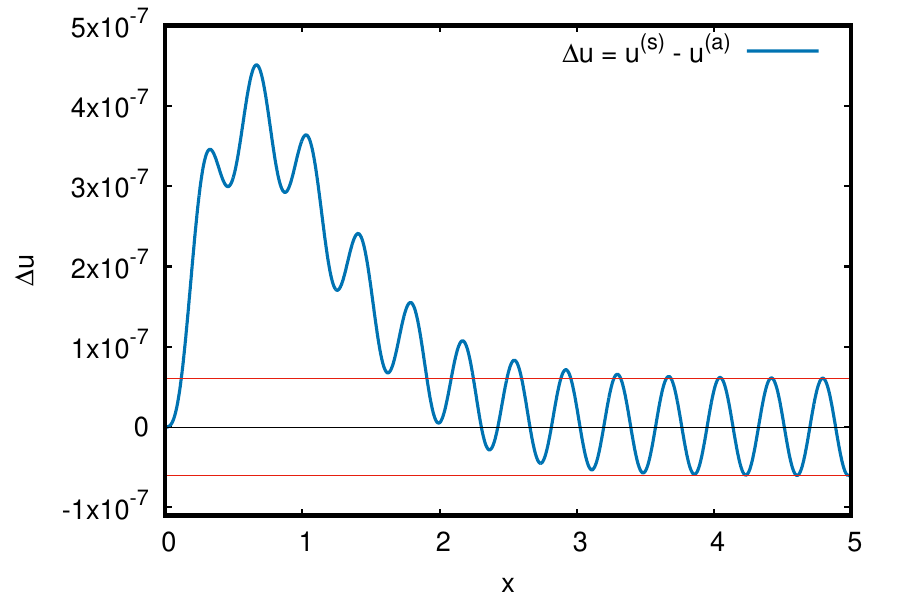}
        \caption{
            A plot of the difference between symmetric and asymmetric solutions for $\epsilon = 0.06$. The thin red horizontal lines represents the minimal-amplitude tail values $\pm \alpha_m$, where $\alpha_m = 6.051923896 \times 10^{-8}$.
    \label{fig:udiff}}
\end{figure}

We subtract the asymmetric solution $u^{(a)}$ from the symmetric solution $u^{(s)}$ for $x>0$ where for large $x$, $u^{(s)}$ has extremely small standing wave oscillations and $u^{(a)}$ decays exponentially to zero. Let us call this difference as $\Delta u = u^{(s)} - u^{(a)}$. The symmetric solution can be written as $u^{(s)} = u^{(c)} + \alpha \ \mathrm{sin}(k x /\epsilon - \delta)$, where $u^{(c)}$ represents the core region of the solution which is a polynomial in $\mathrm{sech}^2(\gamma x)$, where $\gamma$ is related to the amplitude of the KdV 1-soliton solution which in turn is related to the wave speed of the same, $\alpha$ is the amplitude of the standing wave tail oscillations, $k/\epsilon$ is the far field wave number and $\delta$ is the far field phase shift. Hence to the leading order in $\epsilon$ we expect the difference $\Delta u$ as a sine function of amplitude $\alpha$. When $\epsilon$ is very small the minimal-amplitude tail corresponds to $\delta = 0$ (for details, see Chapter (\ref{chapter:wkbsolution})) and hence the linearized symmetric solution can be written as $u^{(s)} = u^{(c)} + \alpha_m \ \mathrm{sin}(kx /\epsilon)$. In this case we expect the difference $\Delta u$ as a sine function with the same amplitude as the minimal-amplitude tail $\alpha_m$ in the far field oscillations of $u^{(s)}$. The plot of the difference $\Delta u$ for $\epsilon = 0.06$ can be seen in Figure (\ref{fig:udiff}).

\begin{figure}[ht!]
    \centering 
    \begin{subfigure}{0.5\textwidth}
    \includegraphics[width=\linewidth]{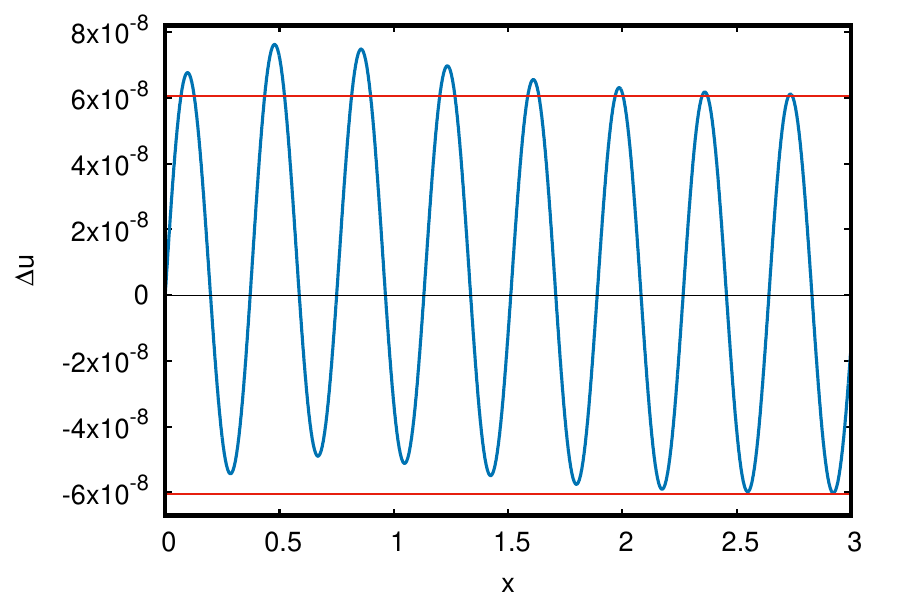}
    \caption{}
    \label{fig:diffsh1}
    \end{subfigure}\hfil 
    \begin{subfigure}{0.5\textwidth}
    \includegraphics[width=\linewidth]{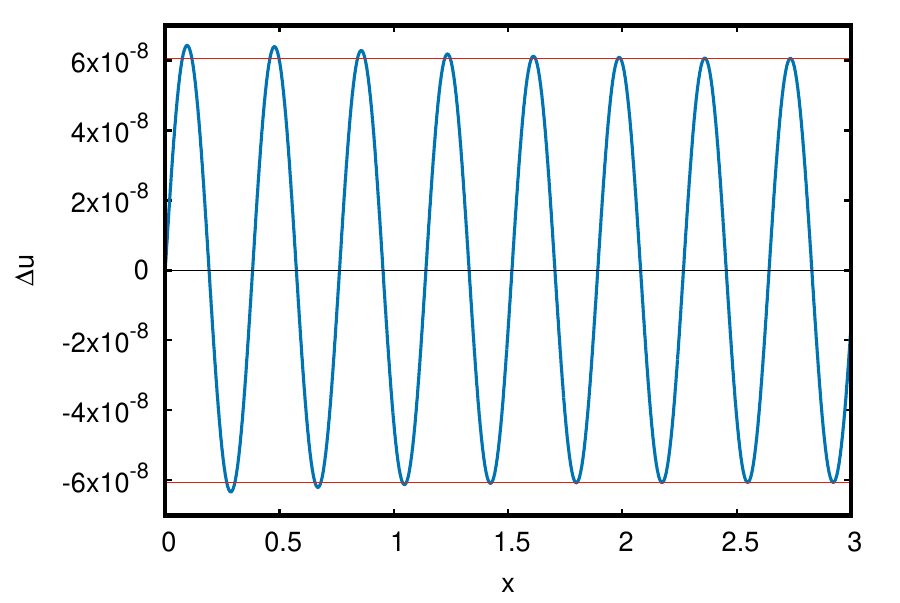}
    \caption{}
    \label{fig:diffsh2}
    \end{subfigure}\hfil 
 \caption{The plots of the difference $\Delta u = u^{(s)} - u^{(a)}$: (\ref{fig:diffsh1}) when only first term is included in the shift $d$ and (\ref{fig:diffsh2}) when first two terms are included in $d$. The thin red horizontal lines represents the minimal- tail amplitude values $\pm \alpha_m$.}
 \label{fig:udiffsh14}
 \end{figure}

The difference function $\Delta u$ in Figure (\ref{fig:udiff}) does not look like the sine function as the amplitude is larger and varies in some region near the origin. However the difference can be made smaller if we include a shift. The magnitude of the shift $d$ which will be calculated in Chapter (\ref{chapter:thder}) is given as 
$$d = \alpha_m\left(-\frac{1}{4\gamma^4\epsilon}+\frac{9}{4\gamma^2}\epsilon+\frac{113}{4}\epsilon^3+\frac{4987}{4}\gamma^2\epsilon^5+\mathcal{O}(\epsilon^7)\right)+\mathcal{O}(\alpha^2_m) \ .$$
Applying  the shift to the asymmetric solution, $u^{(a)}(\tilde x) = u^{(a)}(x - d)$, we expect to get the difference closer to the sine function. A couple of plots of the difference $\Delta u$ when different orders in shift $d$ is applied to $u^{(a)}$ can be seen in Figure (\ref{fig:udiffsh14}). The first graph, Figure (\ref{fig:diffsh1}) already looks quite close to the sine function when including only the leading order term in the shift $d$. Including one more term in the shift $d$, we get the result as shown in Figure (\ref{fig:diffsh2}) which almost looks the sine function. 

\chapter{Analytical results}\label{chapter:analytical_results}
 
In this chapter, we discuss the analytical expansion methods of how to find the extremely small far field oscillatory tail amplitude of \emph{nanopteron} -- stationary symmetric weakly localized solutions $u$ of the fKdV equation, by using the method of matched asymptotics in the complex plane and the Laplace transform (or equivalently the Borel summation). In this chapter, we give the leading order analytical result for the minimal tail amplitude $\alpha_m$. The higher order approximations for tail amplitude $\alpha_m$ are calculated in the latter chapters.

\section{Exponential Asymptotics}\label{section:exponential}

A power series expansion is an asymptotic expansion (also called Poincar{\'e} expansion) to a function $f(\epsilon)$ if, for any fixed integer $n$ and sufficiently small $\epsilon>0$ \cite{boyd, steinruck}
\begin{equation}
    f(\epsilon)-\sum_{j=0}^n a_j\epsilon^j = \mathcal{O}\left(\epsilon^{n+1}\right) \, . \label{eqn:f(x)}
\end{equation}
If equation (\ref{eqn:f(x)}) holds then we can write
\begin{equation}
    f(\epsilon) \sim \sum_{j=0}^n a_j\epsilon^j \, .
\end{equation}
The main point here is that the above asymptotic expansion fails to find terms proportional to $\exp{(-\mathcal{P}/\epsilon)}$, where $\mathcal{P}>0$ is a constant. Such terms cannot be approximated as a power series in the perturbation parameter $\epsilon$ since all the derivatives are zero at $\epsilon = 0$. To find such extremely small terms requires \emph{exponential asymptotics} (also called \emph{asymptotics beyond all orders}). For details of this method, see for instance \cite{boyd, benderorszag, steinruck}.

\begin{figure}[ht!]
\centering
    \begin{tikzpicture}
        \draw [thick, ->] (0.12,0.12) -- ((1.3,1.3)node[right]{dominant singularity};
        \node [black, thick] at (3.3,0.93){above real $x$-axis};
        
        \fill [black!2] (-6,-5.5) rectangle (6,0);
         \draw[->, thick] (-6,-5.5) -- (6,-5.5)node[right]{$\Re(x)$};
        \draw[rotate = 180, green, fill = green!8] (3,0) arc(0:180:3) -- (-3,0);
        \draw[rotate = 180, red, fill = red!8] (1.7,0) arc(0:180:1.7) -- (-1.7,0);
    
        \draw[->,thin] (0,-6.5) -- (0,1.7)node[above]{$\Im(x)$};
        \filldraw[red] (0,0) circle (2.5pt);
        \draw[->,thick, dashed] (-6,0) -- (6,0)node[right]{$\Re(q)$};
        
        \node [red, thick] at (0,-0.75){\emph{Inner region}};
        \node [green, thick] at (0,-2.2){\emph{Matching region}};
        \node [black, thick] at (0,-4.2){\emph{Outer region}};
    \end{tikzpicture}
\caption{A schematic diagram of the outer, inner and matching regions.}
\label{fig:oim}
\end{figure}
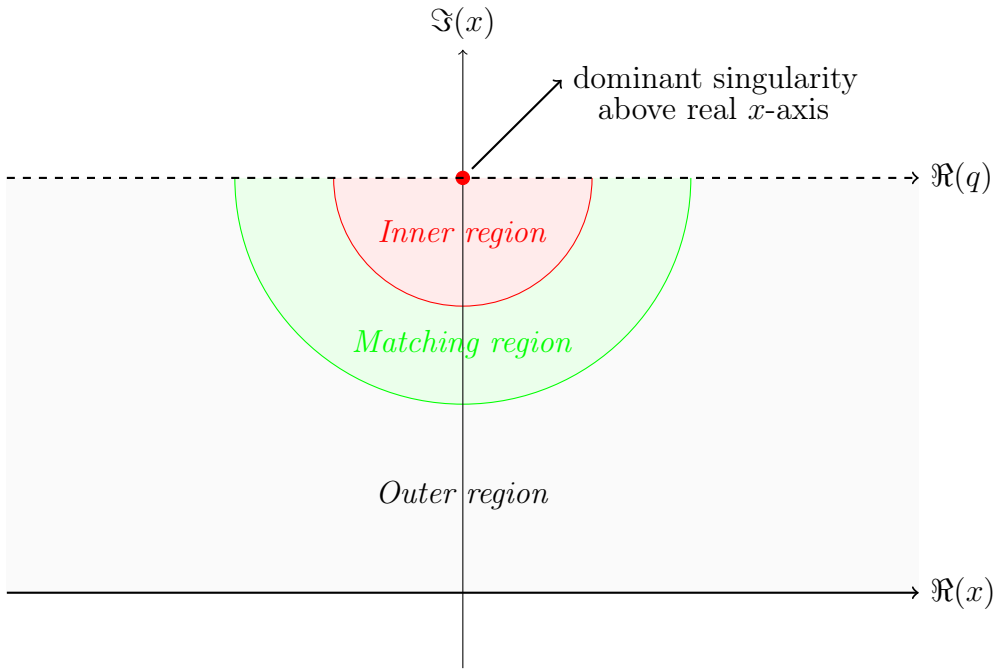

The basic idea of the method of matched asymptotics is the following; we have an asymptotic series expansion (usually this series expansion is a divergent series) in powers of a small parameter $\epsilon$. The optimally truncated version of this expansion approximates well the core region of the solution. This expansion is called the \emph{outer expansion}. Since the oscillatory tail amplitude is beyond all orders small in perturbation theory, i.e. it is $\mathcal{O}(\exp{(-1/\epsilon))}$, this series expansion gives no information regarding these extremely small tail oscillations of the solution. To find the oscillatory tails we allow the independent variable $x$ to take complex values. In the complex plane the series expansion has singularities along the imaginary axis and the outer expansion near these singularities is inaccurate. Hence near the singularities we have to treat this problem as a separate perturbation problem. This new approximation is called the \emph{inner expansion}. We also apply a rescaling in the independent variable $x$ to magnify the inner region. Then we have to combine the outer and inner solutions in a process called \emph{matching} in such a way that an approximate solution for the whole domain is obtained. The schematic diagram of the different regions can be seen in Figure (\ref{fig:oim}).

\subsection{Outer problem}

\begin{figure}[ht!]
    \centering 
    \begin{subfigure}{0.5\textwidth}
    \includegraphics[width=\linewidth]{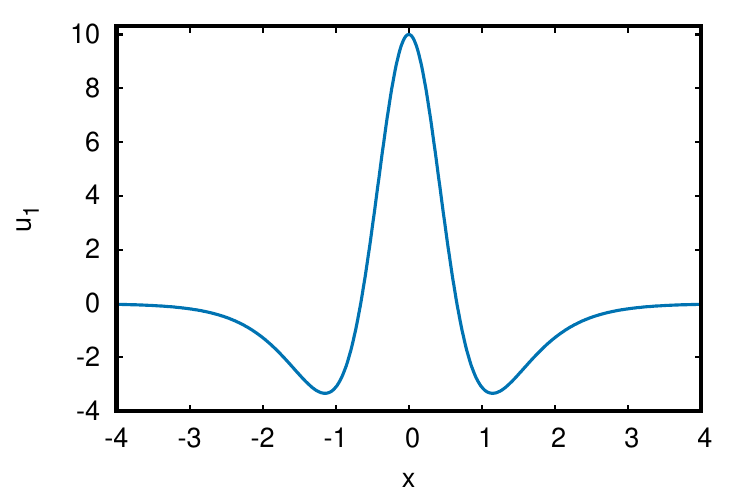}
    \caption{}
    \label{fig:u1}
    \end{subfigure}\hfil 
    \begin{subfigure}{0.5\textwidth}
    \includegraphics[width=\linewidth]{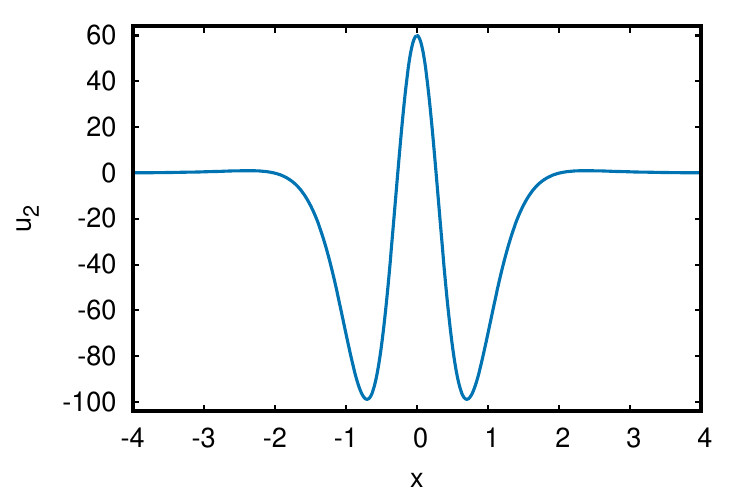}
    \caption{}
    \label{fig:u2}
    \end{subfigure}\hfil 
    \vspace {0.6cm}
    \begin{subfigure}{0.5\textwidth}
    \includegraphics[width=\linewidth]{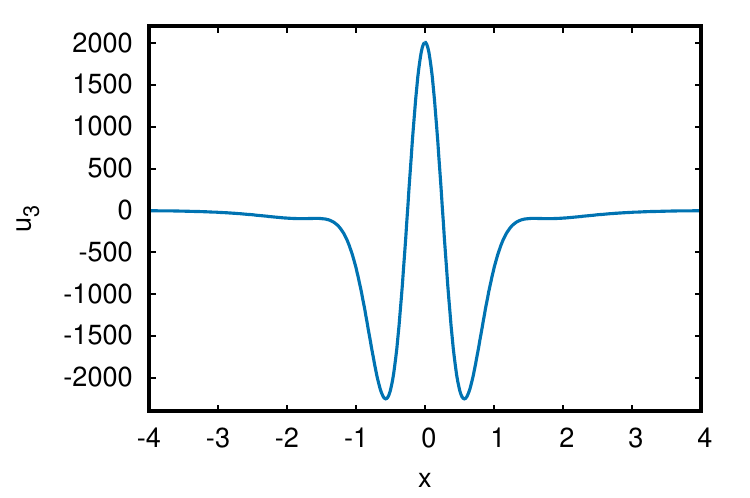}
    \caption{}
    \label{fig:u3}
    \end{subfigure}\hfil 
    \begin{subfigure}{0.5\textwidth}
    \includegraphics[width=\linewidth]{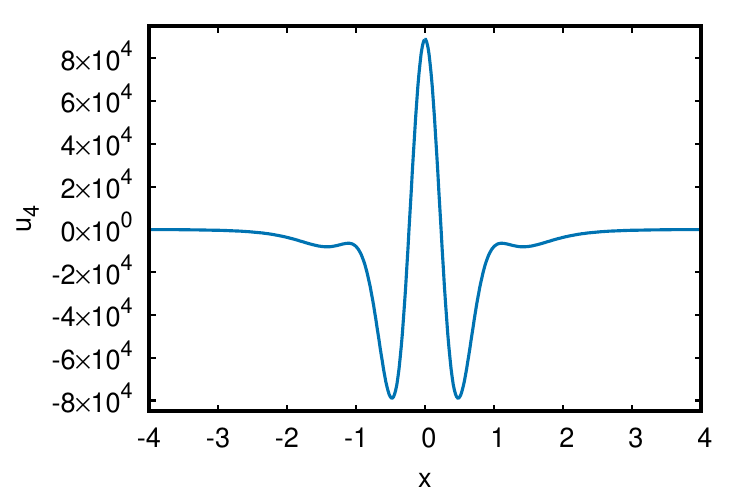}
    \caption{}
    \label{fig:u4}
    \end{subfigure}\hfil 
 \caption{The graph of $u_{1}$, $u_{2}$, $u_{3}$ and $u_{4}$ as a function of $x$. In these plots, the parameter $\gamma$ is set to 1. The function $u_{j}(x)$ oscillate more and more rapidly as the order $j$ increases. All $u_{j}$ with $j > 0$ have at least one pair of roots and the smallest pair moves closer and closer to the origin with increasing $j$, which implies that the length scale of the solution $u_j(x)$ is decreasing with increasing order of the perturbation.}
 \label{fig:u1234}
 \end{figure}

The construction of an asymptotic expansion in powers of the small parameter $\epsilon$ is used to describe the core region of the solution $u$ of the fKdV equation (\ref{eqn:1aintFKDV}). This solution $u$ may be numerically calculated symmetric solution which has the transcendentally small oscillatory tail on both sides of the core or it may be the asymmetric solution which decays exponentially to zero for $x>0$. We seek an asymptotic expansion of the following form
\begin{equation}
    u = \sum_{n=0}^{\infty}\epsilon^{2n}u_{n} \ ,\quad\text{and}\quad c = \sum_{n=0}^{\infty}\epsilon^{2n}c_{n} \, ,
\label{eqn:coreuexp}
\end{equation}
where $u_n$ are functions of $x$ and $c_n$ are numbers. As the series expansion equation (\ref{eqn:coreuexp}) is an asymptotic expansion, we have to truncate it at some positive integer $N$ to get the meaningful finite result. The error of the approximation of the core region is the smallest when the series is truncated at some optimal value $N_{\mathrm{opt}}$, which is expected to increase with decreasing $\epsilon$. Substituting equation (\ref{eqn:coreuexp}) into the once integrated fKdV equation (\ref{eqn:1aintFKDV}), the vanishing of the $\epsilon$ independent part gives the usual KdV equation
\begin{equation}
    u_{0,xx}+3u_0^2-c_0u_0=0 \, ,
\end{equation}
which has the solution $u_0$ given in equation (\ref{eqn:u0}). For $n>0$, the vanishing of the coefficient of $\epsilon^{2n}$ gives
\begin{equation}
    u_{n-1,xxxx}+u_{n,xx}+\sum_{j=0}^n(3u_j - c_j)u_{n-j}=0 \, .
\label{eqn:recurun}
\end{equation}
Equation (\ref{eqn:recurun}) can be considered as an inhomogeneous differential equation which is used to determine the solutions $u_n$, provided we assume that the functions are known up to order $n-1$.
\begin{equation}
    u_{n,xx}+6u_{0}u_{n}-c_{0}u_{n}=R_{n} \, ,
\label{eqn:ujkcnres}
\end{equation}
where 
\begin{equation}
    R_{n}=-u_{n-1,xxxx}-3\sum_{j=1}^{n-1}u_{j}u_{n-j}+\sum_{j=1}^{n}c_{j}u_{n-j} \, 
\label{eqn:res}
\end{equation}
is the source term in the inhomogeneous linear equation  (\ref{eqn:ujkcnres}).  For $n=1$, the source term $R_{1}$ is given as
\begin{equation}
    R_{1}=u_{0,xxxx}-c_{1}u_{0} \, ,
\label{eqn:res1}
\end{equation}
where $u_0$ is given in equation (\ref{eqn:u0}), $u_0=2\gamma^2 \mathrm{sech}^2(\gamma x)$. So, $R_1$ can be written as
\begin{equation}
    R_{1}=(32\gamma^4 - 2 c_{1})\gamma^2 \mathrm{sech}^{2}(\gamma x)-140\gamma^{6}\mathrm{sech}^{4}(\gamma x)+140\gamma^{6}\mathrm{sech}^{6}(\gamma x) \, .
\label{eqn:res1u0}
\end{equation}
If we assume $u_n=a \ \mathrm{sech}^2(\gamma x)+b \ \mathrm{sech}^4(\gamma x)+\cdots+g \ \mathrm{sech}^{2(n+1)}(\gamma x)$, where $a,b,\ldots, g$ are real constants, and operate the linear operator $u_{n,xx}+6u_{0}u_{n}-c_{0}u_{n}$ on it, there will be no terms like $\mathrm{sech}^2(\gamma x)$, but we still get a polynomial in $\mathrm{sech}^2(\gamma x)$. However, we can eliminate these ($\sim \mathrm{sech}^2(\gamma x)$) terms from $R_n$ by a proper choice of the constants $c_{n}$, so that the coefficient of $\mathrm{sech}^2(\gamma x)$ vanishes. For example, for $n=1$, we choose 
\begin{equation}
    c_{1}=(4\gamma^2)^2=c_{0}^2 \, .
\label{eqn:c1}
\end{equation}
Then there will be no $\mathrm{sech}^2(\gamma x)$ term in $R_1$ and hence, we can solve equation (\ref{eqn:res1u0}) to obtain
\begin{equation}
    u_{1}=-20\gamma^{4} \mathrm{sech}^{2}(\gamma x)+30 \gamma^4 \mathrm{sech}^{4}(\gamma x)\, .
\label{eqn:u1}
\end{equation}
It is clear from equation (\ref{eqn:res}) that the source term $R_{n}$ consists of terms proportional to $u_0(x)$ and its integer powers only. By using the $\mathrm{sech}$ identities and recursion, one can show that the general term $u_{n}$ is an $(n+1)$-th order polynomial in $\mathrm{sech}^{2} (\gamma x)$ and the general term $R_{n}$ is an $(n+2)$-th order polynomial in $\mathrm{sech}^{2} (\gamma x)$,
\begin{equation}
    u_{n}=\sum_{j=1}^{n+1}u_{nj}\gamma^{2(n+1)}\mathrm{sech}^{2j}(\gamma x) \, ,
\label{eqn:un}
\end{equation}
and 
\begin{equation}
    R_{n}=\sum_{j=1}^{n+2}R_{nj}\gamma^{2(n+2)}\mathrm{sech}^{2j}(\gamma x) \, .
\label{eqn:Rn}
\end{equation}
Comparing equation (\ref{eqn:un}) with $u_0 = 2\gamma^2\mathrm{sech}^2(\gamma x)$ given in equation (\ref{eqn:u0}), we obtain $u_{01}=2$. We can determine the phase speed corrections $c_{n}$ by keeping in mind that the $j$-th source term should not contain $\mathrm{sech}^2 (\gamma x)$ terms. Mathematically, it can be written as 
\begin{equation}
    R_{n1}=0 \qquad \forall \quad n \, .
\label{eqn:rj1}
\end{equation}
Proceeding with the calculation it turns out that $c_1 = 16\gamma^4$ and $c_j = 0$ for $j\geq 2$, consistent with $c(\epsilon)=4\gamma^2+16\gamma^4\epsilon^2$, see equation (\ref{eqn:exactgamc}). The series expansion for the solution $u$, equation (\ref{eqn:coreuexp}) can be written as 
\begin{equation}
    u(x;\epsilon)\sim \sum_{n=0}^{\infty}\sum_{j=1}^{n+1}u_{nj}\epsilon^{2n}\gamma^{2(n+1)}\mathrm{sech}^{2j} (\gamma x) \, .
\label{eqn:useries}
\end{equation}
We can derive the expressions which can be used to compute the coefficients $u_{jk}$, $c_j$ and $R_{jk}$ by equations (\ref{eqn:ujkcnres}), (\ref{eqn:res}), (\ref{eqn:un}) and (\ref{eqn:Rn}). A similar algorithm to calculate these coefficients is presented in Table 10.5 of Boyd's book \cite{boyd}. Please note that there is a typo in the pseudo-code there, the summation in the phase speed contributions should start from $m=1$, not from $m=0$. The results for the first few $u_{nj}$ coefficients can be found in Table (\ref{table:ujk}). The plots of the $u_n(x)$ functions for first few values of $n$ can also be seen in Figure (\ref{fig:u1234}). 
\begin{table}[ht!]
\centering
\begin{tabular}{c c c c c c c}
\hline
\hline
n$\downarrow$ j$\rightarrow$&       1      &     2       &     3      &      4      &      5    \\
\hline
0                           &       2      &     -       &     -      &      -      &      -     \\
1                           &       -20    &     30      &     -      &      -      &      -      \\
2                           &       60     &   -930      &     930    &      -      &      -      \\
3                           &       -2472  &   21036     &   -66216    &      49662  &     -     \\
4                           &     -$\frac{240780}{7}$ &-$\frac{3177030}{7}$ &$\frac{23319570}{7}$ & -$\frac{48197250}{7}$ &$\frac{28918350}{7}$  \\
\hline
\hline
\end{tabular}
\caption{The value of a few coefficients $u_{nj}$ of the series expansion in equation (\ref{eqn:un}).}
\label{table:ujk}
\end{table}

\begin{figure}[ht!]
\centering
    \includegraphics[width=0.75\linewidth]{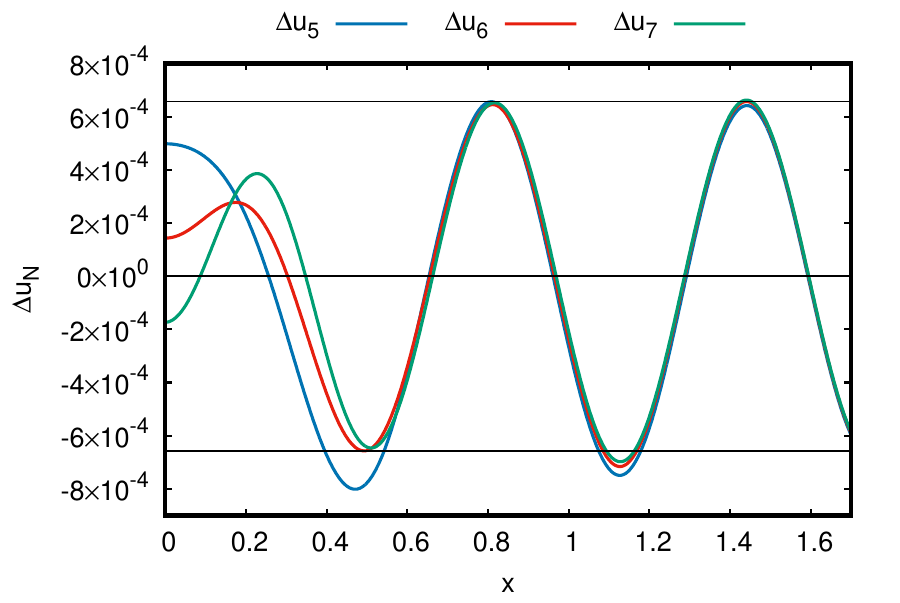}
    \caption{
       The difference of the $N$-th order approximation of the asymptotic series expansion equation (\ref{eqn:coreuexp}) from the numerically calculated minimal tail symmetric solution $u$ for $\epsilon=0.1$. The thin horizontal black lines show the value of the tail amplitude $\alpha_m = 6.571690 \times 10^{-4}$.}
    \label{fig:ucoreexp}
\end{figure}

If the solution $u$ has been already calculated precisely by some numerical method, we can compare it to various orders of the asymptotic series expansion given in equation (\ref{eqn:coreuexp}) where various coefficients are given in Table (\ref{table:ujk}). We define the error of the $N$-th order analytical approximation by
\begin{equation}
    \Delta u_N = u - \sum_{n=0}^N u_n\epsilon^{2n} \, .
\label{eqn:ucoreerror}
\end{equation} 
In Figure (\ref{fig:ucoreexp}), the plot of the error function $\Delta u_N$ is shown for $\epsilon = 0.1$. The plot shows the three values of $N$ for which the error is the smallest. It is clear from the plot that for $\epsilon = 0.1$, the optimal value is at $N_\mathrm{opt} = 6$. The difference $\Delta u_N$ for large $|x|$ should agree with the oscillating tail as all the $u_n$ functions decay exponentially to zero. From the plot, we can see that the difference for the optimal value even agrees in most of the core region. Hence the error of the optimally truncated series expansion is anticipated to be exponentially small in $\epsilon$, just like the oscillating tail at large $|x|$. However please note that the asymptotic series expansion (\ref{eqn:coreuexp}) cannot detect the exponentially small oscillating tail which we will explain a bit later.

Similar plots can be obtained for other $\epsilon$ values. The optimum values for various $\epsilon$ values are given in Table (\ref{table:Nopt}). As can be seen from the table, $N_\mathrm{opt}$ increases as $\epsilon$ decreases as can be expected from the asymptotic series expansion.
\begin{table}[ht!]
\centering
\begin{tabular}{c c|c c|c c}
\hline
\hline
$\epsilon$ &    $N_\mathrm{opt}$ &    $\epsilon$ &    $N_\mathrm{opt}$  &    $\epsilon$ &    $N_\mathrm{opt}$  \\
\hline
0.15       & 3                   &  0.07         &    10                &   0.025       &   30                 \\
0.12       & 7                   &  0.05         &    14                &   0.017       &   45                  \\
0.1        & 6                   &  0.035        &    21                &   0.012       &   64                   \\
\hline
\hline
\end{tabular}
\caption{The optimum values $N_\mathrm{opt}$ for various $\epsilon$ values.}
\label{table:Nopt}
\end{table}

 For real $x$, the series expansion in equation (\ref{eqn:coreuexp}) can be continued in $\epsilon^2$ to any desired order without any oscillatory tail being detected. The reason for this is that the size of the tail is beyond all orders small in perturbation theory, i.e. it is $\mathcal{O}(\exp{(-1/\epsilon)})$, and hence decays faster as $\epsilon\to 0$ than any powers of $\epsilon$. The tail oscillations can be found by solving the fKdV equation (\ref{eqn:1aintFKDV}) in a neighbourhood of the singularities of the function $u$ given in equation (\ref{eqn:coreuexp}) in the complex $x$-plane.
 
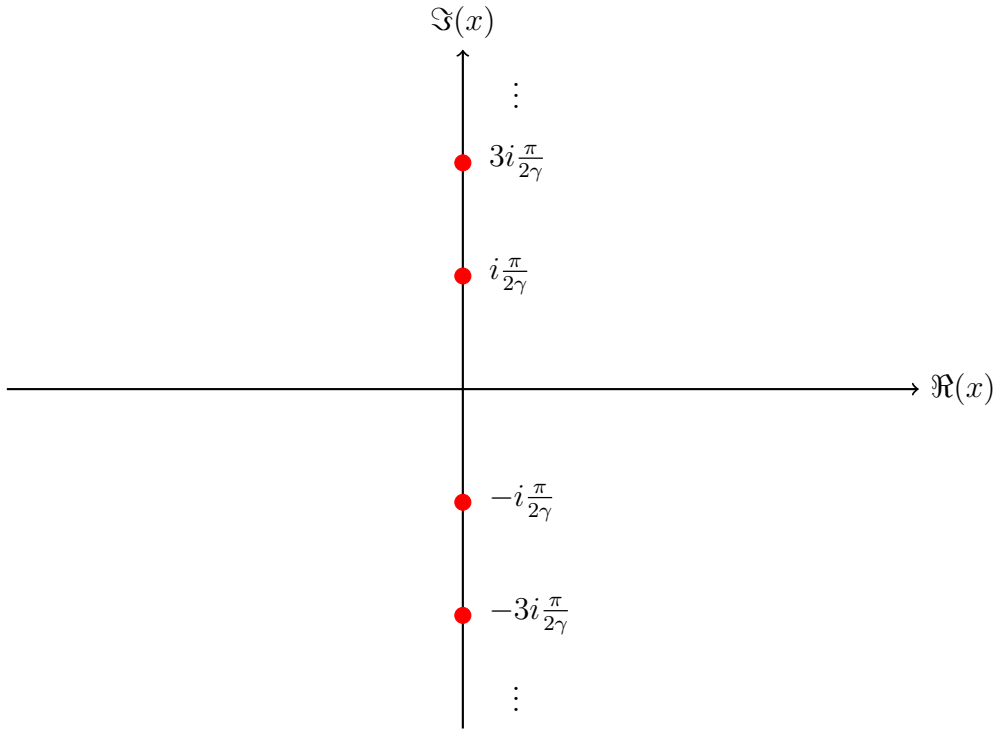
\begin{figure}[ht!]
\centering
    \begin{tikzpicture}
    
        \draw[->,thick] (0,-4.5) -- (0,4.5)node[above]{$\Im(x)$};
        \draw[->,thick] (-6,0) -- (6,0)node[right]{$\Re(x)$};


        \filldraw[red] (0,1.5) circle (3pt);
        \filldraw[red] (0,-1.5) circle (3pt);
        \filldraw[red] (0,3) circle (3pt);
        \filldraw[red] (0,-3) circle (3pt);
        
        \node at (0.2,1.5)[right]{$i\frac{\pi}{2\gamma}$ };
        \node at (0.2,-1.5)[right]{$-i\frac{\pi}{2\gamma}$ };

        \node at(0.2,3)[right]{$3i\frac{\pi}{2\gamma}$ };
        \node at(0.2,-3)[right]{$-3i\frac{\pi}{2\gamma}$ };

        \node at (0.5,4) [right]{$\vdots$ };
        \node at (0.5,-4)[right]{$\vdots$ };
        
    \end{tikzpicture}
\caption{The complex $x$-plane. The red dots represents the singularities in the outer expansion which approximates well the core region of the solution.}
\label{fig:complexplane}
\end{figure}

 To find the tail oscillations, we observe that all $u_j(x)$ functions are singular in the complex plane at $x= (2n+1)i\pi/(2\gamma)$, for integer $n$, see Figure (\ref{fig:complexplane}). 
 
 For the nearest pole above the real axis, $x= i\pi/(2\gamma)$, the Laurent series expansion of $\mathrm{sech}^2(\gamma x)$ is given as
 \begin{equation}
     \mathrm{sech}^2(\gamma x)=-\frac{1}{\gamma^2(x-i\pi/(2\gamma))^2}+\frac{1}{3}-\frac{1}{15}\gamma^2\left(x-\frac{i\pi}{2\gamma}\right)^2+\mathcal{O}\left(x-\frac{i\pi}{2\gamma}\right)^4 \, .
 \label{eqn:sechx}
 \end{equation}
 The expansion of the function $u$, see equation (\ref{eqn:coreuexp}) is actually the summation of different powers of the terms like in equation (\ref{eqn:sechx}). This motivates us to change the variable from $x$ to $q$ as 
 \begin{equation}
    x=\frac{i\pi}{2\gamma}+\epsilon q \, .
  \label{eqn:cordchan}
  \end{equation}
 Then equation (\ref{eqn:sechx}) can be written as
 \begin{equation}
     \mathrm{sech}^2(\gamma x)=-\frac{1}{(\gamma\epsilon q)^2}+\frac{1}{3}-\frac{1}{15}(\gamma\epsilon q)^2+\mathcal{O}(\gamma\epsilon q)^4 \, .
 \label{eqn:sechx1}
 \end{equation}
 Since the function $u$ is growing as $1/\epsilon^2$ near the singularity, it is natural to write the result in terms of the rescaled function $v$ as
 \begin{equation}
     v=\epsilon^2 u \, .
 \label{eqn:vepssq}
 \end{equation}
 Substituting equations (\ref{eqn:cordchan}) and (\ref{eqn:vepssq}) back into the asymptotic expansion of $u$ in equation (\ref{eqn:coreuexp}), we get the expansion of $v$ in $q$ which in general can be written as
 \begin{equation}
    v=\sum^{\infty}_{n=0}\epsilon^{2n}v_{n}(q) \, .
  \label{eqn:vexp}
 \end{equation}
The expansions of the first few $v_n$ functions are given as
\begin{align}
 v_0 &= -\frac{2}{q^2}+\frac{30}{q^4}-\frac{930}{q^6}+\frac{49662}{q^8}-\frac{28918350}{7 q^{10}}+\cdots \ ,
 \label{eqn:v0exp} \\
 v_1 &=\frac{2}{3}\gamma^2 \ ,
 \label{eqn:v1exp} \\
  v_2 &=\left(-\frac{2q^2}{15}+\frac{2}{3}+\frac{64}{q^2}+\frac{5856}{5 q^4}-\frac{827520}{7 q^{6}}+\cdots\right)\gamma^4 \ .
 \label{eqn:v2exp}
\end{align}
It is clear from the above expansions that each $v_n$ function starts with $q^{2(n-1)}$ powers. The function $v_1$ is just a constant which we know exactly, see equation (\ref{eqn:v1exp}). The remaining coefficients are the functions in terms of asymptotic expansion in $1/q$. However, these expansions can be used to approximate the real solutions up to some error. By using the matching procedure, we will use these expansions in the domain (called matching region) where both the ``inner'' and the ``outer'' solutions are valid, i.e. where both $\epsilon q$ and $1/q$ are small and hence $\epsilon^{2n}v_n$ is also small. The matching region is parametrized in the following way:
\begin{align}
    \left\{\epsilon q \ll 1, \ \epsilon \ll 1, \ |q| \gg 1, \ -\pi \leq \mathrm{arg}(q) \leq  0 \right\} \, .
\end{align}

\subsection{Inner problem}

The inner problem is defined by the rescaled variable near the singularity. Using the re-scaled function, equation (\ref{eqn:cordchan}), and the complex coordinate $q$, equation (\ref{eqn:vepssq}) in equation (\ref{eqn:1aintFKDV}), we get
 \begin{equation}
    v_{qqqq}+v_{qq}+3v^2-\epsilon^2c  v=0 \, ,
\label{eqn:vfkdv}
\end{equation}
where $c$ depends on $\epsilon$ through equation (\ref{eqn:exactgamc}). If we search for solutions of this equation in the form of expansion (\ref{eqn:vexp}), then taking the various $\epsilon^n$ contributions, we obtain differential equations for $v_n(q)$ functions.  The first two (zeroth and second order inner) equations are:
\begin{align}
v_{0,qqqq}+v_{0,qq}+3v_0^2 &= 0 \, ,
\label{eqn:v0} \\ 
v_{1,qqqq}+v_{1,qq}+2v_0(3v_1-2\gamma^2) &= 0 \, .
\label{eqn:v1}
\end{align}
Note that $\gamma$ does not appear in the zeroth order inner equation. We look for such solutions of the mode equations of the inner problem which can be matched to the solution of the outer problem.

The differential equation for $v_1$ can be solved independently of $v_0$ by $v_1=2\gamma^2/3$, which also agrees with the result obtained in equation (\ref{eqn:v1exp}). From now on, we will use this as an exact solution for $v_1$. The asymptotic expansion equations (\ref{eqn:v0exp}) through (\ref{eqn:v2exp}) will be used as matching conditions for the corresponding $v_n$ functions. These asymptotic expansions will provide valid boundary conditions when $\Im{(q)}<0$ is fixed and $\Re(q)\to\infty$ (or $\Re(q)\to -\infty$), determining unique inner solutions which we denote by $v_n^{(-)}$ (or $v_n^{(+)}$). As we will see later, these $v_n^{(-)}$ (or $v_n^{(+)}$) functions do not follow the symmetry condition and hence are asymmetric functions which do not have any oscillating tails in the region $\Re{(q)}>0$ (or $\Re{(q)}<0$). They can be associated to the unique asymmetric solutions $u_-$ (or $u_+$) of the once integrated fKdV equation (\ref{eqn:1aintFKDV}) by using the rescaled function given in equation (\ref{eqn:vepssq}). These asymmetric functions $u_-$ (or $u_+$) decay exponentially to zero in the positive (or negative) $x$ direction without any oscillating tail.

\subsubsection{Zeroth order inner problem}

Consider the zeroth order inner equation which is given in equation (\ref{eqn:v0}). We can determine the tail amplitude of the symmetric solution $u$ of equation (\ref{eqn:1aintFKDV}) to the leading order in $\epsilon$ by finding an appropriate precise solution of equation (\ref{eqn:v0}). The equation (\ref{eqn:v0exp}) is the matching condition for $v_0(q)$, which in general can be written as
\begin{equation}
    v_0 = \sum^{\infty}_{n=1}b^{(0)}_nq^{-2n},\quad\text{as}\quad|q|\to\infty\quad\text{in}\quad\Re(q)\geq 0,\quad \Im(q)<0 \, .
\label{eqn:genmatchcondv0}
\end{equation}
Substituting into equation (\ref{eqn:v0}), we obtain the following equation for the $b^{(0)}_n$ coefficients
\begin{align}
    (2n)(2n+1)b^{(0)}_n+(2n-2)(2n-1)(2n)(2n+1)b^{(0)}_{n-1}+3\sum^{n}_{j=1}b^{(0)}_jb^{(0)}_{n-j+1}=0  \, ,
\label{eqn:recurv0a}
\end{align}
for integer $n\geq 2$ and $b^{(0)}_1=-2$. Taking the first and last term out of the summation, we get the recursion relation for the coefficients $b^{(0)}_n$ which is given as 
\begin{equation}
 \begin{aligned}
    (2n-3)(2n+4)b^{(0)}_n+(2n-2)(2n-1)(2n)(2n+1)b^{(0)}_{n-1}+3\sum^{n-1}_{j=2}b^{(0)}_jb^{(0)}_{n-j+1}=0  \, .
\label{eqn:recurv0final}
 \end{aligned}
\end{equation}
This recursion relation gives the same coefficients as those in equation (\ref{eqn:v0exp}). For large $n$, the coefficients diverge as $b^{(0)}_n\sim(-1)^n(2n-1)!$ and hence the series is divergent for all $q$. 

\begin{figure}[ht!]
	\centering
        \includegraphics[width=0.75\linewidth]{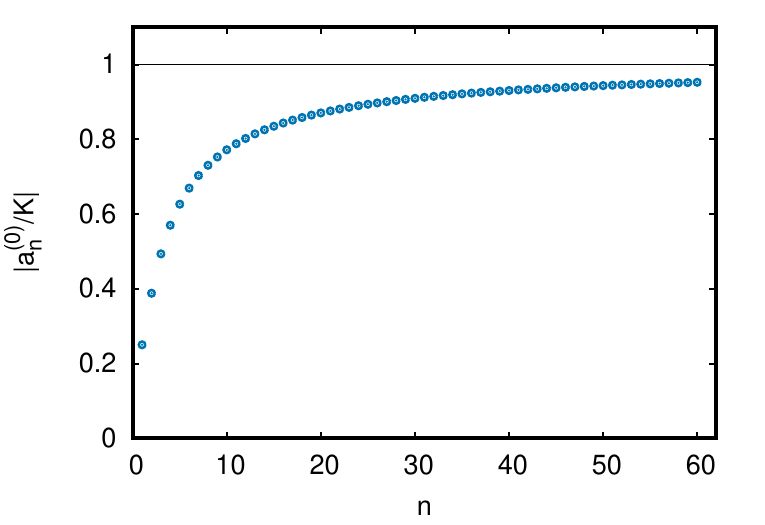}
        \caption{
         The graph of the absolute value of the coefficients $a^{(0)}_n$ of the series in powers of $1/q^2$ for the inner solution of the fKDV non-local soliton, divided by the absolute value of proportionality constant $K$, with respect to the resolution $n$. The top (black) thin horizontal axis in the graph is the limit of $|a^{(0)}_n/K|$ when $n\to\infty$.   
        \label{fig:anbyK}}
\end{figure}

The aim is to sum the series of $v_0$, equation (\ref{eqn:genmatchcondv0}) using the Laplace transform (or equivalently the Borel summation) \cite{gaj, prg}. We seek a solution of equation (\ref{eqn:v0}) in the form of a Laplace transform of the function $V^{\prime}_0(s)$,
\begin{equation}
    v_0=\int_\Gamma I_0(s)\mathrm{d}s \, , \qquad I_0(s)= \exp{(-sq)}V^{\prime}_0(s) \, ,
\label{eqn:laplace}
\end{equation}
where the contour $\Gamma$ runs from 0 to infinity in the complex $s$-plane such that $\Re(sq)>0$. The function $V^{\prime}_0(s)$ represents the derivative of $V_0(s)$ with respect to $s$ which can be obtained by the alternative Borel summation method. The function $V^\prime (s)$ can be expanded by using the identity for the Laplace transform of powers of $s$ as,
\begin{equation}
    V^{\prime}_0(s)=\sum^\infty_{n=0}a^{(0)}_{n}s^{2n+1} \, ,
\label{eqn:Vprs}
\end{equation}
where the relation between the coefficients $a^{(0)}_n$ and $b^{(0)}_n$ is
\begin{equation}
    a^{(0)}_n =\frac{b^{(0)}_{n+1}}{(2n+1)!}  \qquad\text{for}\quad n=0,1,2,\ldots \, .
\label{eqn:anbn}
\end{equation} 
Substituting $b^{(0)}_{n}$ from equation (\ref{eqn:anbn}) into equation (\ref{eqn:recurv0final}), we get the recursion relation for the coefficients $a^{(0)}_n$ as,
\begin{align}
    a^{(0)}_{n-1}+a^{(0)}_n+\frac{3}{(2n+3)!}\sum^{n}_{j=0}(2j+1)!(2n-2j+1)!a^{(0)}_ja^{(0)}_{n-j}=0 \, ,
\label{eqn:recurmain1}
\end{align}
for $n \geq 1$ and $a^{(0)}_0=-2$.  This gives the coefficients $a^{(0)}_n$ which are consistent with equation (\ref{eqn:v0exp}) taking equation (\ref{eqn:anbn}) into account. Taking the first and last term out of the summation and using $a^{(0)}_0 = -2$, we can write equation (\ref{eqn:recurmain1}) as
\begin{align}
    \frac{(2n-1)(2n+6)}{(2n+2)(2n+3)}a^{(0)}_n+a^{(0)}_{n-1}+\frac{3}{(2n+3)!}\sum^{n-1}_{j=1}(2j+1)!(2n-2j+1)!a^{(0)}_j a^{(0)}_{n-j}=0 \, ,
\label{eqn:recurmaina}
\end{align}
for $n\geq 1$. Hence solving the recurrence relation (\ref{eqn:recurmain1}) for the coefficients $a^{(0)}_n$ effectively sums the asymptotic series of $v_0$ given in equation (\ref{eqn:genmatchcondv0}) and yields the solution of equation (\ref{eqn:v0}) for $v_0$ as a Laplace transform. Examination of the recurrence relation (\ref{eqn:recurmain1}) shows that for large $n$, the leading behaviour is $a^{(0)}_0\sim (-1)^n K$, where $K \approx -19.9689$ is a constant found numerically. As we are going to study the higher order $\epsilon$ corrections in the later chapters, the higher precision will be useful for us and for that we have to improve this value of $K$ by including the higher powers of the inverse of $n$ and hence not ignoring all the nonlinear terms. We look for the large $n$ behaviour of the coefficients $a^{(0)}_n$ in the form
\begin{equation}
    a^{(0)}_n=(-1)^n\tilde{a}^{(0)}_n \, , \qquad \tilde{a}^{(0)}_n=KG_0(n) \, , \qquad G_0(n)=1+\sum^{\infty}_{j=1}\frac{g_j}{n^j} \, .
\label{eqn:preciseK}
\end{equation}
The constants $g_j$ should be determined by substituting this expansion into equation (\ref{eqn:recurmain1}). If we are interested in a finite number of terms in $G_0(n)$, then we do not have to take all the $n+1$ terms in the summation in equation (\ref{eqn:recurmain1}). For instance if we take the first $j_m$ and the last $j_m$ terms in the summation, then we get
\begin{align}
    \tilde a_n^{(0)}-\tilde a_{n-1}^{(0)}+\frac{3}{(2n+3)!}\sum_{j=0}^{j_m}(-1)^j(2j+1)!(2n-2j+1)!a_j^{(0)}\tilde a_{n-j}^{(0)} \nonumber \\ +\frac{3}{(2n+3)!}\sum_{j=n-j_m}^{n}(-1)^j(2j+1)!(2n-2j+1)!a_j^{(0)}\tilde a_{n-j}^{(0)}  = 0 \, , \label{eqn:recurmain3a}
\end{align}
where $j_m$ is some positive integer, the value of which is directly related to how many terms in $G_0(n)$ are we intend to determine. The third and the fourth terms in equation (\ref{eqn:recurmain3a}) are the same, so it can also be written as
\begin{equation}
    \tilde{a}^{(0)}_n-\tilde{a}^{(0)}_{n-1}+\sum^{j_m}_{j=0}(-1)^jW^{(0)}_{n,j}\tilde{a}^{(0)}_{n-j}=0 \, ,
 \label{eqn:recurmain3}
 \end{equation} 
where 
\begin{equation}
    W^{(0)}_{n,j}=\frac{6}{(2n+3)!}(2j+1)!(2n-2j+1)!a^{(0)}_j \, ,
\label{eqn:W0}
\end{equation}
where the coefficients $a^{(0)}_n$ in $W_{n,j}^{(0)}$ can be calculated explicitly from equation (\ref{eqn:recurmaina}). The constants $g_j$ can be calculated by substituting the truncated version of equation (\ref{eqn:preciseK}) into equation (\ref{eqn:recurmain3}). For instance if we include first seven terms in the summation of $G_0(n)$, whose coefficients can be computed by using some algebraic manipulation software such as \emph{Mathematica} 
\begin{equation}
    G_0(n) =1-\frac{3}{n}+\frac{39}{4n^2}-\frac{69}{2n^3}+\frac{1929}{16n^4}-\frac{3381}{8n^5}+\frac{46041}{32n^6}-\frac{1089483}{224n^7}+\cdots \ .
 \label{eqn:preciseKvalue}
\end{equation}
For this calculation, we have to take the value of $j_m=3$, which means we have to use at least first four and last four terms from the summation, equation (\ref{eqn:recurmain1}) but the other terms in between can be neglected. The approximate value of the proportionality constant $K$ can be calculated by using the truncated version of the series of $G_0(n)$ as $K\approx (-1)^n a^{(0)}_n/G_0(n)$. In this way, we can calculate the value of the constant $K$ and the result up to 22 digits precision is given as
\begin{equation}
    K=-19.96894735876096051827 \, .
\label{eqn:Kvalue}
\end{equation} 
For $n\to\infty$, the value of $a^{(0)}_n$ approaches the value of $K$,
\begin{equation}
    \lim_{n \to \infty}\left|\frac{a^{(0)}_n}{K}\right|= 1 \, ,
\end{equation}
which can be seen in Figure (\ref{fig:anbyK}).

Hence the series for $V^{\prime}_0(s)$, equation (\ref{eqn:Vprs}) is convergent for $|s|<1$, and for $|s|\geq 1$ the value of $V^{\prime}_0(s)$ can be continued analytically, which shows the uniqueness of the function $V^{\prime}_0(s)$. Depending on how the contour $\Gamma$ is located with respect to the singularities of $V^{\prime}_0(s)$, several different $v_0$ functions can be obtained from it using the Laplace transformation, equation (\ref{eqn:laplace}). The singularities of the function $V^{\prime}_0(s)$ are located at $s=\pm n i$, where $n$ can take any positive integer value. There is no singularity of $V^{\prime}_0(s)$ at $s=0$.

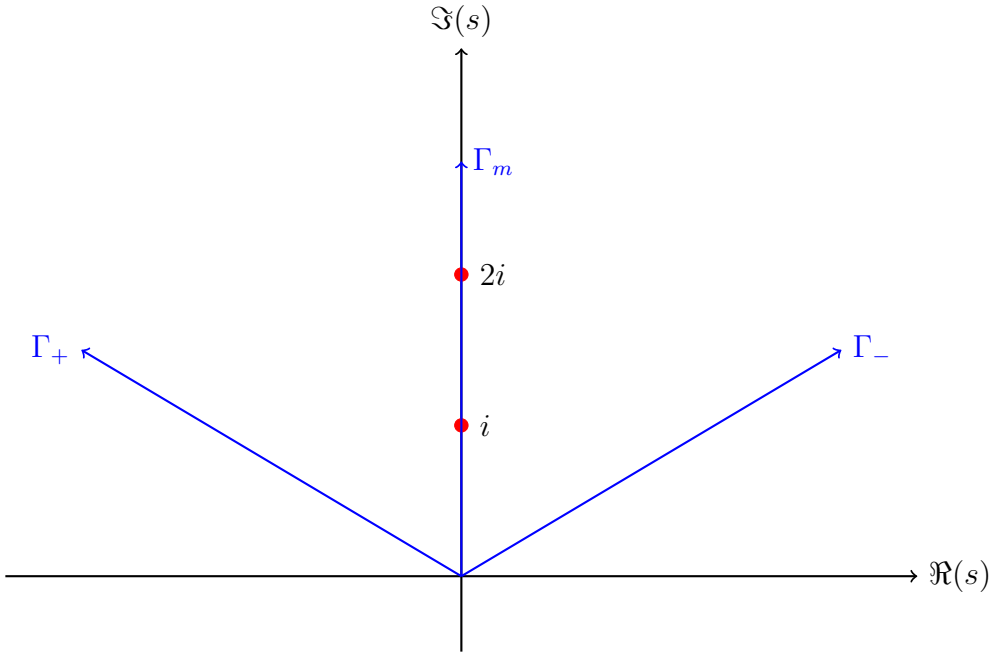
\begin{figure}[ht!]
    \centering
    \begin{tikzpicture}
    
        \draw[->,thick] (0,-1) -- (0,7)node[above]{$\Im(s)$};
        \draw[->,thick] (-6,0) -- (6,0)node[right]{$\Re(s)$};

        \filldraw[red] (0,2) circle (2.5pt);
        \filldraw[red] (0,4) circle (2.5pt);
        
        \draw [->, thick, blue] (0,0) -- (5,3)node[right]{$\Gamma_-$ };
        \draw [->, thick, blue] (0,0) -- (-5,3)node[left]{$\Gamma_+$ };
        \draw [->, thick, blue] (0,0) -- (0,5.5)node[right]{$\Gamma_m$ };

        \node at (0.1,2) [right]{$i$};
        \node at (0.1,4) [right]{$2i$};

    \end{tikzpicture}
    \caption{The complex $s$-plane for the solution $V^\prime(s)$, equation (\ref{eqn:laplace}). The red dots represents the singularities in the complex $s$-plane which are at $s = n i$, for positive integer $n$.}
    \label{fig:complexsplane}
\end{figure}

We can calculate the function $v_0$ from equation (\ref{eqn:laplace}). Depending on the chosen contour $\Gamma$, we can define two asymmetric functions, $v^{(-)}_0$ and $v^{(+)}_0$ which tends to zero exponentially for $\Re{(q)}\to\infty$ and $\Re{(q)}\to -\infty$ respectively. In the former case the contour is in the first quadrant, i.e. $0\leq\mathrm{arg}(s)<\pi/2$, while for the later, it is in the second quadrant, i.e. $\pi/2<\mathrm{arg}(s)\leq\pi$. Let us call these corresponding contours as $\Gamma_-$ and $\Gamma_+$, see Figure (\ref{fig:complexsplane}). 

The asymmetric functions $v^{(-)}_0$ and $v^{(+)}_0$ are valid for $-\pi<\mathrm{arg}(q)\leq 0$ and $-\pi\leq\mathrm{arg}(q)< 0$ respectively. The function $v^{(-)}_0$ cannot be extended by equation (\ref{eqn:laplace}) to $\mathrm{arg}(q)=-\pi$ and same is true for $v^{(+)}_0$ when $\mathrm{arg}(q)=0$. In both the cases, we need to shift the contour to the other side of positive imaginary $q$ axis and collect the contributions from the singularities $s= n i$, where $n$ is any positive integer. These asymmetric solutions have a core similar to the minimal tail symmetric solution, but generally diverges in the negative and positive $x$ direction respectively.

The asymmetric functions $v^{(-)}_0$ and $v^{(+)}_0$ correspond to the asymmetric solutions $u_-$ and $u_+$ of once integrated fKdV equation (\ref{eqn:1aintFKDV}) respectively which tends to zero exponentially for $x\to\infty$ and $x\to -\infty$. The difference of the asymmetric solutions, $v^{(-)}_0$ and $v^{(+)}_0$ can be calculated by using the residue theorem. The dominant contribution to this difference will be given by the singularity closest to the real $s$ axis, i.e. at $s= i$. So we need to determine the behaviour of $V^{\prime}_0(s)$ close to this singularity. The higher order singularities $s=2i, 3i,\cdots$ are subdominant with respect to $s= i$ and hence can be neglected.

The residue at $s=i$ (dominant pole above the real $s$ axis) of the function $I_0(s)$ given in equation (\ref{eqn:laplace}) will be determined by substituting equation (\ref{eqn:Vprs}) and leading order in equation (\ref{eqn:preciseK}) which corresponds to $G_0(n)=1$ into the equation (\ref{eqn:laplace}). The result can be summed as
\begin{equation}
    I_0(s)\approx\sum^{\infty}_{n=0}\exp(-sq)K(-1)^ns^{2n+1}=\exp(-sq)\frac{Ks}{1+s^2} \, .
\label{eqn:sumresidue}
\end{equation}
Since the residue of $s/(1+s^2)$ is $1/2$, it follows that
\begin{equation}
     \underset{s=i}{\mathrm{Res}} \, I_0(s)=\frac{1}{2}K\exp{(-iq)} \, .
\label{eqn:resI0}
\end{equation}

In addition to the two asymmetric functions $v^{(-)}_0$ and $v^{(+)}_0$, we can define the symmetric function $v^{(m)}_0$ if we move the contour $\Gamma$ exactly on the upper half of imaginary $s$ axis, call this contour as $\Gamma_m$, see Figure (\ref{fig:complexsplane}). The symmetry of the function $v^{(m)}_0$ can be seen by putting, $q=i Q$, $Q>0$ in equation (\ref{eqn:laplace}). From equation (\ref{eqn:Vprs}), $i V^{\prime}(iQ)$ is purely real since the coefficients $a^{(0)}_n$ are all real valued. The function $v^{(m)}_0$ corresponds to a symmetric $u_m$ solution on the real axis. The integral in equation (\ref{eqn:laplace}) also has an imaginary part and hence becomes asymmetric function $v^{(-)}_0$, when we add the pole contributions by applying the Residue theorem by choosing a curve going from $s=0$ to infinity in the domain $0\leq\mathrm{arg(s)}<\pi/2$, then coming back along $\Gamma_m$,
\begin{equation}
\begin{aligned}
    v^{(-)}_0 &= \int_{\Gamma_m}\exp{(sq)}V^\prime(s)\mathrm{d}s+ i\frac{\pi K}{2}\exp{(-iq)}+\mathcal{O}\left(\exp{(-2|q|)}\right) \, , \\
    &= v^{(m)}_0 + i\frac{\pi K}{2}\exp{(-iq)}+\mathcal{O}\left(\exp{(-2|q|)}\right) \, ,
\label{eqn:v0res}
\end{aligned}
\end{equation}
for $\Re{(q)}>0$ and $\Im{(q)}<0$. Similarly if we chose the contour going from $s=0$ to infinity in the domain $\pi/2<\mathrm{arg(s)}\leq\pi$, then coming back along $\Gamma_m$, we get asymmetric function $v^{(+)}_0$,
\begin{equation}
\begin{aligned}
    v^{(+)}_0=v^{(m)}_0 - i\frac{\pi K}{2}\exp{(-iq)}+\mathcal{O}\left(\exp{(-2|q|)}\right) \, ,
\label{eqn:v0resa}
\end{aligned}
\end{equation}
for $\Re{(q)}<0$ and $\Im{(q)}<0$. Subtracting equation (\ref{eqn:v0resa}) from equation (\ref{eqn:v0res}), we get the difference of the two asymmetric functions $v^{(-)}_0$ and $v^{(+)}_0$ which is given as
\begin{equation}
    v^{(-)}_0-v^{(+)}_0=\left[\pi iK\exp(-iq)+\mathcal{O}\left(\frac{\exp{(-|q|)}}{|q|}\right)\right]+\mathcal{O}\left(\exp{(-2|q|)}\right) \, ,
\label{eqn:difvmvp}
\end{equation}
which is valid for any $q$ satisfying $\Im(q)<0$. The error term $\exp{(-|q|)}/|q|$ is there because we neglected the $1/n^j$ contribution from equation (\ref{eqn:preciseK}) and the error term $\exp{(-2|q|)}$ is there because we neglected the contribution from the subdominant singularities $s=ni$, for $n\geq2$. As we are interested in the far field oscillations as $|q|\to\infty$, these error terms will not be a problem for us.

\begin{figure}[ht!]
\centering
    \includegraphics[width=0.75\linewidth]{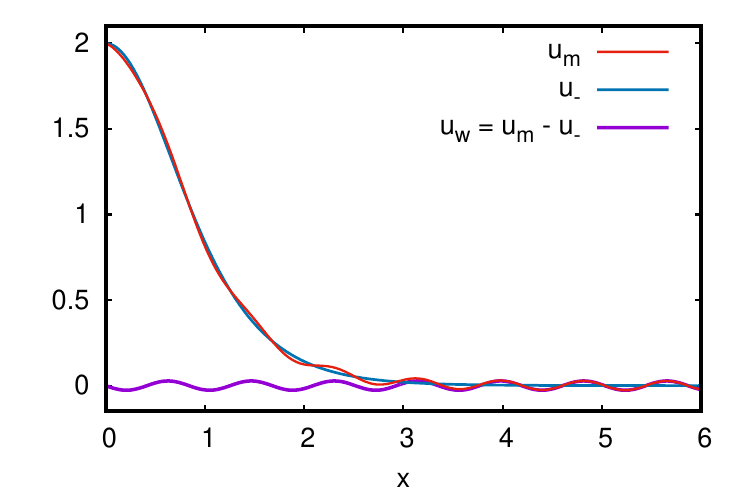}
    \caption{
       A schematic diagram of the minimal-amplitude tail symmetric solution $u_m$, asymmetric function $u_-$ which decay exponentially to zero in the positive $x$ direction and the function $u_w$ which is the difference of $u_m$ and $u_-$.}
    \label{fig:usuauw}
\end{figure}

Equation (\ref{eqn:v0res}) can also be written as,
\begin{equation}
\begin{aligned}
    v^{(m)}_0 = v_0^{(-)}-\frac{i\pi K}{2}\exp{(-iq)}, \quad\text{for}\quad \Im{(q)}<0 \, ,
\label{eqn:v0res2}
\end{aligned}
\end{equation}
Now, it remains to bring the solution back to the real $x$-axis. Using equation (\ref{eqn:cordchan}), $x=i\pi/(2\gamma)+\epsilon q$ in equation (\ref{eqn:v0res2}), we get
\begin{equation}
    v^{(m)}_0=v^{(-)}_0-\frac{i\pi K}{2}\exp\left(-\frac{\pi}{2\epsilon\gamma}-\frac{ix}{\epsilon}\right) \, .
\label{eqn:vmvm1}
\end{equation}
As we are bringing back the solution to the real $x$-axis, we let $\Im{(x)}\to 0$ and use equation (\ref{eqn:vepssq}), $v=\epsilon^2 u$. Here we must also collect a similar contribution from the singularity in the lower half of the complex $x$-plane at $x=-i\pi/(2\gamma$). For $x>0$, we obtain
\begin{equation}
    u_m=u_- -\frac{\pi K}{\epsilon^2}\exp\left(-\frac{\pi}{2\epsilon\gamma}\right)\mathrm{sin}\left(\frac{x}{\epsilon}\right) \, .
\label{eqn:umum1}
\end{equation}
The symmetric function $u_m$ corresponds to the minimal-amplitude tail solution in the $x$-plane. The asymmetric function $u_-$ decays exponentially to zero for $x>0$. Both functions have a similar core that can be approximated by the asymptotic expansion given in equation (\ref{eqn:coreuexp}). The difference of the minimal-amplitude tail symmetric solution $u_m$ and the asymmetric  solution $u_-$ is small not only in the far field region ($x\gg 0$) but also in the core region. Let us call this difference as $u_w = u_m -u_-$. The schematic diagram of the functions $u_m$, $u_-$ and $u_w$ can be seen in Figure (\ref{fig:usuauw})

From equation (\ref{eqn:umum1}), the function $u_w$ is given as
\begin{equation}
    u_w=-\frac{\pi K}{\epsilon^2}\exp\left(-\frac{\pi}{2\epsilon\gamma}\right)\mathrm{sin}\left(\frac{x}{\epsilon}\right) \ .
\label{eqn:uw0}
\end{equation}
Far from the core region ($x\gg 0$), the function $u_w$ gives the minimal-amplitude oscillatory tail. To the leading order in $\epsilon$ the minimal tail amplitude $\alpha^{(0)}_m$ is given as
\begin{equation}
    \alpha_m^{(0)}=-\frac{\pi K}{\epsilon^2}\exp\left(-\frac{\pi}{2\epsilon\gamma}\right)  \, ,
\label{eqn:alphaeps}
\end{equation}
where zero in the superscript of $\alpha_m$ denotes that the result is correct up to $\epsilon^0$ order. This result corresponds to the result of Pomeau et al. \cite{prg} for $\delta=0$. However, there is an incorrect additional factor of 2 and a missing division sign in the exponential there.

\begin{figure}[ht!]
\centering
    \includegraphics[width=0.75\linewidth]{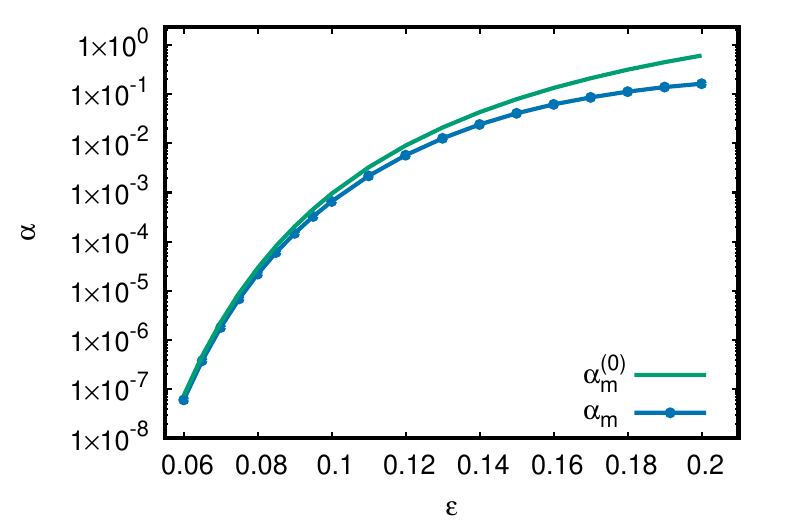}
    \caption{Comparison of the leading order analytical result of the minimal tail amplitude $\alpha^{(0)}_m$ with the more precise numerical results.}
    \label{fig:lead}
\end{figure}

The comparison of the leading order result of minimal tail amplitude $\alpha^{(0)}_m$, equation (\ref{eqn:alphaeps}) with the more precise numerical results for several $\epsilon$ values can be seen in Figure (\ref{fig:lead}). This plot shows that for $\epsilon \geq 0.08$, we certainly need a better approximation for $\alpha_m$ to make the result more precise. This can be done by including the higher order corrections to the minimal amplitude tail $\alpha_m$.

The higher order generalization of equation (\ref{eqn:alphaeps}) which is valid up to $\mathcal{O}(\epsilon^j)$ order can be written in the form as
\begin{equation}
    \alpha_m^{(j)}=-\frac{\pi K}{\epsilon^2}\exp\left(-\frac{\pi}{2\epsilon\gamma}\right)[1+\zeta_1\epsilon+\zeta_2\epsilon^2+\cdots+\zeta_j\epsilon^j+\mathcal{O}(\epsilon^{j+1})]  \, ,
\label{eqn:alphamhigher}
\end{equation}
where the constant coefficients $\zeta_1, \, \zeta_2, \, \cdots, \, \zeta_j$ are yet to be determined. We will calculate these coefficients up to $\zeta_5$ in Chapter (\ref{chapter:higher_orders}).

\chapter{WKB solution} \label{chapter:wkbsolution}

We intend to compute analytically the higher order approximations for the leading order result of the minimal tail amplitude $\alpha_m$ by extending the procedure described in the previous chapter to the next higher orders in $\epsilon$. Before considering this extension, we intend to linearize the once integrated fKdV equation (\ref{eqn:1aintFKDV}) around the function $u$ which may be some numerically calculated solution which can be approximated well by some appropriately truncated version of equation (\ref{eqn:coreuexp}). We seek a new solution of the form
\begin{equation}
    u\sim u+u_w \, ,
\label{eqn:uuw}
\end{equation}
where the function $u$ may be the symmetric solution with extremely small amplitude tail in both directions, or the asymmetric solution that decays exponentially to zero for $x>0$. For $x \geq 0$ the function $u_w$ turns out to be exponentially small in terms of the small parameter $\epsilon$ everywhere including the core region. In the asymptotic region, the $u_w$ function will give the oscillatory tail of amplitude $\alpha$ which is exponentially small in terms of $\epsilon$. Substituting equation (\ref{eqn:uuw}) into equation (\ref{eqn:1aintFKDV}) and linearizing, we get
\begin{equation}
    \epsilon^2u_{w,xxxx}+u_{w,xx}+6uu_w-cu_w=0 \, .
\label{eqn:uweqn}
\end{equation}
When $u$ is small, this equation is discussed in Section (\ref{subsection:linearasymp}), and the solutions are given by equation (\ref{eqn:ualpha}). Because of the high wave number $k/\epsilon$, to leading order this solution is valid even in the core region. The generalized $u_w$ solution of the linearized problem, equation (\ref{eqn:uweqn}) is expected to look like
\begin{equation}
    u_w=\beta\left(1+\sum^\infty_{n=1}\Theta_n(x)\epsilon^{n}\right)\mathrm{sin}\left(\frac{k x}{\epsilon}+\sum^\infty_{n=1}\Phi_n(x)\epsilon^{n}\right) \, ,
\label{eqn:genuw}
\end{equation}
where the coefficients $\Theta_n(x)$ and $\Phi_n(x)$ are yet to be determined. The tail oscillations have a spatial frequency proportional to $1/\epsilon$, see equation (\ref{eqn:ualpha}). Hence, we apply the WKB method to solve linear differential equation (\ref{eqn:uweqn}) to obtain the solution of the form of equation (\ref{eqn:genuw}) and to calculate the functions $\Theta_n(x)$ and $\Phi_n(x)$. 

We seek the solution $u_w$ of equation (\ref{eqn:uweqn}) of the form 
\begin{equation}
    u_w=\beta_c\exp{(A)} \, ,
\label{eqn:uwA}
\end{equation}
where $A$ is a function of $x$ and $\beta_c$ is a complex constant. Substituting equation (\ref{eqn:uwA}) into equation (\ref{eqn:uweqn}), we get
\begin{equation}
\begin{aligned}
    \epsilon^2\left[A_{xxxx}+4A_xA_{xxx}+3(A_{xx})^2+6(A_x)^2A_{xx}+(A_x)^4\right]+\\+A_{xx}+(A_x)^2+6u_s-c=0 \, .
\label{eqn:Aeqn}
\end{aligned}
\end{equation}
The next step is to expand the function $A$ in powers of $\epsilon$, starting with $1/\epsilon$,
\begin{equation}
    A=\sum^{\infty}_{n=-1}A_n\epsilon^n \, ,
\label{eqn:Asum}
\end{equation}
and solve equation (\ref{eqn:Aeqn}) order-by-order in $\epsilon$. The function $A(x)$ and consequently all $A_n$ appears only in the differential form, there will be an additive complex scalar freedom at any of these functions and all these can be absorbed into the complex valued $\epsilon$ dependent factor $\beta_c$. Hence, we do not write out these unspecified constants.

To the leading $\epsilon^{-2}$ order, we obtain
\begin{equation}
    (A_{-1,x})^2[(A_{-1,x})^2+1]=0 \, .
\label{eqn:aminus1x}
\end{equation}
There are three solutions to this equation. The solution zero corresponds to the absence of the high frequency oscillations in the asymptotic region and hence we are not interested in such kind of solutions here. The other two solutions are $A_{-1,x}=\pm i$. The solutions obtained from these two values turns out to be the complex conjugate of each other to all orders in $\epsilon$. Moving the constant freedom to the amplitude, we set $A_{-1}=-ix$ and proceed order-by-order in $\epsilon$. At each order, we obtain a condition determining $A_{n,x}$, which then can be relatively easily integrated for $A_n$. The first several $A_n$ functions are:
\begin{align}
 A_{-1} &= -ix  \, , \label{eqn:aminus1} \\
 A_{0} &=  0  \, , \label{eqn:a0} \\
 A_{1} &= -2i\gamma^2 x+6i\gamma \mathrm{tanh}(\gamma x)  \, , \label{eqn:a1} \\
 A_{2} &= 15\gamma^2\mathrm{sech}^2(\gamma x) \, , \label{eqn:a2} \\
 A_{3} &= 2i\gamma^4 x + 111i\gamma^3\mathrm{sech}^2(\gamma x)\mathrm{tanh}(\gamma x) \, , \label{eqn:a3} \\
 A_{4} &= \frac{525}{2}\gamma^4\mathrm{sech}^2(\gamma x)[3\mathrm{sech}^2(\gamma x)-2] \, , \label{eqn:a4}  \\
A_5 &= -4i\gamma^6x+\frac{3}{5}i\gamma^5[12267\mathrm{sech}^4(\gamma x)-4089\mathrm{sech}^2(\gamma x)+632]\mathrm{tanh}(\gamma x) \, , \label{eqn:a5}  \\
A_6 &= \frac{3}{2}\gamma^6\mathrm{sech}^6(\gamma x)[49317\mathrm{sech}^4(\gamma x)-49317\mathrm{sech}^2(\gamma x)+8050] \, . \label{eqn:a6}
\end{align}
This calculation can be continued, relatively easily to much higher orders in $\epsilon$ by using some algebraic manipulation software such as \emph{Mathematica}. The odd functions $A_{2n-1}$ are all purely imaginary for real $x$ and hence they will contribute to the phase of the linearized solution $u_w$.

Since the asymptotic spatial frequency is $k/\epsilon$, where $k$ depends on $\epsilon$ defined in equation (\ref{eqn:kappa}), it is natural to insert this factor $k$ into the linear term. The terms in the odd indexed $A_n$ functions which are proportional to $x$ can be absorbed into the $A_{-1}/\epsilon=-ix/\epsilon$ factor if we replace it by $-ikx/\epsilon$. So, we can write the phase of the linearized solution $u_w$ as
\begin{equation}
\begin{aligned}
    -\frac{ix}{\epsilon}+\sum^{\infty}_{\substack{n=1 \\ \text{odd}}}A_n\epsilon^n &=-\frac{ikx}{\epsilon}+\left(\frac{ix}{\epsilon}(k-1)+\sum^{\infty}_{\substack{n=1 \\ \text{odd}}}A_n\epsilon^n\right) \ ,\\
    &= -\frac{ikx}{\epsilon}+\sum^{\infty}_{\substack{n=1 \\ \text{odd}}}i\tilde A_n\epsilon^n \, .
\label{eqn:summation}
\end{aligned}
\end{equation}
The functions $\tilde A$ have a finite limit at infinity which will determine the asymptotic phase shift of the minimal tail configuration. We denote this asymptotic phase shift by
\begin{equation}
    \tilde \delta_{2n-1}=\lim_{x\to\infty}\tilde{A}_{2n-1} \, .
\label{eqn:tildedelta}
\end{equation}
The first few values are given as
\begin{align}
     \tilde{\delta}_{1} &= 6\gamma \, , \label{eqn:d1} \\
     \tilde{\delta}_{3} &= 0 \, , \label{eqn:d3} \\
     \tilde{\delta}_{5} &= \frac{1896}{5}\gamma^5 \, , \label{eqn:d5} \\
     \tilde{\delta}_{7} &= \frac{67140}{7}\gamma^7 \, ,  \label{eqn:d7} \\
     \tilde{\delta}_{9} &= \frac{2662320}{7}\gamma^9 \, .  \label{eqn:d9}
\end{align}
For real $x$, all even indexed $A_n$ functions are real and hence they will contribute to the amplitude of the linearized solution $u_w$. The general solution of the linearized problem which takes real values on the real $x$ axis can be obtained by a linear combination of the solutions belonging to $A_{-1}=-ix$ and $A_{-1}=+ix$,
\begin{equation}
    u_w=\beta\exp{\left(\sum^{\infty}_{\substack{n=2 \\ \text{even}}}A_n\epsilon^n\right)}\mathrm{sin}\left(\frac{kx}{\epsilon}-\delta_w-\sum^{\infty}_{\substack{n=1 \\ \text{odd}}}\tilde{A}_n\epsilon^n\right) \ ,
\label{eqn:uwevenodd}
\end{equation}
where $\beta$ and $\delta_w$ are real constants that can have arbitrary $\epsilon$ dependence and are related to the magnitude and phase of the complex constant $\beta_c$. This can be used to get equation (\ref{eqn:genuw}).

\section{Amplitude and phase}

As can be see from equations (\ref{eqn:aminus1}) through (\ref{eqn:a6}) the even indexed $A_{2n}$ functions tend to zero at infinity. In the asymptotic region $(x\gg 0)$, the exponential in equation (\ref{eqn:uwevenodd}) tends (but never equal) to 1. Hence the asymptotic behaviour of the function $u_w$ is
\begin{equation}
    u_w=\beta\ \mathrm{sin}\left(\frac{kx}{\epsilon}-\delta_w-\delta_m\right) \, ,
\label{eqn:uwdm}
\end{equation}
where the complex constant $\beta$ gives the asymptotic amplitude of the oscillations represented by $u_w$ and
\begin{equation}
    \delta_m=\lim_{x\to\infty}\sum^{\infty}_{\substack{n=1 \\ \text{odd}}}\tilde{A}_n\epsilon^n=\sum^{\infty}_{\substack{n=1 \\ \text{odd}}}\tilde{\delta}_n\epsilon^n \, .
\label{eqn:tildedelta1}
\end{equation}
The first few $\tilde{\delta}_n$ are given in equations (\ref{eqn:d1}) through (\ref{eqn:d9}). From equation (\ref{eqn:uwdm}), we can see that the phase with respect to the sine function comes from two contributions, $\delta_w$ and $\delta_m$. Since $\tilde{A}_{2n-1}=0$ at $x=0$, see equations (\ref{eqn:aminus1}) through (\ref{eqn:a4}), the part $\delta_w$ gives the phase near the centre. The function $u_w$ in a small region around the centre behaves as
\begin{equation}
    u_w\sim\mathrm{sin}\left(\frac{kx}{\epsilon}-\delta_w\right) \, .
\label{eqn:uwcenter}
\end{equation}

The other part of the phase which is $\delta_m$, is $\epsilon$ dependent and gives the phase shift between the centre and infinity. This phase shift is known in terms of an asymptotic series expansion (\ref{eqn:tildedelta1}) with coefficients $\tilde{\delta}_n$, see equations (\ref{eqn:d1}) through (\ref{eqn:d9}) for the first few values.

We can decompose the function $u_w$ given in equation (\ref{eqn:uwdm}) to a sine part with $\delta_w=0$ and a cosine part corresponding to $\delta_w=\pi/2$, both with arbitrary amplitudes ,
\begin{equation}
    u_w= \beta\exp{\left(\sum^{\infty}_{\substack{n=2 \\ \text{even}}}A_n\epsilon^n\right)}\left[\beta_{\mathrm{sin}} \ \mathrm{sin}\left(\frac{kx}{\epsilon}-\sum^{\infty}_{\substack{n=1 \\ \text{odd}}}\tilde{A}_n\epsilon^n\right)+\beta_{\mathrm{cos}} \ \mathrm{cos}\left(\frac{kx}{\epsilon}-\sum^{\infty}_{\substack{n=1 \\ \text{odd}}}\tilde{A}_n\epsilon^n\right)\right] \, .
\label{eqn:uwsincos}
\end{equation}
Since $\tilde A_{2n-1}=0$ at $x=0$, the sine part is antisymmetric at the center $x=0$, while the cosine part is symmetric. We assumed that the tail region of the symmetric solution $u$ of the once integrated fKdV equation is given by equation (\ref{eqn:ualpha}), $u = \alpha \, \mathrm{sin}(kx/\epsilon - \delta)$. Adding and subtracting $\delta_m$ in the argument of sine function in this equation and using the identity, $\mathrm{sin}(a-b)=\mathrm{sin}(a)\ \mathrm{cos}(b)-\mathrm{cos}(a)\ \mathrm{sin}(b)$, we can write it as
\begin{equation}
    u=\alpha \ \mathrm{cos}(\delta-\delta_m) \ \mathrm{sin}\left(\frac{kx}{\epsilon}-\delta_m\right)-\alpha \ \mathrm{sin}(\delta-\delta_m) \ \mathrm{cos}\left(\frac{kx}{\epsilon}-\delta_m\right) \, .
\label{eqn:usdelta}
\end{equation}
Adding equation (\ref{eqn:usdelta}) and equation (\ref{eqn:uwsincos}) with $\beta_\mathrm{sin}=0$ and $\beta_\mathrm{cos}=\alpha \ \mathrm{sin}(\delta-\delta_m)$, we get that in the tail region
\begin{equation}
    u+u_w=\alpha \ \mathrm{cos}(\delta-\delta_m) \ \mathrm{sin}\left(\frac{kx}{\epsilon}-\delta_m\right) \, .
\label{eqn:uuw1}
\end{equation}
In this way we obtain the minimal tail symmetric solution $u_m=u+u_w$ and for any given $\epsilon$, this solution is unique. The asymptotic behaviour of the minimal-amplitude tail solution $u_m$ for $x>0$ is given as
\begin{equation}
    u_m=\alpha_m \ \mathrm{sin}\left(\frac{kx}{\epsilon}-\delta_m\right) \, ,
\label{eqn:umam}
\end{equation}
where $\alpha_m=\alpha \ \mathrm{cos}(\delta-\delta_m)$. The minimal amplitude tail symmetric solution $u_m$ necessarily has the asymptotic phase $\delta_m$ which has been already calculated in terms of asymptotic series expansion, see equation (\ref{eqn:tildedelta1}). 

\begin{figure}[ht!]
\centering
    \includegraphics[width=0.75\linewidth]{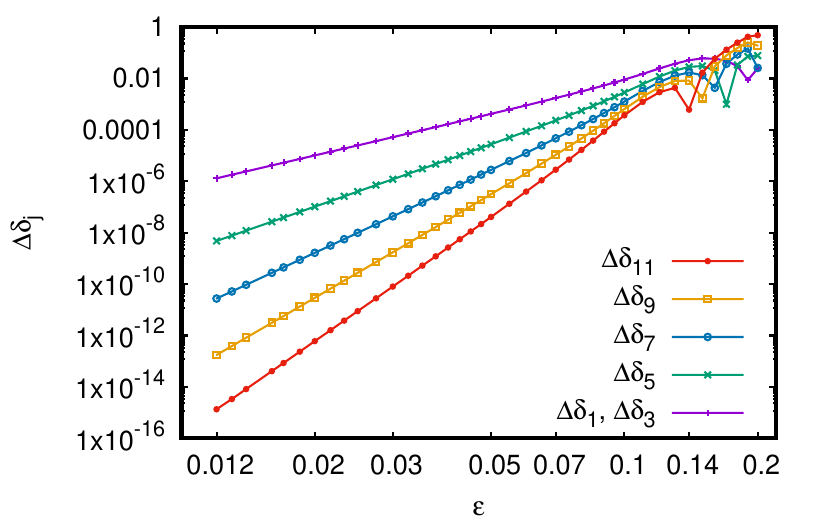}
    \caption{
       A log-log plot of $\Delta\delta_j=|(\delta_m-\delta_m^{(j)})/\delta_m|$, which shows the relative difference of the numerically calculated minimum phase $\delta_m$ and the various analytical approximations $\delta_m^{(j)}$.
    \label{fig:deltamr}}
\end{figure}

We compare the numerically calculated minimal amplitude asymptotic phase $\delta_m$ and its various analytical approximations provided by equation (\ref{eqn:tildedelta1}),
\begin{equation}
    \delta_m^{(j)}=\sum^{j}_{\substack{n=1 \\ \text{odd}}}\tilde \delta_n\epsilon^n \, .
\label{eqn:deltamr}
\end{equation}
Since the numerical simulations are more precise, we define the relative error $\Delta\delta_j$ of the $j$-th order analytical expansion result as 
\begin{equation}
    \Delta\delta_j=\left|\frac{\delta_m-\delta_m^{(j)}}{\delta_m}\right| \, .
\label{eqn:relerrdelm}
\end{equation}
Figure (\ref{fig:deltamr}) shows logarithmically the relative difference, equation (\ref{eqn:relerrdelm}) for $j\leq 11$.

The relative difference $\Delta\delta_1$ is equal to  $\Delta\delta_3$, since the third order correction $\tilde \delta_3$ is zero. The plot shows that the result is 16 digits precise when we add just six corrections. The precision for this agreement can be further improved for not so small $\epsilon$ values by including the higher order $\tilde \delta_n$ corrections. This precision of the numerical result shows the remarkable power of the exponentially convergent pseudo-spectral method when combined with arbitrary precision arithmetic. For larger $\epsilon$, the result gets worsen which is clear on the right upper corner in the Figure (\ref{fig:deltamr}). The reason for this is that the series expansion for $\delta_m$, see equation (\ref{eqn:tildedelta1}) is an asymptotic expansion.  In general, an asymptotic series expansion is expected to give the best approximation when the contribution from the next term starts becoming larger than the contribution from the previous term. For example for $\epsilon=0.13$, the best approximation of the numerical result is given by $\delta_m^{(13)}$ and adding higher order terms gives larger and larger error.

The important consequence of equation (\ref{eqn:usdelta}) is that for a given $\epsilon$ the tail amplitude $\alpha$ of any symmetric solution $u$ with phase $\delta$ is related to the minimal tail amplitude $\alpha_m$ by
\begin{equation}
    \alpha=\frac{\alpha_m}{\mathrm{cos}(\delta-\delta_m)} \ .
\label{eqn:alphamdeltam}
\end{equation}
In order to check equation (\ref{eqn:alphamdeltam}) numerically, we have calculated the minimum phase $\delta_m$, and the corresponding minimum tail amplitude $\alpha_m$ for various $\epsilon$ values. We have also calculated $\alpha$ for different $\delta$ phase shift values, for instance at 100 points in the interval $[0,\pi]$.

\begin{figure}[ht!]
\centering
    \includegraphics[width=0.75\linewidth]{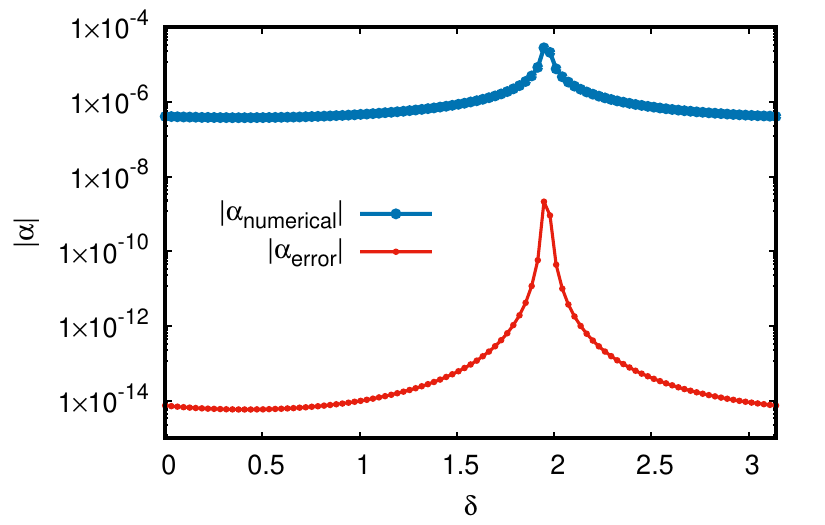}
    \caption{
       A plot of the dependence of the absolute value of the oscillatory tail amplitude $\alpha$ on the phase shift $\delta$ in the interval $[0,\pi]$ for $\epsilon=0.065$. The error is the difference of the numerically calculated $\alpha$ value and the analytical result given in equation (\ref{eqn:alphamdeltam}).}
    \label{fig:err}
\end{figure}

Figure (\ref{fig:err}) shows the comparison of the numerically calculated tail amplitude $\alpha$ and the analytical result which is given in equation (\ref{eqn:alphamdeltam}) for $\epsilon=0.065$ as a function of the phase shift $\delta$. The error in the tail amplitude as a function $\delta$ is also shown in the plot which turns out to be of the order $\alpha^2$. This error is similar to the error in the tail amplitude $\alpha$ caused by the linearization of the fKdV equation in the asymptotic region.

For the currently chosen $\epsilon$ there is a huge peak at around $\delta\approx 1.98$ and it occurs when the denominator of equation (\ref{eqn:alphamdeltam}) becomes zero, i.e. when $\delta=\pi/2+\delta_m$. Generally if we calculate the tail amplitude as a function of $\delta$ in the larger interval, then there will be more than one peak and the position of these peaks will be at around $\delta=\pm(n+1/2)\pi+\delta_m$, where $n$ is any positive integer.

\section{Complex extension} \label{section:complexext}

The function $u_w$ is singular at the same places as the original $u$ function is, i.e. at $x=\pm (2n+1)i\pi/(2\gamma)$ for positive integer $n$. It is because, all the functions $A_n$ contains various powers of $\mathrm{sech}(\gamma x)$. Close to the singularity above the real $x$ axis, i.e. near $x=i\pi/(2\gamma)$, we substitute equation (\ref{eqn:cordchan}), $x=i\pi/(2\gamma) + \epsilon q$ in equation (\ref{eqn:uwevenodd}) to obtain the $u_w$ function depending on $q$ which is given as 
\begin{equation}
    u_w=\beta\exp{\left(\sum^{\infty}_{\substack{n=2 \\ \text{even}}}A_n\epsilon^n\right)}\mathrm{sin}\left(i\frac{k\pi}{2\epsilon\gamma}+kq-\delta_w-\sum^{\infty}_{\substack{n=1 \\ \text{odd}}}\tilde{A}_n\epsilon^n\right) \, .
\label{eqn:uwq}
\end{equation}
Expressing the sine function in terms of the exponential functions and neglecting the term proportional to $\exp{[-k\pi/(2\epsilon\gamma)]}$, equation (\ref{eqn:uwq}) can be written as
\begin{equation}
\begin{aligned}
    u_w=\frac{i\beta}{2}\exp{\left(\sum^{\infty}_{\substack{n=2 \\ \text{even}}}A_n\epsilon^n\right)}\exp{\left(\frac{k\pi}{2\epsilon\gamma}-ikq+i\delta_w+i\sum^{\infty}_{\substack{n=1 \\ \text{odd}}}\tilde{A}_n\epsilon^n\right)} \, .
\label{eqn:uwexp1}
\end{aligned}
\end{equation}
Now, in order to bring back the original $A_n$, we use equation (\ref{eqn:summation}) in equation (\ref{eqn:uwexp1}), hence
\begin{equation}
\begin{aligned}
    u_w = \frac{i\beta}{2}\exp{\left(\frac{\pi k}{2\epsilon\gamma}\right)}\exp{(-iq+i\delta_w)}\exp{\left(-\frac{(k-1)\pi}{2\epsilon\gamma}+\sum^{\infty}_{n=1}A_n\epsilon^n\right)} \, .
\label{eqn:uwexp2}
\end{aligned}
\end{equation}
Substituting the value of $k(\epsilon)=\sqrt{1+4\gamma^2\epsilon^2}$ into the third exponential in equation (\ref{eqn:uwexp2}), we see that the term $(k-1)/\epsilon$ is small. Hence we proceed by expanding the argument of the third exponential in powers of $1/q$, and then expand the exponential of the result in $\epsilon$. The result correct up to $\mathcal{O}(\epsilon^4)$ order is given as
\begin{equation}
    u_w =\frac{i\beta}{2}\exp{\left(\frac{\pi k}{2\epsilon\gamma}\right)}\exp{(-iq+i\delta_w)}\left[(1+5\gamma^2\epsilon^2) F_0(q)+\gamma^4\epsilon^4 F_2(q)+\mathcal{O}(\epsilon^5)\right] \, ,
\label{eqn:result1}
\end{equation}
where
\begin{align}
    F_0(q) &= 1+\frac{6i}{q}-\frac{33}{q^2}-\frac{237i}{q^3}+\frac{1890}{q^4}+\frac{17028 i}{q^5}+\cdots \label{eqn:F0} \, , \\
    F_2(q) &= -\frac{2iq^3}{15}-\frac{q^2}{5}+\frac{39iq}{5}-25+\frac{234i}{q}-\frac{14343}{5q^2}+\cdots  \label{eqn:F2} \, .
\end{align}
For $\delta_w=0$ and $q=\pm iQ$, $Q>0$, the function $u_w$ is purely imaginary and hence correspond to the complex extension of an antisymmetric function, while for $\delta_w=\pi/2$ and $q=\pm iQ$, $Q>0$, the function $u_w$ is purely real and hence correspond to the complex extension of a symmetric function. Please note that the result in the paper of Grimshaw and Joshi \cite{gaj} is not really valid up to $\epsilon^2$ order as it is claimed there, due to the missing $(1+5\gamma^2\epsilon^2)$ factor in their equation.

\chapter{Higher orders}\label{chapter:higher_orders}

In this chapter we are going to discuss the higher order approximations of the leading order result of minimal tail amplitude $\alpha_m^{(0)}$ given in equation (\ref{eqn:alphaeps})
$$\alpha_m^{(0)}=-\frac{\pi K}{\epsilon^2}\exp\left(-\frac{\pi}{2\epsilon\gamma}\right)\left(1+\mathcal{O}(\epsilon)\right)  \, ,$$
where $K \approx -19.9689$ and $\gamma$ is related to the amplitude of the KdV 1-soliton solution, $u_0 = 2\gamma^2\mathrm{sech}^2(\gamma x)$ which in turn is related to the phase speed of the same, see equation (\ref{eqn:c0gam}). The higher order approximations of the above result can be achieved by solving the higher order inner problems. In addition to the method of matched asymptotics in the complex plane and the Laplace transform (or equivalently the Borel summation), we will also use the complex extension of the $u_w$ function which we have already discussed in Section (\ref{section:complexext}) to get the higher order approximations of $\alpha_m$. In this chapter we also compare different orders of the analytical result to the more precise numerical simulations.

We will return now to the higher order inner problems, which can be obtained from equation (\ref{eqn:vfkdv}),
$$v_{qqqq}+v_{qq}+3v^2-\epsilon^2cv = 0 \, .$$
If we search for the solutions to the above equation in the form of equation (\ref{eqn:vexp})
\begin{align*}
    v=\sum_{n=0}^\infty \epsilon^{2n}v_n \, ,
\end{align*}
and solve order-by-order in $\epsilon$, we get different orders of the inner problems. We have already studied the leading (zeroth) order inner problem in Chapter (\ref{chapter:analytical_results}). Next we intend to examine the effect of the next terms $v_1$ and $v_2$ in this chapter which can be done by considering the (inner) differential equations satisfied by $v_1$ and $v_2$.

The function $u_w$ in equation (\ref{eqn:result1}) with $\beta=\alpha_m$ and $\delta_w=0$ can be written as
\begin{equation}
    u_w =\frac{i\alpha_m}{2}\exp{\left(\frac{\pi k}{2\epsilon\gamma}\right)}\exp{(-iq)} \left[(1+5\gamma^2\epsilon^2) F_0(q)+\gamma^4\epsilon^4 F_2(q)+\mathcal{O}(\epsilon^5)\right] \, ,
\label{eqn:result1a}
\end{equation}
where the functions $F_0(q)$ and $F_2(q)$ are given in equations (\ref{eqn:F0}) and (\ref{eqn:F2}) respectively and $k=\sqrt{1+4\gamma^2\epsilon^2}$, see equation (\ref{eqn:kappa}).

There are two important solutions of the stationary fKdV equation (\ref{eqn:1aintFKDV}), the symmetric solution $u_m$ that has a minimal-amplitude tail, and the asymmetric solution $u_-$ that has no tail and decays exponentially to zero for $x>0$. There is no accurate analytical solutions known for either of them, but the difference $u_m-u_-$ can be determined very precisely by using the higher order WKB method which we have already discussed in Chapter (\ref{chapter:wkbsolution}). This difference is exponentially small in $\epsilon$ not only in the far field region, but also in the core region. Hence the function $u_w$ in equation (\ref{eqn:uwdm}) with $\beta=\alpha_m$ and $\delta_w=0$ can be used to represent the difference $u_m - u_-$,
\begin{align}
    u_w = u_m - u_-  \qquad \text{for}\quad \beta=\alpha_m \, , \quad \delta_w=0 \, .
\label{eqn:uwumumi}
\end{align}
Because of its symmetry, $u_m$ is purely real on the imaginary axis, both on the complex $x$ and $q$ planes. Hence the imaginary parts of $u_-$ and $-u_w$ have to agree on the imaginary $q$-axis,
\begin{equation}
    \Im{(u_-)}=-\Im{(u_w)} \ ,\quad\text{for}\quad\Re{(q)}=0\quad\text{and}\quad\Im{(q)}<0 \, ,
\label{eqn:Imumuw}
\end{equation}
which is valid to any order in $\epsilon$ and $1/q$. We will use this equation for the calculation of the minimal tail amplitude $\alpha_m$.  The imaginary part of $u_w$ can be obtained from equation (\ref{eqn:result1a}). Next, our main aim is to calculate the imaginary part of the asymmetric $v^{(-)}_n$ functions. Then by using the rescaled function $v = \epsilon^2 u$, see equation (\ref{eqn:vepssq}), we can calculate the imaginary part of the corresponding asymmetric solution $u_-$. 

From equation (\ref{eqn:v0res}), the imaginary part of the $u_-$ function to the leading order in $\epsilon$ is given as
\begin{equation}
    \Im{(u_-)}=\frac{\pi K}{2\epsilon^2}\exp{(-iq)}\left(1+\mathcal{O}(\epsilon)\right) \, ,
\label{eqn:imv0minus}
\end{equation}
for $\Re{(q)}=0$ and $\Im{(q)}<0$.

\section{Second order}

The second order inner equation is given in equation (\ref{eqn:v1}),
\begin{align*}
    v_{1,qqqq}+v_{1,qq}+2v_0(3v_1-2\gamma^2) = 0 \, .
\end{align*}
The appropriate precise solution of the above $\epsilon$ independent equation can be used to determine the tail amplitude of the symmetric solution $u$ of the once integrated fKdV equation (\ref{eqn:1aintFKDV}) up to $\mathcal{O}(\epsilon^3)$ order.

 The above differential equation has an appropriate exact solution $v_1=2\gamma^2/3$ because it satisfies the boundary condition given in equation (\ref{eqn:v2}). So we have $\Im{(v_1)} = 0$ on the imaginary $q$-axis. To this approximation, equation (\ref{eqn:vexp}) gives $v\approx v_0+\epsilon^2 v_1$. Hence equation (\ref{eqn:imv0minus}) is also valid up to $\mathcal{O}(\epsilon^2)$ order,
\begin{equation}
    \Im{(u_{-})}=\frac{\pi K}{2\epsilon^2}\exp{(-iq)}\left(1+\mathcal{O}(\epsilon^4)\right) \, , 
\label{eqn:imuminus}
\end{equation}
for $\Re{(q)}=0$ and $\Im{(q)}<0$. In order to use equation (\ref{eqn:Imumuw}), we also need $\Im{(u_w)}$ up to $\mathcal{O}(\epsilon^2)$ order which can be obtained from equation (\ref{eqn:result1a}),
\begin{equation}
\begin{aligned}
    \Im{(u_w)}=\frac{\alpha_m}{2}\exp{\left(\frac{\pi k}{2\epsilon\gamma}\right)}\exp{(-iq)}\left(1+5\gamma^2\epsilon^2+\mathcal{O}(\epsilon^4)\right) \, ,
\label{eqn:mImuw2a}
\end{aligned}
\end{equation}
for $\Re{(q)}=0$ and $\Im{(q)}<0$. This is valid for large $|q|$ when according to equation (\ref{eqn:F0}), the function $F_0(q)$ tends to $1$.

\subsection{Oscillatory tail amplitude up to third order}

Using equations (\ref{eqn:imuminus}) and (\ref{eqn:mImuw2a}) in equation (\ref{eqn:Imumuw}), we obtain the expression for the minimal tail amplitude which is valid up to $\mathcal{O}(\epsilon^3)$ order,
\begin{equation}
    \alpha_m=-\frac{\pi K}{\epsilon^2}\exp{\left(-\frac{\pi k}{2\epsilon\gamma}\right)}\left(1-5\gamma^2\epsilon^2+\mathcal{O}(\epsilon^4)\right) \, .
\label{eqn:alpham3}
\end{equation}
If we denote $\alpha^{(k,j)}_m$ as the tail amplitude valid up to $\mathcal{O}(\epsilon^j)$ order when $k = \sqrt{1+4\gamma^2\epsilon^2}$ is kept inside the exponential instead of expanding it in $\epsilon$ then the expressions for the minimal tail amplitude which are valid up to $\mathcal{O}(\epsilon^1)$ and $\mathcal{O}(\epsilon^3)$ order are given as
\begin{align}
    \alpha^{(k,1)}_m &=-\frac{\pi K}{\epsilon^2}\exp{\left(-\frac{\pi k}{2\epsilon\gamma}\right)} \, ,
\label{eqn:alphak1} \\
\alpha^{(k,3)}_m &=-\frac{\pi K}{\epsilon^2}\exp{\left(-\frac{\pi k}{2\epsilon\gamma}\right)}\left(1-5\gamma^2\epsilon^2\right) \, .
\label{eqn:alphak3}
\end{align}
Since the next order correction would be proportional to $\mathcal{O}(\epsilon^4)$, the tail amplitude $\alpha^{(k,3)}_m$ is really valid up to $\mathcal{O}(\epsilon^3)$ order. 

The series expansion of the exponential term in equation (\ref{eqn:alpham3}) is given as
\begin{equation}
 \begin{aligned}
    \exp{\left(-\frac{\pi k}{2\gamma\epsilon}\right)} = \exp{\left(-\frac{\pi}{2\gamma\epsilon}\right)}\left[1-\pi\gamma\epsilon+\frac{1}{2}\pi^2\gamma^2\epsilon^2-\frac{1}{6}\pi(\pi^2-6)\gamma^3\epsilon^3+ \right. \\ \left. +\frac{1}{24}\pi^2(\pi^2-24)\gamma^4\epsilon^4-\frac{1}{120}\pi(\pi^4-60\pi+240)\gamma^5\epsilon^5+\mathcal{O}(\epsilon^6)\right] \, .
\label{eqn:exponential1}
 \end{aligned}
\end{equation}
If we denote $\alpha^{(j)}_m$ as the tail amplitude valid up to $\mathcal{O}(\epsilon^j)$ order when $k$ is not in the exponential, i.e. when we use the expansion given in equation (\ref{eqn:exponential1}) in equation (\ref{eqn:alpham3}), then the minimal tail amplitude valid up to different orders in $\epsilon$ are given as 
 \begin{align}
    \alpha^{(0)}_m &=-\frac{\pi K}{\epsilon^2}\exp{\left(-\frac{\pi}{2\gamma \epsilon}\right)} \, , \label{eqn:alpham0} \\
    \alpha^{(1)}_m &=-\frac{\pi K}{\epsilon^2}\exp{\left(-\frac{\pi}{2\gamma \epsilon}\right)}\left(1-\pi\gamma\epsilon\right)  \, , \label{eqn:alpham1}  \\
    \alpha^{(2)}_m &=-\frac{\pi K}{\epsilon^2}\exp{\left(-\frac{\pi}{2\gamma \epsilon}\right)}\left[1-\pi\gamma\epsilon+\left(\frac{\pi^2}{2}-5\right)\gamma^2\epsilon^2\right]  \, , \label{eqn:alpham2}  \\
    \alpha^{(3)}_m &=-\frac{\pi K}{\epsilon^2}\exp{\left(-\frac{\pi}{2\epsilon\gamma}\right)}\left[1-\pi\gamma\epsilon+\left(\frac{\pi^2}{2}-5\right)\gamma^2\epsilon^2-\left(\frac{\pi^2}{6}-6\right)\pi\gamma^3\epsilon^3\right] \, .
\label{eqn:alpham4}
\end{align}
So, the inclusion of $k$ in the exponential improves the result to make minimal tail amplitude valid to odd powers in $\epsilon$ also. In order to get $\alpha^{(3)}_m$ correctly, we have to use the series expansion of $\exp{\left[-\pi k/(2\gamma\epsilon)\right]}$, equation (\ref{eqn:exponential1}) at least up to $\mathcal{O}(\epsilon^3)$ order.

\begin{figure}[ht!]
\centering
    \includegraphics[width=0.75\linewidth]{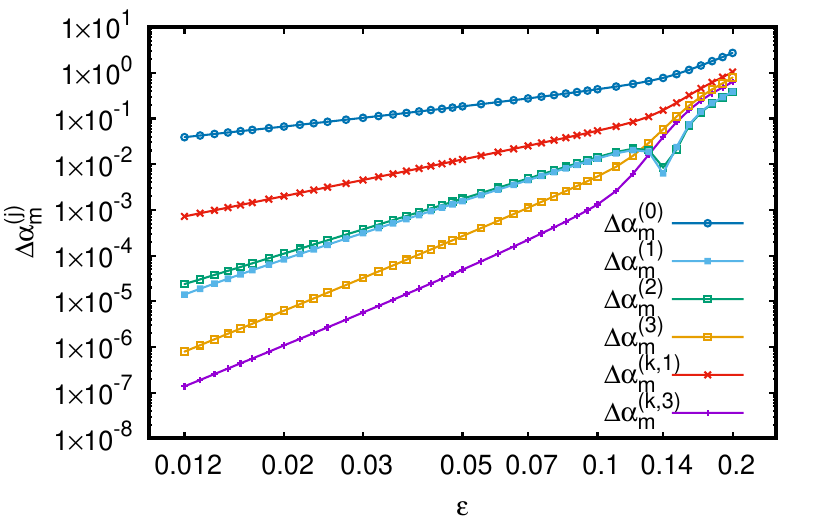}
    \caption{
       A log-log plot of the absolute value of the relative error, $\Delta\alpha_m^{(j)}=|(\alpha_m - \alpha^{(j)}_m)/\alpha_m |$, where $\alpha_m$ is the numerically calculated minimal tail amplitude and $\alpha^{(j)}_m$ is the analytical expansion of the minimal tail amplitude valid up to order $\epsilon^j$. Here we plot the relative difference for $0\leq j\leq 3$.}
    \label{fig:alphamerr}
\end{figure}

Next we compare the various orders of the analytical result $\alpha^{(j)}_m$ and $\alpha^{(k,j)}_m$ to the numerically calculated minimal tail amplitude $\alpha_m$. Since the numerical simulations are more precise, we define the relative error $\Delta\alpha_m^{(j)}$ of the $j$-th order analytical result as
\begin{equation}
    \Delta\alpha_m^{(j)}=\left|\frac{\alpha_m - \alpha^{(j)}_m}{\alpha_m}\right| \, .
\label{eqn:Deltaalpha}
\end{equation}
In Figure (\ref{fig:alphamerr}) we plot the relative error $\Delta\alpha_m^{(j)}$ logarithmically as a function of $\epsilon$. It is clear from the figure, $\Delta\alpha^{(1)}_m$ is much lower than $\Delta\alpha^{(k,1)}_m$. The reason for this is that the $\alpha^{(k,1)}_m$ expansion contains an $\epsilon^2$ term as well which has the coefficient $\pi^2\gamma^2 /2\approx 4.935\gamma^2$. This value is much larger than the correct second order coefficient in $\epsilon$ which is $(\pi^2 / 2-5)\gamma^2 \approx -0.0652\gamma^2$ as can be seen from equation (\ref{eqn:alpham2}). The relative error $\Delta\alpha^{(2)}_m$ is slightly greater than $\Delta\alpha^{(1)}_m$ ,which means that the result $\alpha^{(1)}_m$ is slightly more precise than $\alpha^{(2)}_m$. The reason for this is the neglected third order contribution  which is $(-\pi^2/6 + 6)\pi\gamma^3 \approx 13.682\gamma^3$, still larger than the second order contribution which is only $(\pi^2/2 - 5)\approx -0.0652$. However, when we consider $\alpha^{(3)}_m$ which includes the contributions up to $\epsilon^3$ order, the relative error $\Delta\alpha^{(3)}_m$ decreases significantly and is very close to $\Delta\alpha^{(k,3)}_m$. The result $\alpha^{(k,3)}_m$ is seven digits precise for the smallest $\epsilon$ value considered which is $\epsilon = 0.012$.  

The result in the paper of Grimshaw and Joshi \cite{gaj} is equivalent to $\alpha^{(k,1)}_m$ which is valid up to linear order in $\epsilon$. The correct expression for $\alpha_m$ which is valid up to $\mathcal{O}(\epsilon^3)$ order is $\alpha^{(k,3)}_m$ given in equation (\ref{eqn:alphak3}). The difference clearly is the factor $(1-5\gamma^2\epsilon^2)$ in $\alpha^{(k,3)}_m$ which is missing in \cite{gaj}. The corrected version is also supported by the numerical results.

\section{Fourth order}

In this section, we are going to compute the minimal tail amplitude up to fifth order in $\epsilon$ by solving precisely the fourth order inner problem. The fourth-order inner equation can be obtained simply by substituting equation (\ref{eqn:vexp}) in equation (\ref{eqn:vfkdv}) and use equation (\ref{eqn:exactgamc}).

The equation for the $v_2$ function, which is an $\epsilon$ independent fourth order differential equation is given as
\begin{equation}
\begin{aligned}
    v_{2,qqqq}+v_{2,qq}+6v_0v_2-16\gamma^4v_0+v_1(3v_1-4\gamma^2)=0 \, .
\label{eqn:v2a}
\end{aligned}
\end{equation}
Here we assume that the functions $v_0$ and $v_1$ are known. Although we do not know the exact $v_0$ solution, we can use the power series expansion given in equation (\ref{eqn:genmatchcondv0}) which provides a valid boundary condition for the function $v^{(-)}_0$ when $\Im{(q)}<0$ is fixed and $\Re{(q)}\to\infty$. Using $v_1=2\gamma^2/3$, we can write equation (\ref{eqn:v2a}) as
\begin{equation}
    v_{2,qqqq}+v_{2,qq}+6v_0v_2-16\gamma^4v_0-\frac{4}{3}\gamma^4=0 \, .
\label{eqn:v2}
\end{equation}
The matching condition for $v_2$ function is given in equation (\ref{eqn:v2exp}),
    $$v_2 =\left(-\frac{2q^2}{15}+\frac{2}{3}+\frac{64}{q^2}+\frac{5856}{5 q^4}-\frac{827520}{7 q^{6}}+\cdots\right)\gamma^4 \, .$$
The above (asymptotic) expansion of $v_2$ function is not convergent, since the coefficients of the  $1/q$ power series increases too fast. We are going to apply the Laplace transform method that we have already used to solve the zeroth order inner problem in Chapter (\ref{chapter:analytical_results}). However, for any integer $j\geq 0$, we cannot obtain $q^j$ as the Laplace transform of a smooth function. We approach this problem by defining a second function $\tilde{v}_2$ so that we can write the original $v_2$ function as the sum of the new $\tilde{v}_2$ function and the terms which can not be obtained from the Laplace transform,
\begin{equation}
    v_2=-\frac{2\gamma^4}{15}q^2+\frac{2\gamma^4}{3}+\tilde{v}_2 \, .
\label{eqn:v2tilde}
\end{equation}
The differential equation for the new function $\tilde{v}_2$ which is obviously different from the one given in equation (\ref{eqn:v2}) can be obtained by substituting equation (\ref{eqn:v2tilde}),
\begin{equation}
    \tilde{v}_{2,qqqq}+\tilde{v}_{2,qq}+6v_0\tilde{v}_2-12\gamma^4 v_0-\frac{4\gamma^4}{5}q^2 v_0-\frac{8\gamma^4}{5}=0 \, .
\label{eqn:v2tildeeqn}
\end{equation}
The matching condition for $\tilde{v}_2$ is
\begin{equation}
\begin{aligned}
    \tilde{v}_2 = \left(\frac{64}{q^2}+\frac{5856}{5 q^4}-\frac{827520}{7 q^{6}}+\frac{11936160}{q^{8}}+\cdots\right)\gamma^4 \, ,
\label{eqn:matchcond2}
\end{aligned}
\end{equation}
as $|q|\to\infty$ in $\Re(q)\geq 0$, $\Im(q)<0$, which in general can be written as
\begin{equation}
    \tilde{v}_2 = \sum^{\infty}_{n=1}b^{(2)}_n q^{-2n} \, .
\label{eqn:tildev2sum}
\end{equation}
Substituting into equation (\ref{eqn:v2tildeeqn}) gives the recursion relation for the coefficients $b^{(2)}_n$ which is given as
\begin{equation}
\begin{aligned}
    (2n-2)(2n-1)(2n)(2n+1)b^{(2)}_{n-1}+(2n)(2n+1)b^{(2)}_n+6\sum^{n}_{j=1}b^{(0)}_j b^{(2)}_{n-j+1} \\ -12\gamma^4 b^{(0)}_{n+1}-\frac{4\gamma^4}{5}b^{(0)}_{n+2}=0 \ ,\quad\text{for}\quad n \geq 2 \, ,
\label{eqn:b2nrecur}
\end{aligned}
\end{equation}
where $b^{(2)}_1=64\gamma^4$ and the coefficients $b^{(0)}_n$ satisfies the recursion relation given in equation (\ref{eqn:recurv0final}). The recursion relation, equation (\ref{eqn:b2nrecur}) gives the same coefficients as those in equation (\ref{eqn:matchcond2}).

Next, we will solve the equation (\ref{eqn:v2tildeeqn}) for $\tilde{v}_2$ by using the Laplace transform of the function $V^{\prime}_2(s)$ as,
\begin{equation}
    \tilde{v}_2=\int_\Gamma I_2(s)\mathrm{d}s \ , \qquad I_2(s)=\exp{(-sq)}V^{\prime}_2(s) \, ,
\label{eqn:laplacev2}
\end{equation}
where the contour $\Gamma$ runs from $s = 0$ to infinity in the complex $s$ plane such that $\Re{(sq)}>0$. The function $V^{\prime}_2(s)$ denotes the derivative of $V_2(s)$ with respect to $s$. The function $V^{\prime}_2(s)$ can be determined by using the identity for the Laplace transform of powers of $s$,
\begin{equation}
     V^\prime_2(s)=\sum^{\infty}_{n=0}a^{(2)}_ns^{2n+1} \ ,
\label{eqn:V2prs}
\end{equation}
where
\begin{equation}
    a^{(2)}_n = \frac{b^{(2)}_{n+1}}{(2n+1)!} \, , \quad\text{for}\quad n \geq 0 \ .
\label{eqn:b2a2}
\end{equation}
Substituting into equation (\ref{eqn:b2nrecur}), we get the recursion relation for the coefficients $a^{(2)}_n$ which is given as
\begin{equation}
  \begin{aligned}
    a^{(2)}_n+a^{(2)}_{n-1}+\frac{6}{(2n+3)!}\sum^{n}_{j=0}(2j+1)!(2n-2j+1)!a^{(0)}_j a^{(2)}_{n-j}-12\gamma^4 a^{(0)}_{n+1}-\\-\frac{4\gamma^4}{5}(2n+5)(2n+4) a^{(0)}_{n+2}=0 \, , \quad \text{for}\quad n \geq 1 \ ,
\label{eqn:aprimerecur1}
  \end{aligned}
\end{equation}
where $a^{(2)}_0=64\gamma^4$ and the coefficients $a^{(0)}_n$ satisfy the recursion relation given in equation (\ref{eqn:recurmain1}). The recursion relation in equation (\ref{eqn:aprimerecur1}) gives the coefficients which are consistent with equation (\ref{eqn:matchcond2}), taking equation (\ref{eqn:b2a2}) into account. 

For large $n$, the leading order behaviour of equation (\ref{eqn:aprimerecur1}) is $a^{(2)}_n\sim(-1)^n n^3$. Hence the series for $V^{\prime}_2(s)$ is convergent. The singularities are again located at $s= \pm ni$, where $n$ can take any positive integer value. There is no singularity of $V^\prime_2(s)$ at $s=0$. In order to study the behaviour of the function $V^{\prime}_2(s)$ near the singularity $s=i$, we need to study the large $n$ behaviour of the coefficients $a^{(2)}_n$. We look for the large $n$ behaviour of the coefficients $a^{(2)}_n$ in the form
\begin{equation}
    a^{(2)}_n=(-1)^n\tilde{a}^{(2)}_n \, , \qquad \tilde{a}^{(2)}_n= \sum^{\infty}_{j=-3}\frac{g^{(2)}_j}{n^j} \ .
\label{eqn:a2c2}
\end{equation}
Using equation (\ref{eqn:preciseK}) for the expansion of $a^{(0)}_n$, we can write equation (\ref{eqn:aprimerecur1}) as
\begin{equation}
 \begin{aligned}
    \tilde{a}^{(2)}_n-\tilde{a}^{(2)}_{n-1}+\sum^{j_m}_{j=0}(-1)^jW^{(0)}_{n,j}\tilde{a}^{(2)}_{n-j}+\sum^{j_m}_{j=0}(-1)^jW^{(2)}_{n,j}\tilde{a}^{(0)}_{n-j} \\ -\frac{4\gamma^4}{5}(2n+5)(2n+4) \tilde{a}^{(0)}_{n+2} +12\gamma^4 \tilde{a}^{(0)}_{n+1}=0 \ ,
\label{eqn:anwn2}
 \end{aligned}
\end{equation}
where $j_m$ is some positive integer, $W_{n,j}^{(0)}$ is given in equation (\ref{eqn:W0}) and 
\begin{equation}
    W^{(2)}_{n,j}=\frac{6}{(2n+3)!}(2j+1)!(2n-2j+1)!a^{(2)}_j \ .
\label{eqn:Wm}
\end{equation}

The homogeneous part of the equation (\ref{eqn:anwn2}) can be obtained by setting $\tilde{a}^{(0)}_n=0$, and it agrees exactly with equation (\ref{eqn:recurmain3}). Hence it has the solution $G_0(n)$ multiplied by some arbitrary constant, it follows that the general solution of equation (\ref{eqn:anwn2}) can be written as 
\begin{equation}
    \tilde{a}^{(2)}_n=K[G_2(n)+K_2G_0(n)] \, ,
\label{eqn:tildea2}
\end{equation}
where $K_2$ is an arbitrary constant and $G_2(n)$ is a particular solution of equation (\ref{eqn:anwn2}). The expansion of the function $G_2(n)$ can be obtained by using some algebraic manipulation software such as \emph{Mathematica}, similarly as we did before for $G_0(n)$, see equation (\ref{eqn:preciseKvalue}),
\begin{equation}
 \begin{aligned}
    G_2(n) = \left(\frac{16}{15}n^3+\frac{28}{5}n^2+\frac{368}{15}n-\frac{132}{n}+\frac{4122}{5n^2}-\frac{50833}{10n^3}+\frac{3144915}{112n^4}+\cdots\right)\gamma^4 \, .
 \end{aligned}
\label{eqn:G2}
\end{equation}
 Here, we have chosen the unique particular solution $G_2(n)$ for which there is no $n^0$ term. Similarly as before in calculating the value of the constant $K$, see equation (\ref{eqn:Kvalue}), we can obtain the value of constant $K_2$, by calculating the coefficients $a^{(2)}_n$ for sufficiently large $n$. Using equations (\ref{eqn:a2c2}) and (\ref{eqn:tildea2}), we can write $K_2\approx[(-1)^na^{(2)}_n/K-G_2(n)]/G_0(n)$. The value of constant $K_2$ up to 20 digits is given as
\begin{equation}
    K_2= -36.544068193583744293 \ \gamma^4 \, .
\label{eqn:K2value}
\end{equation}
The next task is to calculate the imaginary part of the asymmetric function $\tilde v_2$ ($\equiv$ imaginary part of the asymmetric function $v_2$) on the lower part of the imaginary $q$-axis. The function $\tilde{v}_2$ can be calculated in the domain $-\pi<\mathrm{arg}(q)\leq 0$ by the integral given in equation (\ref{eqn:laplacev2}), using the same arguments as for $v^{(-)}_0$ in the few paragraphs before equation (\ref{eqn:sumresidue}). The leading order imaginary part of the $\tilde{v}_2$ function on the imaginary $q$-axis can be determined by using the residue of the function $I_2(s)$ at $s=i$. Substituting equations (\ref{eqn:V2prs}), (\ref{eqn:a2c2}) and (\ref{eqn:tildea2}) into equation (\ref{eqn:laplacev2}), we get
\begin{equation}
    I_2(s)\approx \exp{(-sq)}K\sum^{\infty}_{n=0}(-1)^n[G_2(n)+K_2G_0(n)]s^{2n+1} \, . \label{eqn:I2s}
\end{equation}
From the expansions of $G_0(n)$ and $G_2(n)$ given in equations (\ref{eqn:preciseKvalue}) and (\ref{eqn:G2}) respectively, the terms of the form $(-1)^nn^js^{2n+1}$ with $j\geq 0$ will give the contribution to the residue. These terms can be summed, for instance for $j=1$ and $j=2$,
\begin{equation}
    \sum^{\infty}_{n=0}(-1)^n ns^{2n+1}=-\frac{s^3}{(1+s^2)^2} \ ,\quad\text{and}\quad \sum^{\infty}_{n=0}(-1)^n n^2s^{2n+1}=\frac{s^3(s^2-1)}{(1+s^2)^3} \, .
\end{equation}
Such kinds of sum can be calculated for any arbitrary integer $j\geq 0$. But, unfortunately, we could not find a general formula for this sum which is valid for arbitrary $j\geq 0$. However, the residue turns out to be very simple,
\begin{equation}
    \underset{s=i}{\mathrm{Res}} \, \sum^{\infty}_{n=0}(-1)^n n^js^{2n+1}=\frac{1}{2}(-1)^j\quad\text{for}\quad j\geq 0 \, .
\label{eqn:residue2}
\end{equation}
Using equations (\ref{eqn:preciseKvalue}) and (\ref{eqn:G2}) in equation (\ref{eqn:I2s}), we get
\begin{equation}
 \begin{aligned}
    \underset{s=i}{\mathrm{Res}} \, I_2(s) = \frac{1}{2}K\exp{(-iq)}(K_2-20\gamma^4) \, .
\label{eqn:residueI2}
 \end{aligned}
\end{equation}
The next step is to define the symmetric function $v^{(m)}_2$ by taking the contour $\Gamma_m$ along the positive imaginary $s$-axis. The difference of the asymmetric function $v^{(-)}_2$ and the symmetric function $v^{(m)}_2$ can be calculated by using the residue theorem which can be applied by choosing a curve going from $s=0$ to infinity in the domain $0 < \mathrm{arg(s)} < \pi/2$ and coming back along the upper half of the imaginary $s$-axis. In this way we have to take into account half of the residue at $s=i$, so we get
\begin{equation}
   v^{(-)}_2 - v^{(m)}_2=\frac{1}{2}\pi i K\exp{(-iq)}(K_2-20\gamma^4) \, .
 \label{eqn:vminusm}
\end{equation}
Here we do not consider the finite number of divergent terms proportional to $q^n\exp{(-iq)}$ with $n >0$. The coefficients of these terms are completely fixed already by the leading order constant $K$.
Since $\Im(v^{(m)}_2) = 0$ because to its symmetry, the imaginary part of $v^{(-)}_2$ on the lower part of imaginary $q$-axis is given by
\begin{equation}
    \Im{(v^{(-)}_2)}\approx\frac{1}{2}\pi K\exp{(-iq)}(K_2-20\gamma^4) \, .
\label{eqn:imv2minus}
\end{equation}
Using the rescaled function (\ref{eqn:vepssq}), $v=\epsilon^2 u$, we get the imaginary part of the asymmetric function $u_-$ which up to $\mathcal{O}(\epsilon^{4})$ order is given as
\begin{equation}
 \begin{aligned}
    \Im{(u_-)} \approx \frac{\pi K}{2\epsilon^2}\exp{(-iq)}\left[1+\left(\frac{K_2}{\gamma^4}-20\right)\gamma^4\epsilon^4+\mathcal{O}(\epsilon^6)\right] \, ,
    \label{eqn:Imum4a1}
 \end{aligned}
\end{equation}
for $\Re{(q)}=0$ and $\Im{(q)}<0$.

\subsection{Oscillatory tail amplitude up to fifth order}

The generalized expression for the minimal tail amplitude $\alpha_m$ given in equation (\ref{eqn:alpham3}) can be obtained by adding a fourth order contribution,
\begin{equation}
    \alpha^{(k,5)}_m=-\frac{\pi K}{\epsilon^2}\exp{\left(-\frac{\pi k}{2\gamma\epsilon}\right)}[1-5\gamma^2\epsilon^2-\zeta_4\gamma^4\epsilon^4+\mathcal{O}(\epsilon^6)] \, ,
\label{eqn:alphak5}
\end{equation}
where the constant coefficient $\zeta_4$ is yet to be determined. Considering only the $q^0\exp{(-iq)}$ terms up to $\mathcal{O}(\epsilon^4)$ order from the $u_w$ function in equation (\ref{eqn:result1a}), 
\begin{equation}
    u_w=\frac{i\alpha_m}{2}\exp{\left(\frac{\pi k}{2\gamma\epsilon}\right)}\exp{(-iq)}[1+5\gamma^2\epsilon^2-25\gamma^4\epsilon^4+\mathcal{O}(\epsilon^6)] \, .
\label{eqn:uweps4}
\end{equation}
Substituting equation (\ref{eqn:alphak5}) into equation (\ref{eqn:uweps4}), we get
\begin{equation}
    \Im{(u_w)}=-\frac{\pi K}{2\epsilon^2}\exp{(-i q)}[1-(\zeta_4+50)\gamma^4\epsilon^4+\mathcal{O}(\epsilon^6)] \ .
\label{eqn:Imuweps4a}
\end{equation}
Using equation (\ref{eqn:Imumuw}), $\Im{(u_-)}=-\Im{(u_w)}$, we obtain
\begin{equation}
    \zeta_4=-\left(\frac{K_2}{\gamma^4}+30\right)\approx 6.544068193583744293 \, .
\label{eqn:A2}
\end{equation}

Since the next correction would be proportional to $\mathcal{O}(\epsilon^6)$, equation (\ref{eqn:alphak5}) is really valid up to $\mathcal{O}(\epsilon^5)$ order. Using the series expansion of $\exp[-\pi k/(2\gamma\epsilon)]$ given in equation (\ref{eqn:exponential1}), the expressions for the minimal tail amplitude $\alpha_m$ which are valid up to the $\mathcal{O}(\epsilon^4)$ and $\mathcal{O}(\epsilon^5)$ orders are given as
\begin{equation}
\begin{aligned}
    \alpha^{(4)}_m = -\frac{\pi K}{\epsilon^2}\exp{\left(-\frac{\pi}{2\epsilon\gamma}\right)}\left[1-\pi\gamma\epsilon+\left(\frac{\pi}{2}-5\right)\gamma^2\epsilon^2-\left(\frac{\pi^2}{6}-6\right)\pi\gamma^3\epsilon^3+ \right. \\ \left. +\left(\frac{\pi^4}{24}-\frac{7\pi^2}{2}-\zeta_4\right)\gamma^4\epsilon^4\right] \, ,
\label{alpham4} 
\end{aligned}
\end{equation}
\begin{equation}
 \begin{aligned}
     \alpha^{(5)}_m = -\frac{\pi K}{\epsilon^2}\exp{\left(-\frac{\pi}{2\epsilon\gamma}\right)}\left[1-\pi\gamma\epsilon+\left(\frac{\pi}{2}-5\right)\gamma^2\epsilon^2-\left(\frac{\pi^2}{6}-6\right)\pi\gamma^3\epsilon^3+ \right. \\ \left. +\left(\frac{\pi^4}{24}-\frac{7\pi^2}{2}-\zeta_4\right)\gamma^4\epsilon^4-\left(\frac{\pi^4}{120}-\frac{4\pi^2}{3}-7-\zeta_4\right)\gamma^5\epsilon^5\right] \, ,
\label{eqn:alpham4c}
 \end{aligned}
\end{equation}
where the value of $\zeta_4$ is given in equation (\ref{eqn:A2}). To get $\alpha^{(5)}_m$ correctly, we have to expand $\exp{[-\pi k/(2\gamma\epsilon)]}$ at least up to $\mathcal{O}(\epsilon^5)$ order before using it in equation (\ref{eqn:alphak5}).

\begin{figure}[ht!]
\centering
    \includegraphics[width=0.75\linewidth]{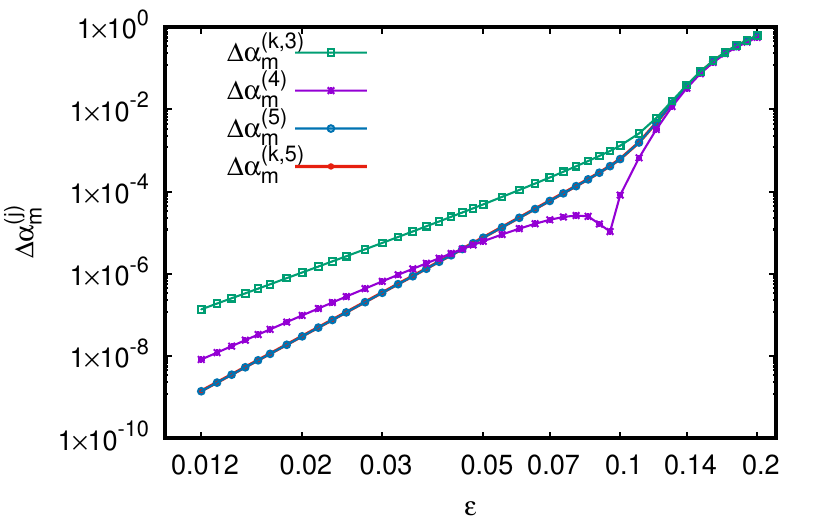}
    \caption{
       A log-log plot of the relative error, $\Delta\alpha_m^{(j)}=|(\alpha_m - \alpha^{(j)}_m)/\alpha_m |$, where $\alpha_m$ is the numerically calculated minimal tail amplitude and $\alpha^{(j)}_m$ is the analytical expansion of minimal tail amplitude valid up to $\mathcal{O}(\epsilon^j)$ order. Here we plot the relative difference for $3\leq j\leq 5$.}
    \label{fig:alphamerrex}
\end{figure}

Figure (\ref{fig:alphamerrex}) shows the relative error $\Delta\alpha_m^{(j)}=|(\alpha_m - \alpha^{(j)}_m)/\alpha_m |$ of the analytical expansion $\alpha^{(j)}_m$ when compared with the more precise numerically computed $\alpha_m$. The results $\alpha^{(3)}_m$ and $\alpha^{(4)}_m$ are correct to seven and eight digits of precision respectively for the smallest $\epsilon$ value considered. The two fifth order results $\alpha^{(k,5)}_m$ and $\alpha^{(5)}_m$ are so close to each other that the relative error $\Delta\alpha^{(5)}_m$ and $\Delta\alpha^{(k,5)}_m$ are indistinguishable on the logarithmic plot. The results $\alpha^{(5)}_m$ and $\alpha^{(k,5)}_m$ are correct to nine digits of precision for the smallest $\epsilon$ value considered which is $\epsilon = 0.012$.

\chapter{Asymmetric solutions}\label{chapter:thder}

The stationary (once integrated) fKdV equation is given in equation (\ref{eqn:1aintFKDV}),
\begin{align}
    \epsilon^2 u_{xxxx}+u_{xx}+3u^2-cu = 0 \, . \nonumber
\end{align}
There are two important solutions of this equation, the symmetric solution $u_m$ with minimal amplitude tail, and the asymmetric solution $u_-$ that has no tail, decays exponentially to zero for $x>0$ and blows up to the negative values at finite $x<0$. Both these solutions have a central core which can be well described by the appropriately truncated asymptotic expansion given in equation (\ref{eqn:coreuexp}), but neither the oscillatory tail of the symmetric solution nor the blow up behaviour of the asymmetric solution can be explained by this series expansion. However both the solutions can be calculated numerically. In the previous chapters we calculated the oscillatory minimal tail amplitude $\alpha_m$ of the symmetric solution which turns out to be exponentially small in $\epsilon$. In this chapter we are going to study the asymmetry of the solution $u_-(x)$ which can be quantified by calculating its third derivative at the origin. We fix the translational invariance of the solutions by requiring that the first derivative is zero at $x=0$. Like the oscillatory tail amplitude $\alpha_m$, the third derivative of the asymmetric function at the origin also turns out $\mathcal{O}(\exp(-1/\epsilon)$, i.e. it also turns out beyond all orders small in perturbation theory (but of course with different perturbation coefficients). We also compare the more precise numerical simulations with the analytical expansion.

Let us study the behaviour of some solution which blow up at some finite $x$. Let this solution be of the form $u = A(x-x_0)^b$, where $A$ and $b$ are the constants yet to be determined. For such large $|u|$ solutions, it can be shown that the fourth derivative term is dominant. So we solve $\epsilon^2 u_{xxxx}+3u^2=0$ and get
\begin{equation}
    \epsilon^2 b(b-1)(b-2)(b-3)(x-x_0)^{b-4}+3A(x-x_0)^{2b} = 0 \, ,
\label{eqn:uxxxx3u2}
\end{equation}
which gives $ b = -4$ and $A = -280 \epsilon^2$. Hence the solution $u_-(x)$ in the blow-up region can be approximated as 
\begin{equation}
    u_-(x) = -\frac{280}{(x-x_0)^4}\epsilon^2 \, ,
\label{eqn:ublowup}
\end{equation}
which shows that the function $u_-$ is blowing up to the negative values, see Figure (\ref{fig:uasym}) and the blow up rate of the solution is $1/(x-x_0)^4$.

As we have already discussed in Chapter (\ref{chapter:higher_orders}) the difference of the symmetric and the asymmetric solutions, $u_m - u_-$ is exponentially small in terms of the small parameter $\epsilon$ everywhere for $x>0$ including the core region. For $x>0$ the minimal tail amplitude $\alpha_m$ of the symmetric solution $u_m$ can be compensated by the linearized solution $u_w$ in equation (\ref{eqn:uwdm}) with $\beta=\alpha_m$ and $\delta_w=0$ to obtain the asymmetric solution $u_-$. Both functions $u_-(x)$ and $u_m(x)$ has zero derivative at the origin, hence the function $u_-(x)$ must be necessarily shifted,
\begin{equation}
    u_-(x-d) = u_m(x) - u_w(x) \, ,
\label{eqn:uminus}
\end{equation}
where $d$ is a very small constant which turns out to be of the order of minimal tail amplitude $\alpha_m$, and hence neglecting the $\mathcal{O}(d^2)$ terms will be justified. The asymptotic series expansion in powers of the small parameter $\epsilon$, equation (\ref{eqn:coreuexp}) approximates well the core region of the functions $u_-(x)$ and $u_m(x)$,
\begin{align}
    u^{(c)}_m \approx u^{(c)}_- \approx \sum_{n=0}^\infty \epsilon^{2n}u_n \, , 
\end{align}
where $(c)$ in the superscript denotes the core region, and $u_n$ functions are given in equation (\ref{eqn:un}),
$$u_n=\sum_{j=1}^{n+1}u_{nj}\gamma^{2(n+1)}\mathrm{sech}^{2j}(\gamma x) \, ,$$
where the first few $u_{n,j}$ coefficients can be seen in Table (\ref{table:ujk}).

The function $u_w$ which is a difference of $u_m$ and $u_-$ is given as
\begin{equation}
     u_w=\alpha_m\exp{\left(\sum^{\infty}_{\substack{n=2 \\ \text{even}}}A_n\epsilon^n\right)}\mathrm{sin}\left(\frac{kx}{\epsilon}-\sum^{\infty}_{\substack{n=1 \\ \text{odd}}}\tilde{A}_n\epsilon^n\right) \ ,
\label{eqn:uwas}
\end{equation}
where the first few $A_n$ values can be seen in equations (\ref{eqn:aminus1})-(\ref{eqn:a6}), $\delta_m=6\gamma\epsilon+\cdots$, and the minimal tail amplitude $\alpha_m$ up to $\mathcal{O}(\epsilon^5)$ order is given in equation (\ref{eqn:alphak5}). Differentiate equation (\ref{eqn:uminus}) once with respect to $x$,
\begin{equation}
    u^{(1)} _-(x-d) = u^{(1)} _m(x) - u^{(1)} _w(x) \, ,
\label{eqn:1der}
\end{equation}
where $u^{(j)}$ denotes the $j$-th derivative of a function $u$ with respect to $x$. Since $u^{(1)}_-(0)$ is zero which is a boundary condition in our numerical simulations to make the solution unique, hence from equation (\ref{eqn:1der}), we get
\begin{equation}
    u^{(1)} _m(d) - u^{(1)} _w(d) = 0 \ .
\label{eqn:1derumuw}
\end{equation}
By using the Taylor series expansion, $f(x) = f(x_0)+(x-x_0)f^\prime(x_0)+\mathcal{O}\left((x-x_0)^2\right)$, we can write
\begin{align}
    u^{(1)}_m(d) &= u^{(1)}_m(0)+ d \cdot u^{(2)}_m(0)+\mathcal{O}(d^2) \, , \label{eqn:umxx} \\
     u^{(1)}_w(d) &= u^{(1)}_w(0)+ d \cdot u^{(2)}_w(0)+\mathcal{O}(d^2) \, . \label{eqn:uwxx}
\end{align}
The different orders of the derivatives of the given functions $u_m$ and $u_w$ can be calculated to any desirable order in $\epsilon$ by using some algebraic manipulation software such as \emph{Mathematica}, 
\begin{align}
    u_m(0) &= 2\gamma^2[1+5\gamma^2\epsilon^2+30\gamma^4\epsilon^4+1005\gamma^6\epsilon^6+\mathcal{O}(\epsilon^8)] \, , \label{eqn:um0} \\
    u^{(1)}_m(0) &= 0 \, , \label{eqn:1derum} \\
    u^{(2)}_m(0) &= -4\gamma^4[1+20\gamma^2\epsilon^2+495\gamma^4\epsilon^4+19800\gamma^6\epsilon^6+\mathcal{O}(\epsilon^8)] \, , \label{eqn:2derum} \\
    u^{(3)}_m(0) &= 0 \, , \label{eqn:3derum} \\
     u^{(4)}_m(0) &= 16\gamma^6[2+85\gamma^2\epsilon^2+3780\gamma^4\epsilon^4+220140\gamma^6\epsilon^6+\mathcal{O}(\epsilon^8)] \, , \label{eqn:4derum} 
\end{align}
\begin{align}
    u_w(0) &= 0 \, , \label{eqn:uw01}\\
    u^{(1)}_w(0) &= \alpha_m\left(\frac{1}{\epsilon}+11\gamma^2\epsilon+202\gamma^4\epsilon^3+8098\gamma^6\epsilon^5+\mathcal{O}(\epsilon^7)\right) \, , \label{eqn:1deruw} \\
    u^{(2)}_w(0) &= 0 \, , \label{eqn:2deruw} \\
    u^{(3)}_w(0) &= -\alpha_m\left(\frac{1}{\epsilon^3}+\frac{3\gamma^2}{\epsilon}-18\gamma^4\epsilon+734\gamma^6\epsilon^3+\mathcal{O}(\epsilon^5)\right) \, , \label{eqn:3deruw} \\
    u^{(4)}_w(0) &= 0 \, . \label{eqn:4deruw} 
\end{align}
The shift $d$ to the leading order in $\alpha_m$ can be obtained from equation (\ref{eqn:1derumuw}),
\begin{equation}
    d = \alpha_m\left(-\frac{1}{4\gamma^4\epsilon}+\frac{9}{4\gamma^2}\epsilon+\frac{113}{4}\epsilon^3+\frac{4987}{4}\gamma^2\epsilon^5+\mathcal{O}(\epsilon^7)\right)+\mathcal{O}(\alpha^2_m) \, .
\label{eqn:shift}
\end{equation}
Equation (\ref{eqn:shift}) shows that for the small $\epsilon$ values, the value of the shift $d$ is always negative which means that the maximum of the function $u_-$ is shifted to the left with respect to maximum of $u_m$. The third derivative of the function $u_-$ at $x = 0$ can be calculated similarly and is given as
\begin{equation}
    u^{(3)}_-(0) = u^{(3)}_m(d) - u^{(3)}_w(d) = d \cdot u^{(4)}_m(0) - u^{(3)}_w(0) \, ,
\label{eqn:thder}
\end{equation}
where the shift $d$ is given in equation (\ref{eqn:shift}), and the fourth derivative of the function $u_m$ and third derivative of the function $u_w$ at the origin are given in equations (\ref{eqn:4derum}) and (\ref{eqn:3deruw}) respectively. Using equation (\ref{eqn:thder}), the third derivative of the function $u_-$ at the origin is given as
\begin{equation}
    \partial^3_x u_-(0) = \frac{\alpha_m}{\epsilon^3}\left[1-5\gamma^2\epsilon^2-286\gamma^4\epsilon^4-10422\gamma^6\epsilon^6-666164\gamma^8\epsilon^8+\mathcal{O}(\epsilon^{10})\right]+\mathcal{O}(\alpha_m^2) \, ,
\label{eqn:u30}
\end{equation}
where $\alpha_m$ is minimal tail amplitude of the symmetric solution which up to $\mathcal{O}(\epsilon^5)$ order is given in equation (\ref{eqn:alphak5}). Since $\alpha_m > 0$, equation (\ref{eqn:u30}) shows that for small $\epsilon$ values the third derivative of the function $u_-$ at the origin is always positive.

\section{Third derivative up to fifth-order}

Substituting equation (\ref{eqn:alphak5}) for the minimal tail amplitude $\alpha_m$ into equation (\ref{eqn:u30}), we get different order analytical expansions for the third derivative of a function $u_-$ at the origin which are given as
\begin{align}
    \partial^3_xu^{(k,1)}_-(0) &= -\frac{\pi K}{\epsilon^5}\exp{\left(-\frac{\pi k}{2\gamma\epsilon}\right)} \ , \label{eqn:d3k1} \\
    \partial^3_xu^{(k,3)}_-(0) &= -\frac{\pi K}{\epsilon^5}\exp{\left(-\frac{\pi k}{2\gamma\epsilon}\right)}\left(1-10\gamma^2\epsilon^2\right) \, , \label{eqn:d3k3} \\
    \partial^3_x u^{(k,5)}_-(0) &= -\frac{\pi K}{\epsilon^5}\exp{\left(-\frac{\pi k}{2\gamma\epsilon}\right)}\left(1-10\gamma^2\epsilon^2-(\zeta_4+261)\gamma^4\epsilon^4\right) \, , \label{eqn:d3k5} 
\end{align}
where $K$, $k$, and $\zeta_4$ are given in equations (\ref{eqn:Kvalue}), (\ref{eqn:kappa}), and (\ref{eqn:A2}) respectively. The superscript $(k,j)$ denotes the result is valid up to $\mathcal{O}(\epsilon^j)$ order when $k$ is kept in the argument of the exponential term. If we use the expansion of the exponential given in equation (\ref{eqn:exponential1}) in equation (\ref{eqn:d3k5}), then the third derivative of the asymmetric function $u_-$ at the center valid up to different orders in $\epsilon$ are given as
\begin{align}
    \partial^3_xu^{(0)}_-(0) &= -\frac{\pi K}{\epsilon^5}\exp{\left(-\frac{\pi}{2\gamma\epsilon}\right)} \, , \\
    \partial^3_xu^{(1)}_-(0) &= -\frac{\pi K}{\epsilon^5}\exp{\left(-\frac{\pi}{2\gamma\epsilon}\right)}(1-\pi\gamma\epsilon) \, , \\
    \partial^3_xu^{(2)}_-(0) &= -\frac{\pi K}{\epsilon^5}\exp{\left(-\frac{\pi}{2\gamma\epsilon}\right)}\left[1-\pi\gamma\epsilon+\left(\frac{\pi^2}{2}-10\right)\gamma^2\epsilon^2\right] \, , \\
    \partial^3_xu^{(3)}_-(0) &= -\frac{\pi K}{\epsilon^5}\exp{\left(-\frac{\pi}{2\gamma\epsilon}\right)}\left[1-\pi\gamma\epsilon+\left(\frac{\pi^2}{2}-10\right)\gamma^2\epsilon^2-\left(\frac{\pi^2}{6}-11\right)\pi\gamma^3\epsilon^3\right] \, , \\
    \partial^3_xu^{(4)}_-(0) &= -\frac{\pi K}{\epsilon^5}\exp{\left(-\frac{\pi}{2\gamma\epsilon}\right)}\left[1-\pi\gamma\epsilon+\left(\frac{\pi^2}{2}-10\right)\gamma^2\epsilon^2-\left(\frac{\pi^2}{6}-11\right)\pi\gamma^3\epsilon^3 \right. \nonumber \\& \left. -\left(\zeta_4-\frac{\pi^4}{24}+6\pi^2+261\right)\gamma^4\epsilon^4\right] \, , \\
    \partial^3_xu^{(5)}_-(0) &= -\frac{\pi K}{\epsilon^5}\exp{\left(-\frac{\pi}{2\gamma\epsilon}\right)}\left[1-\pi\gamma\epsilon+\left(\frac{\pi^2}{2}-10\right)\gamma^2\epsilon^2-\left(\frac{\pi^2}{6}-11\right)\pi\gamma^3\epsilon^3 \right. \nonumber \\& \left. -\left(\zeta_4-\frac{\pi^4}{24}+6\pi^2+261\right)\gamma^4\epsilon^4+\left(\zeta_4-\frac{\pi^4}{120}+\frac{13}{6}\pi^2+249\right)\pi\gamma^5\epsilon^5\right] \, ,
\end{align}
where the superscript $(j)$ denotes that the result is valid up to $\mathcal{O}(\epsilon^j)$ order.

\begin{figure}[ht!]
	\centering
        \includegraphics[width=0.75\linewidth]{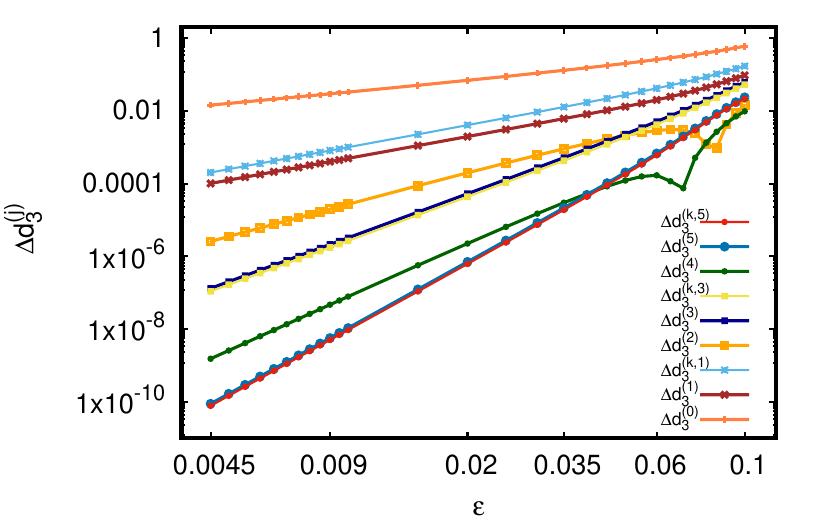}
        \caption{
            A log-log plot of $\Delta \mathrm{d}^{(j)}_3 = \left|\left(\partial^{3}_x u_-(0) -\partial^{3}_x u^{(j)}_-(0)\right)/\partial^{3}_x u_-(0)\right|$, showing the relative difference of the numerically calculated $\partial^3_x u_-(0)$ from its various order analytical approximations $\partial^3_x u_-^{(j)}(0)$ up to order five in $\epsilon$.
    \label{fig:eps_3dermix}}
\end{figure}

The comparison of the more precise numerical calculations and the various order analytical results can be seen in Figure (\ref{fig:eps_3dermix}) which shows that the $\mathcal{O}(\epsilon^5)$ order analytical result is more than 10 digits precise for the smallest $\epsilon$ considered which is $\epsilon = 0.0045$. 

\chapter{Conclusions}\label{chapter:conclusions}

In this thesis, we have solved the time independent fifth-order Korteweg-de Vries (fKdV) equation numerically as well as analytically. This equation models the capillary-gravity waves and exhibits both symmetric and asymmetric solutions. The symmetric solutions that we are studying are the weakly localized solutions which have a well defined central core similar to the KdV 1-soliton solution, but accompanied with extremely small oscillatory tail to infinity on both sides of the core. These stationary symmetric solutions form a one parameter family for fixed $\epsilon$ characterized by the phase shift $\delta$ of the oscillating tail. The investigated asymmetric solutions do not have any oscillatory tail either side of the asymmetric core, instead they decay exponentially to zero on one side of the core and blow up to negative values on the other side of the core.

For the numerical calculations, we have used the exponentially convergent pseudo-spectral method to search for the stationary symmetric solutions of the fKdV equation. We have calculated the extremely small tail amplitude $\alpha$ for relatively small $\epsilon$ values by using the \verb !C++! library called CLN. We have also used the \verb !C!  library called the ARB which turns out to be about twenty times faster due to its more advanced matrix manipulation methods. The speed of the codes allows us to search numerically for the minimum phase $\delta_m$ by using the Brent's minimization method. The minimum phase $\delta_m$ gives the corresponding minimal tail amplitude $\alpha_m$. Since we have used the linear approximation in the far field analysis, the matching to this linear approximation of the tail brings an error of the order $\alpha_m^2$. The numerical values of $\alpha_m$ to correct digits of precision for various $\epsilon$ values can be seen in Table (\ref{table:alpdel}). We have also investigated the asymmetry of the $u_-(x)$ solution that decays to zero on one side by computing the third derivative of the function $u_-$ at the origin. This computation does not involve a minimization process and hence the CLN code was fast enough for this calculation. The numerical values of the third derivative $\partial^3_x u_-(0)$ for several $\epsilon$ values can be seen in Table (\ref{table:thirdder}).

We have used the method of matched asymptotics in the complex plane in the limit $\epsilon\to 0$ for the analytical calculations of $\alpha_m$. We have calculated the minimal tail amplitude $\alpha_m$ up to $\mathcal{O}(\epsilon^5)$ order. Furthermore, we have used the complex extension of the antisymmetric solution $u_w$ for the calculation of the third derivative of the asymmetric function $u_-$ at the center. We have also calculated the third derivative of the function $u_-$ up to $\mathcal{O}(\epsilon^5)$ order. In the limit $\epsilon\to 0$, both the minimal tail amplitude $\alpha_m$ and the third derivative $\partial^3_xu_-(0)$ decreases exponentially with $1/\epsilon$, see Figure (\ref{fig:invepsal}) and Figure(\ref{fig:inv3der}). The main analytical result for the minimal tail amplitude $\alpha_m$ and the third derivative $\partial^3_x u_-$ can be presented as
\begin{align}
     \alpha_m &= \frac{-\pi K}{\epsilon^2}\exp{\left(-\frac{\pi k}{2\gamma\epsilon}\right)}\left[1-5\gamma^2\epsilon^2-\zeta_4\gamma^4\epsilon^4+\mathcal{O}(\epsilon^6)\right] \ , \nonumber \\
    \partial^3_x u_- &= \frac{-\pi K}{\epsilon^5}\exp{\left(-\frac{\pi k}{2\gamma\epsilon}\right)}\left[1-10\gamma^2\epsilon^2-(\zeta_4+261)\gamma^4\epsilon^4+\mathcal{O}(\epsilon^6)\right] \ , \nonumber
\end{align}
where $K \approx -19.9689$, $\zeta_4 \approx 6.54407$, $\gamma$ characterizes the amplitude which in turn is directly related to the speed of the KdV 1-soliton solution, and $k = \sqrt{1+4\gamma^2\epsilon^2}$.

We have compared the more precise numerically calculated minimal tail amplitude $\alpha_m$ and the third derivative $\partial^3_x u_-$ with the various order analytical results $\alpha^{(j)}_m$ and $\partial^3_xu^{(j)}_-$, where $(j)$ in the superscript denotes that the result is calculated up to $\mathcal{O}(\epsilon^j)$ order. We did this comparison by showing graphically the logarithmic dependence of the relative differences of $\alpha_m$ and $\partial^3_x u_-$ and their various orders of analytical approximations with respect to $\epsilon$. It turns out that the value of $\alpha^{(5)}_m$ is correct to nine digits of precision for the smallest $\epsilon$ considered which is $\epsilon=0.012$, see Figure (\ref{fig:alphamerrex}) and $\partial^3_xu^{(5)}_-$ is correct to $14$ digits of precision for the smallest $\epsilon$ considered which is $\epsilon = 0.0045$, see Figure (\ref{fig:eps_3dermix}). This shows the excellent agreement between the analytical and the numerical results.

Furthermore, we have calculated analytically the asymptotic expansion for the phase shift $\delta_m$ corresponding to the minimal tail, see equation (\ref{eqn:deltamr}). The optimally truncated asymptotic series for $\delta_m$ also results in extremely good agreement with the numerical results. The numerical values of $\delta_m$ to correct digits of precision for various $\epsilon$ values can be seen in Table (\ref{table:alpdel}). We have also computed the relation between the tail amplitude $\alpha$ of any symmetric solution $u$ of phase $\delta$ and the minimal tail amplitude $\alpha_m$, which is given by $\alpha = \alpha_m/\mathrm{cos}(\delta-\delta_m)$. The error in this case also turns out to be of the $\mathcal{O}(\alpha^2)$ order, similar to the error caused by the linearization of the fKdV equation in the asymptotic region.

There was a long-standing discrepancy between the second order perturbation result of Grimshaw and Joshi \cite{gaj} and the multiple precision numerical results of Boyd \cite{boyd2}. The numerical calculations of Boyd showed that the second order perturbation coefficient is between $-0.1< \zeta_2 < 0$, for $\gamma = 1$, while Grimshaw and Joshi predicted it as $\pi^2/2\approx 4.935$, for $\gamma=1$. It turns out that this difference was due to a missing factor $(1+5\gamma^2 \epsilon^2)$ in the expression which describes the linear perturbation around the intended solution in \cite{gaj}, see equation (\ref{eqn:result1}). Including this factor, our perturbation result shows that the second order perturbation coefficient is $(\pi/2-5) \approx -0.0652$, for $\gamma=1$, which is also consistent with Boyd's numerical results. The fitted (numerical) value of not only of the second order perturbation coefficient but of all the perturbation coefficients up to the fifth order in $\epsilon$ agrees well with the theoretical result to several digits of precision, see Figures (\ref{fig:alphamerr}) and (\ref{fig:alphamerrex}).

\appendix

\chapter{Minimal tail amplitude of the symmetric solution and third derivative of the asymmetric solution for arbitrary parameter independent wave speed}

The $\epsilon$ dependent wave speed $c$ is given in equation (\ref{eqn:exactgamc}),
$$c = 4\gamma^2+16\gamma^4\epsilon^2 \, .$$ 
The minimal oscillatory tail amplitude of the symmetric function $u_m$ and the third derivative of the asymmetric function $u_-$ at the center valid up to $\mathcal{O}(\epsilon^5)$ order are given in equations (\ref{eqn:alphak5}) and (\ref{eqn:d3k5}),
\begin{align}
    \alpha^{(k,5)}_m &= \frac{-\pi K}{\epsilon^2}\exp{\left(-\frac{\pi k}{2\gamma\epsilon}\right)}\left[1-5\gamma^2\epsilon^2-\zeta_4\gamma^4\epsilon^4+\mathcal{O}(\epsilon^6)\right] \, , \nonumber \\
    \partial^3_xu_-^{(k,5)} &= \frac{-\pi K}{\epsilon^5}\exp{\left(-\frac{\pi k}{2\gamma\epsilon}\right)}\left[1-10\gamma^2\epsilon^2-(\zeta_4+261)\gamma^4\epsilon^4+\mathcal{O}(\epsilon^6)\right] \, , \nonumber
\end{align}
where $K \approx -19.9689$, $\zeta_4 \approx 6.54407$, see equations (\ref{eqn:Kvalue}) and (\ref{eqn:A2}).
Here we are going to discuss the expressions of minimal oscillatory tail amplitude $\alpha_m$ of the symmetric solution $u_m$ and the third derivative of the asymmetric solution $u_-$ at the center when we choose the wave speed $c$ as the parameter independent constant. In this way the parameter $\gamma$ has to depend on $\epsilon$. To get the wave speed dependent expressions for $\alpha_m$ and $\partial^3_xu_-$ we have to expand the factor $\gamma$ in terms of $\epsilon$ in addition to $k$ in the argument of exponential and then expand the resulting exponential.

When $c \, (>0)$ is an arbitrary $\epsilon$ independent constant, then $\gamma$ is given as
\begin{equation}
    \gamma = \frac{\sqrt{-1+\sqrt{1+4c\epsilon^2}}}{2\sqrt{2}\epsilon} \, .
\label{eqn:gammac}
\end{equation}
For $\epsilon = 0$, $\gamma = \sqrt{c}/2$. The wave number $k$ is given in equation (\ref{eqn:kappa}),
$$k = \sqrt{1+4\gamma^2\epsilon^2} \, ,$$
where $\gamma$ is now given in equation (\ref{eqn:gammac}). Since there is $k/\gamma$ factor in the argument of exponential, we expand this ratio by using some algebraic manipulation software such as \emph{Mathematica},
\begin{equation}
    \frac{k}{\gamma} = \frac{2}{\sqrt{c}}\left[1+ c \ \epsilon^2-c^2\epsilon^4+2c^3\epsilon^6-5c^4\epsilon^8+\mathcal{O}(\epsilon^{10})\right]  \ .
\label{eqn:kbyg1}
\end{equation}
Since we are interested in the minimal oscillatory tail amplitude  and the third derivative of a function at the origin up to $\mathcal{O}(\epsilon^5)$, we have to expand the $k/\gamma$ at least up to $\mathcal{O}(\epsilon^6)$ order because of the presence of $\epsilon$ in the denominator in the argument of the exponential term. Using equation (\ref{eqn:kbyg1}), we can expand the exponential term as
\begin{align}
    \exp{\left(-\frac{\pi k}{2\gamma\epsilon}\right)}  =& \exp{\left(-\frac{\pi}{\sqrt{c} \, \epsilon}\right)}\left[1+\frac{1}{2}c\pi^2\epsilon^2-\frac{1}{6}c^{3/2}\left(\pi^2-6\right)\pi\epsilon^3+\frac{1}{24}c^2\left(\pi^2-24\right)\pi^2\epsilon^4 \right. \nonumber \\& \left. -\frac{1}{120}c^{5/2}\left(\pi^4-60\pi^2+240\right)\pi\epsilon^5+\mathcal{O}(\epsilon^6)\right] \label{eqn:expc} \, .
\end{align}

\section{Oscillatory tail amplitude}

Using equation (\ref{eqn:expc}) in $\alpha_m^{(k,5)}$, we get the oscillatory tail amplitude $\alpha_m$ in terms of the parameter independent wave speed $c$. The result up to $\mathcal{O}(\epsilon^5)$ order is given as
\begin{align}
    \alpha^{(c,5)}_m =& -\frac{\pi K}{\epsilon^2}\exp{\left(-\frac{\pi}{\sqrt{c} \, \epsilon}\right)}\left[1-c^{1/2}\pi\epsilon+\frac{1}{4}c\left(2\pi^2-5\right)\epsilon^2 \right. \nonumber \\& \left. -\frac{1}{12}c^{3/2}\left(2\pi^2-27\right)\pi\epsilon^3+\frac{1}{48}c^2\left(2\pi^4-78\pi^2-3(\zeta_4-20)\right)\epsilon^4 \right. \nonumber \\& \left. -\frac{1}{240}c^{5/2}\left(2\pi^4-170\pi^2-15(\zeta_4-72)\right)\pi\epsilon^5+\mathcal{O}(\epsilon^6)\right] \ ,
    \label{eqn:alphamarbc}
\end{align}
where $K$ and $\zeta_4$ are given in equations (\ref{eqn:Kvalue}) and (\ref{eqn:A2}) respectively.

\section{Third derivative at the origin}

Using equation (\ref{eqn:expc}) in $\partial^3_xu^{(k,5)}_-$, we get the third derivative of the asymmetric function at the origin in terms of the parameter independent wave speed $c$. The result up to $\mathcal{O}(\epsilon^5)$ order is given as
\begin{align}
    \partial^3_xu^{(c,5)}_-(0) =& -\frac{\pi K}{\epsilon^5}\exp{\left(-\frac{\pi}{\sqrt{c} \, \epsilon}\right)}\left[1-c^{1/2}\pi\epsilon+\frac{1}{2}c\left(\pi^2-5\right)\epsilon^2 \right. \nonumber \\& \left. +\frac{1}{6}c^{3/2}(\pi^2-21)\pi\epsilon^3+\frac{1}{48}c^2\left(2\pi^4-108\pi^2-3(\zeta_4-221\right)\epsilon^4 \right. \nonumber \\& \left. -\frac{1}{240}c^{5/2}\left(2\pi^4-220\pi^2-15(\zeta_4-149)\right)\pi\epsilon^5+\mathcal{O}(\epsilon^6)\right] \ ,
    \label{eqn:3derc}
\end{align}
where $K$ and $\zeta_4$ are given in equations (\ref{eqn:Kvalue}) and (\ref{eqn:A2}) respectively.

\newpage

\addcontentsline{toc}{chapter}{Bibliography}
\bibliographystyle{abbrv}
\bibliography{references}

\begin{thebibliography}{10}

\bibitem{asym}
E.~Benilov, R.~Grimshaw, and E.~Kuznetsova.
\newblock The generation of radiating waves in a singularly-perturbed
  {Korteweg-de Vries} equation.
\newblock {\em Physica D: Nonlinear Phenomena}, 69(3-4):270--278, 1993.

\bibitem{boyd1}
J.~P. Boyd.
\newblock Weakly non-local solitons for capillary-gravity waves: fifth-degree
  {Korteweg-de Vries} equation.
\newblock {\em Physica D: Nonlinear Phenomena}, 48(1):129--146, 1991.

\bibitem{boyd2}
J.~P. Boyd.
\newblock Multiple precision pseudospectral computations of the radiation
  coefficient for weakly nonlocal solitary waves: Fifth-order {Korteweg--De
  Vries} equation.
\newblock {\em Computers in Physics}, 9(3):324--334, 1995.

\bibitem{boydcheb}
J.~P. Boyd.
\newblock {\em Chebyshev and Fourier spectral methods}.
\newblock Courier Corporation, 2001.

\bibitem{boyd}
J.~P. Boyd.
\newblock {\em Weakly nonlocal solitary waves and beyond-all-orders
  asymptotics: generalized solitons and hyperasymptotic perturbation theory},
  volume 442.
\newblock Springer Science \& Business Media, 2012.

\bibitem{solitons}
T.~Dauxois and M.~Peyrard.
\newblock {\em Physics of solitons}.
\newblock Cambridge University Press, 2006.

\bibitem{jager}
E.~De~Jager.
\newblock On the origin of the {Korteweg-de Vries} equation.
\newblock {\em arXiv preprint math/0602661}, 2006.

\bibitem{Muneeb}
G.~Fodor, P.~Forg\'acs, and M.~Mushtaq.
\newblock Higher order corrections to beyond-all-order effects in a fifth order
  {Korteweg--de Vries} equation.
\newblock {\em Phys. Rev. D}, 107:105002, May 2023.

\bibitem{gard}
C.~S. Gardner, J.~M. Greene, M.~D. Kruskal, and R.~M. Miura.
\newblock Method for solving the {Korteweg-de Vries} equation.
\newblock {\em Physical review letters}, 19(19):1095, 1967.

\bibitem{gaj}
R.~Grimshaw and N.~Joshi.
\newblock Weakly nonlocal solitary waves in a singularly perturbed
  {Korteweg--de Vries} equation.
\newblock {\em SIAM Journal on Applied Mathematics}, 55(1):124--135, 1995.

\bibitem{CLN}
B.~Haible and R.~Kreckel.
\newblock Cln, a class library for numbers.
\newblock {\em URL: http://www. ginac. de/CLN}, 21, 1996.

\bibitem{hs}
J.~K. Hunter and J.~Scheurle.
\newblock Existence of perturbed solitary wave solutions to a model equation
  for water waves.
\newblock {\em Physica D: Nonlinear Phenomena}, 32(2):253--268, 1988.

\bibitem{arb}
F.~Johansson.
\newblock Arb: efficient arbitrary-precision midpoint-radius interval
  arithmetic.
\newblock {\em IEEE Transactions on Computers}, 66(8):1281--1292, 2017.

\bibitem{lax}
P.~D. Lax.
\newblock Integrals of nonlinear equations of evolution and solitary waves.
\newblock {\em Communications on pure and applied mathematics}, 21(5):467--490,
  1968.

\bibitem{miura}
R.~M. Miura.
\newblock {Korteweg-de Vries} equation and generalizations. {I}. a remarkable
  explicit nonlinear transformation.
\newblock {\em Journal of Mathematical Physics}, 9(8):1202--1204, 1968.

\bibitem{miura1}
R.~M. Miura.
\newblock The {Korteweg--de Vries} equation: a survey of results.
\newblock {\em SIAM review}, 18(3):412--459, 1976.

\bibitem{benderorszag}
S.~Orszag and C.~M. Bender.
\newblock {\em Advanced mathematical methods for scientists and engineers}.
\newblock McGraw-Hill New York, 1978.

\bibitem{prg}
Y.~Pomeau, A.~Ramani, and B.~Grammaticos.
\newblock Structural stability of the {Korteweg-de Vries} solitons under a
  singular perturbation.
\newblock {\em Physica D: Nonlinear Phenomena}, 31(1):127--134, 1988.

\bibitem{numeric}
W.~H. Press, S.~A. Teukolsky, W.~T. Vetterling, and B.~P. Flannery.
\newblock {\em Numerical recipes 3rd edition: The art of scientific computing}.
\newblock Cambridge university press, 2007.

\bibitem{kruskal1987}
H.~Segur and M.~Kruskal.
\newblock Nonexistence of small-amplitude breather solutions in $\varphi^4$
  theory.
\newblock {\em Phys. Rev. Lett}, 58(8):747--750, 1987.

\bibitem{steinruck}
H.~Steinr{\"u}ck.
\newblock Asymptotic methods in fluid mechanics: Survey and recent advances.
\newblock {\em CISM Courses and Lectures, vol. 523}, 2012.

\bibitem{sun}
S.~Sun.
\newblock On the oscillatory tails with arbitrary phase shift for solutions of
  the perturbed {Korteweg--de Vries} equation.
\newblock {\em SIAM Journal on Applied Mathematics}, 58(4):1163--1177, 1998.

\bibitem{yang}
J.~Yang.
\newblock Dynamics of embedded solitons in the extended {Korteweg--de Vries}
  equations.
\newblock {\em Studies in Applied Mathematics}, 106(3):337--365, 2001.

\bibitem{zabusky1}
N.~J. Zabusky.
\newblock Solitons and bound states of the time-independent {Schr{\"o}dinger}
  equation.
\newblock {\em Physical review}, 168(1):124, 1968.

\bibitem{zabusky}
N.~J. Zabusky and M.~D. Kruskal.
\newblock Interaction of "solitons" in a collisionless plasma and the
  recurrence of initial states.
\newblock {\em Physical review letters}, 15(6):240, 1965.

\end{thebibliography}

\end{document}